\documentstyle[aps,prl,epsfig,preprint,floats]{revtex}
\begin{document}
\epsfysize3cm
\epsfbox{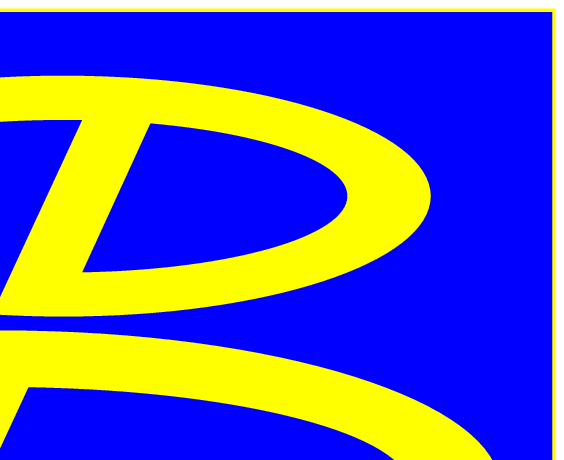}

\vskip -3cm
\noindent
\hspace*{12cm}KEK Preprint 2002-70 \\
\hspace*{12cm}Belle Preprint 2002-24 \\

\vskip 1cm

\renewcommand{\baselinestretch}{1.2}

%

\begin{center}
{\Large\bf
Charmless Hadronic Two-Body $B$ Meson Decays
\footnote{Submitted to PRD.}}
\vskip 0.5cm
The Belle Collaboration
\vskip 0.2cm
\end{center}

\begin{abstract}
We report the results of a study of two-body $B$ meson decays to the
 complete set of $K\pi$, $\pi\pi$, and $K\bar{K}$ final
 states. The study is performed on a
data sample of $31.7 \pm 0.3$ million $B\bar{B}$ events recorded on
 the $\Upsilon$(4S) resonance by the Belle experiment at KEKB.  
We observe significant signals in all $K\pi$ final states and in the
 $\pi^+\pi^-$ and $\pi^+\pi^0$ final states. We set limits on the
 $\pi^0\pi^0$ and $K\bar{K}$ final states.  A search is performed for
 partial-rate asymmetries between conjugate states for flavor-specific
 final states.
\end{abstract}

\pacs{PACS numbers: 13.25.Hw, 11.30.Er, 14.40.Nd, 12.15.Hh}

{\normalsize

\centerline{(The Belle Collaboration)}

\begin{center}
  B.~C.~K.~Casey$^{8}$,       
  K.~Abe$^{9}$,               
  T.~Abe$^{44}$,              
  I.~Adachi$^{9}$,            
  Byoung~Sup~Ahn$^{16}$,      
  H.~Aihara$^{45}$,           
  M.~Akatsu$^{23}$,           
  Y.~Asano$^{50}$,            
  T.~Aso$^{49}$,              
  V.~Aulchenko$^{2}$,         
  T.~Aushev$^{13}$,           
  A.~M.~Bakich$^{40}$
% ...
  M.~Yokoyama$^{45}$,         
  Y.~Yuan$^{11}$,             
  Y.~Yusa$^{44}$,             
  Z.~P.~Zhang$^{37}$,         
  V.~Zhilich$^{2}$,           
and
  D.~\v Zontar$^{50}$         
\end{center}

\small
\begin{center}
% ...
$^{2}${Budker Institute of Nuclear Physics, Novosibirsk}\\
% ...
$^{8}${University of Hawaii, Honolulu HI}\\
$^{9}${High Energy Accelerator Research Organization (KEK), Tsukuba}\\
% ...
$^{11}${Institute of High Energy Physics, Chinese Academy of Sciences,
Beijing}\\
% ...
$^{13}${Institute for Theoretical and Experimental Physics, Moscow}\\
% ...
$^{16}${Korea University, Seoul}\\
% ...
$^{23}${Nagoya University, Nagoya}\\
% ...
$^{37}${University of Science and Technology of China, Hefei}\\
% ...
$^{40}${University of Sydney, Sydney NSW}\\
% ...
$^{44}${Tohoku University, Sendai}\\
$^{45}${University of Tokyo, Tokyo}\\
% ...
$^{49}${Toyama National College of Maritime Technology, Toyama}\\
$^{50}${University of Tsukuba, Tsukuba}\\
% ...
\end{center}

\normalsize

}

\normalsize

\maketitle
%
%
%

\section{Introduction}

There exists a gold mine of weak and hadronic physics in two-body $B$ meson
decays to the $K\pi$, $\pi\pi$, and $K\bar{K}$ final states.  Indeed, if one
assumes unitarity of the quark mixing matrix~\cite{bib:km}, 
these modes contain enough information to measure
all angles of the Unitarity Triangle~\cite{bib:quinn}.  Methods to
extract weak-sector physics
from these decays are complicated by hadronic
uncertainties. However, if enough final states are measured, we will
have sufficient information to constrain the sizes of hadronic
amplitudes and strong phases, a necessity in disentangling the unitarity
angles from measurements of flavor asymmetries and the relative
size of the partial widths among these modes~\cite{bib:fleischer,bib:gronau,bib:beneke,bib:keum,bib:hou,bib:atwood,bib:agashe,bib:gronau2}.

We have previously reported measurements of, or 
limits on, the branching fractions of $B$ mesons
to the $K\pi$, $\pi\pi$, and $K\bar{K}$ final states excluding the
$\pi^0\pi^0$ and $K^0\bar{K}^0$ final states~\cite{bib:kazuhito} as well as a search for charge
asymmetries in the flavor-specific $K\pi$ final
states~\cite{bib:casey},
 based on a data
sample of $11.1$ million $B\bar{B}$ events.  The results presented
here include the previous data and supersede all previous
results. Similar studies have been performed by other
experiments~\cite{bib:aleph,bib:delphi,bib:cleo1,bib:cleo2,bib:sld,bib:babar1,bib:cleo3,bib:babar2}.  

Here, we present measurements of, or limits on, the branching fractions of $B$
mesons to the $K\pi$, $\pi\pi$, and $K\bar{K}$ final states including all
combinations of
charged and neutral kaons and pions. We refer to these final states
collectively as $B\rightarrow hh$, including charge conjugate states
unless explicitly stated.  For final states where the charge of the
kaon or pion specifies the flavor of the parent $B$ meson, known as
flavor-specific final states, 
we present limits on the partial-rate asymmetries defined as
 $${\cal A}_{\rm CP}(f) = {N(\bar{B}\rightarrow \bar{f}) -
N(B\rightarrow f)  \over N(\bar{B}\rightarrow \bar{f}) +
N(B\rightarrow f) },$$
where $B$ represents either a $B^0$ or $B^+$ meson, $f$ represents a
flavor-specific final state, and $\bar{B}$ and $\bar{f}$ are their
conjugates.  

\section{Apparatus and Data Set}

The analysis is based on data taken by the Belle
detector~\cite{bib:belle}  at the KEKB
$e^+e^- $ storage ring~\cite{bib:kekb}.  The data set consists of $29.1$ $\rm fb^{-1}$ on
the $\Upsilon$(4S) resonance corresponding to $31.7\pm 0.3$ million $B\bar{B}$
events.  An off-resonance data set of $4.4$ $\rm fb^{-1}$ was taken $60$ MeV
below the $\Upsilon$(4S) resonance to perform systematic studies of the
continuum $e^+e^-\rightarrow q\bar{q}$ background where $q$ is either
a $u$, $d$, $s$, or $c$ quark.
KEKB collides $8$ GeV electrons and $3.5$ GeV positrons that are
stored in separate
rings, producing an
$\Upsilon$(4S) system that is boosted by $\gamma\beta = 0.425$ along the beam axis.
In this analysis, all variables are calculated in the center-of-mass
frame of the electron and positron beams unless explicitly stated.

The Belle detector is a general purpose magnetic spectrometer with a
$1.5$ T axial magnetic field.  Charged tracks are reconstructed using a $50$
layer central drift chamber (CDC) and a $3$ layer double-sided Silicon vertex
vetector (SVD). 
Candidate electrons and photons are identified using an $8736$ crystal
CsI(T{\it l}) calorimeter (ECL) inside the magnet.  
Muon and $K^0_L$ candidates are identified using resistive plate
chambers embedded in the iron magnetic flux return (KLM). Hadron and
auxiliary lepton identification is provided by an array of $1188$
Silica aerogel \v{C}erenkov threshold counters (ACC) and a barrel of
$128$ time-of-flight (TOF) plastic scintillator modules. 

\section{Event Reconstruction}

Event triggers based on fast signals from the CDC, ECL, TOF, and
KLM~\cite{bib:belle}. 
Hadronic events are selected using event multiplicity and total
energy variables~\cite{bib:chulsu}. For signal events that pass all
$B\rightarrow hh$ selection
criteria, the triggering and hadronic event selection
efficiencies range from $99\%$ for $B^0\rightarrow h^+h^-$ modes to
$76\%$ for the $B^0\rightarrow \pi^0\pi^0$ final state.    

Charged $\pi$ and $K$ mesons are
identified by their energy loss ($dE/dx$) in the CDC and their
\v{C}erenkov light yield in the ACC.  
For each hypothesis ($K$ or $\pi$), 
the $dE/dx$ and ACC probability density functions are
combined to form 
likelihoods, ${\cal L}_K$ and ${\cal L}_\pi$. 
$K$ and $\pi$ mesons are distinguished by a cut
on the likelihood ratio ${\cal L}_K/({\cal L}_K + {\cal
L}_\pi)$. A similar likelihood ratio including calorimeter
information is used to identify electrons.  
 All charged tracks that originate from the interaction point and are
not positively identified as electrons are
considered as kaon or pion candidates.

\begin{figure}[htbp]
\centerline{\epsfysize 2.5
truein\epsfbox{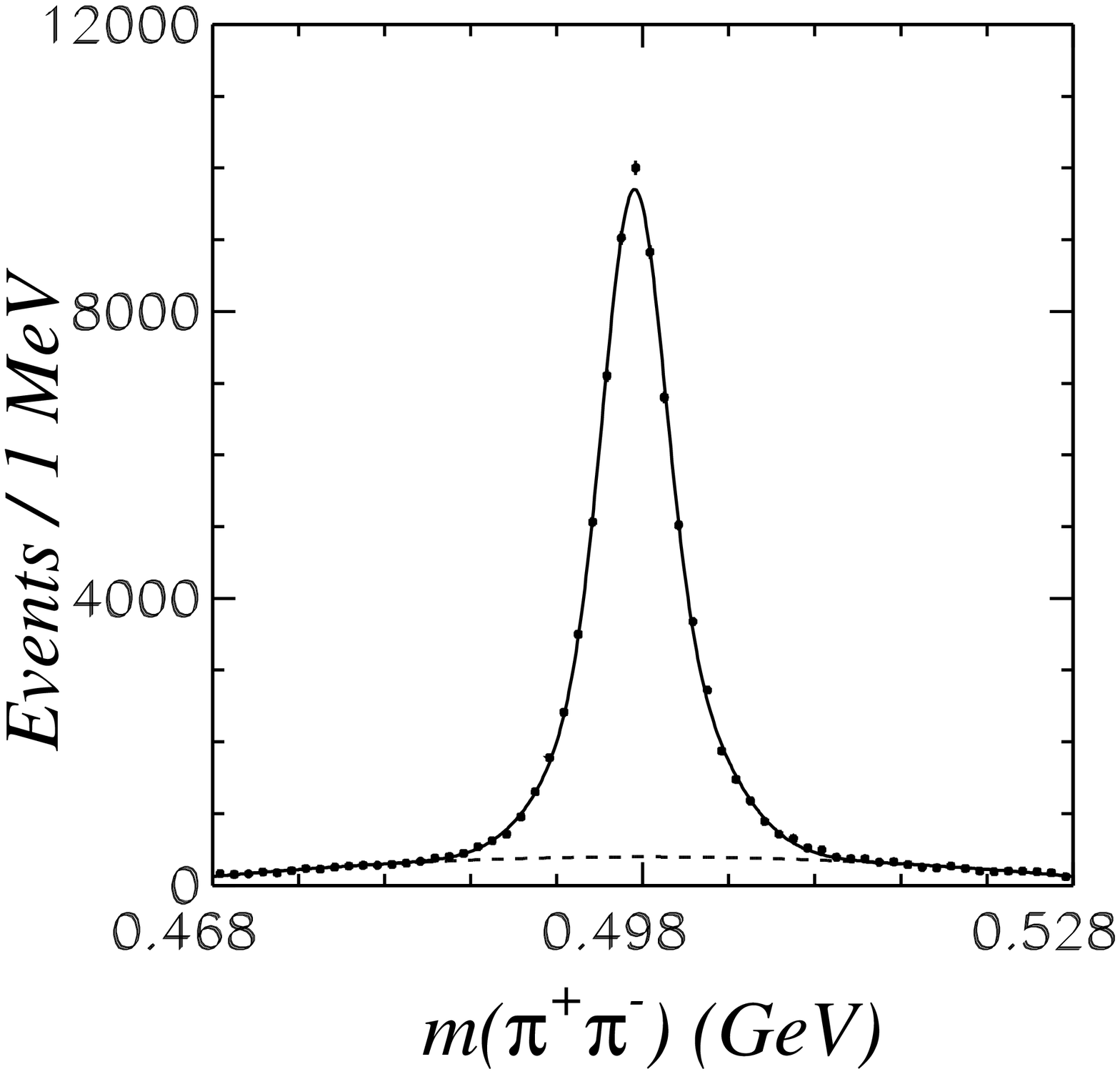}\epsfysize 2.5 truein\epsfbox{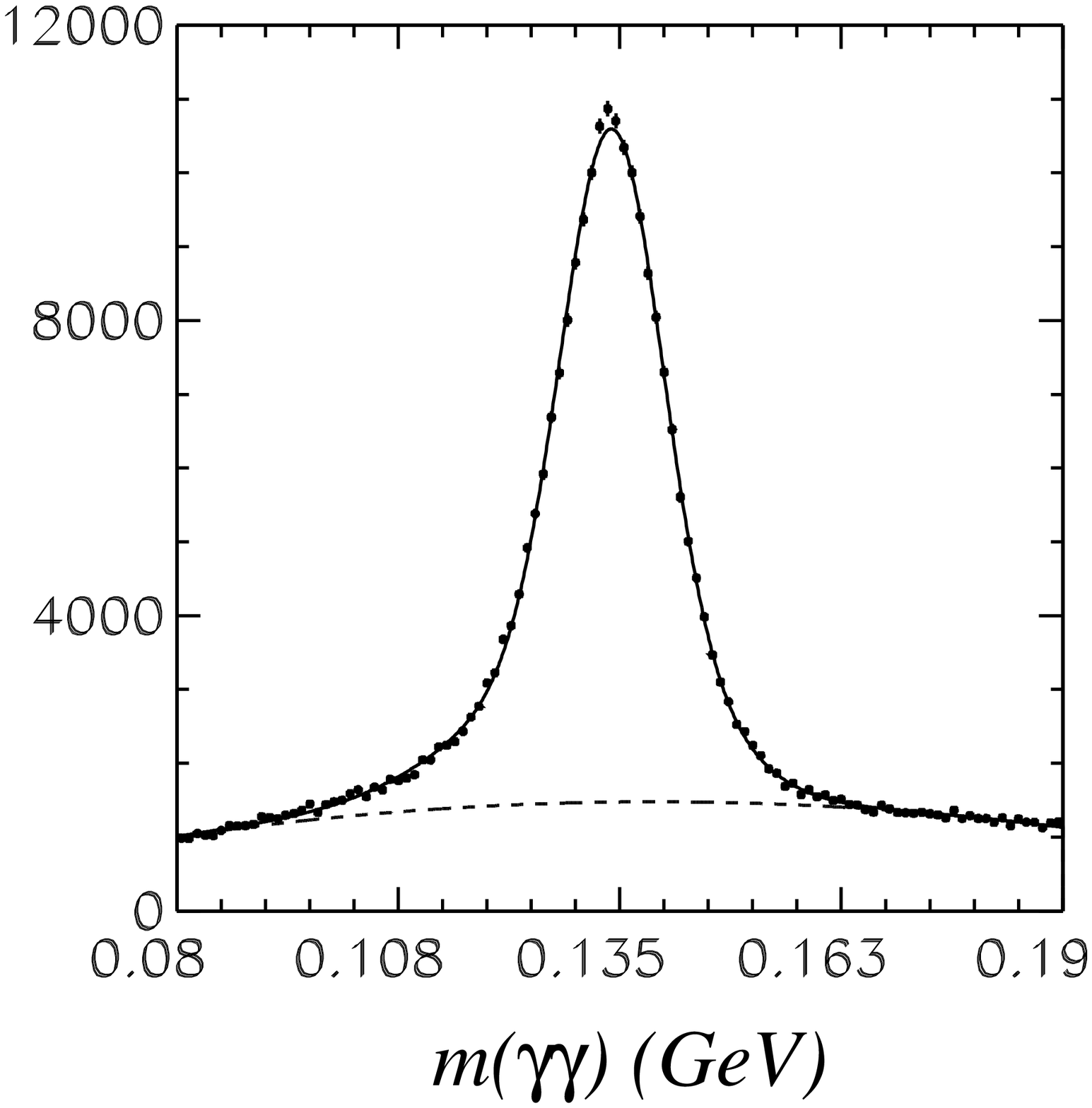} }\nopagebreak
\caption{The $\pi^+\pi^-$ (left) and  $\gamma\gamma$ (right) mass
spectra for $B^+\rightarrow K^0_Sh^+$ and
$h^+\pi^0$ candidates in the beam constrained mass sideband data sample.  The
$\pi^+\pi^-$ distribution is modeled as the sum of two Gaussians for
true $K^0_S$ candidates while the background is modeled as a
second-order 
polynomial. The weighted average resolution of the two Gaussians
is $3.4$ MeV for the $K^0_S$ mass peak. The $\gamma\gamma$
distribution is modeled as the sum of a primary symmetric Gaussian and
a secondary asymmetric Gaussian [24] for $\pi^0$ candidates and a
second-order polynomial for background. The  weighted average
resolution of the two Gaussians is $8.9$ MeV for the $\pi^0$
mass peak.  In both distributions the solid curves are the sum of
signal and background components while the dashed curve is the
background component.}
\label{fig:ks}
\end{figure}

Candidate $K^0_S$ mesons are reconstructed using
pairs of oppositely charged tracks that have an invariant mass in the range
$480$ MeV$< m(\pi^+\pi^-)<516$ MeV~\cite{bib:c}.  The candidate must have a
displaced vertex and flight direction consistent with a $K^0_S$ originating from the interaction point. Candidate $\pi^0$
mesons are formed from pairs of photons with an invariant mass in the range
$ 114$ MeV$< m(\gamma\gamma)< 150$  MeV and $E_{\gamma(\rm
lab)}>70$ MeV.  The $\pi^+\pi^-$ and $\gamma\gamma$ mass spectra are
shown in Fig.~\ref{fig:ks} for $B^+\rightarrow K^0_Sh^+$ and
$h^+\pi^0$ candidates in the beam constrained mass sideband data sample
defined below.

Continuum background is reduced using event shape variables.
We quantify the event topology with modified
Fox-Wolfram moments~\cite{bib:fox} defined as
$$h^{so}_l = \sum_{i,j}{p_ip_jP_l(\cos\theta_{ij})},$$
$$h^{oo}_l = \sum_{j,k}{p_jp_kP_l(\cos\theta_{jk})},$$
where $i$ enumerates $B$ {\it signal} candidate particles ($s$
particles) and  $j$ and $k$ enumerate the {\it remaining} particles in the
event ($o$ particles); $p_i$ is the $i$th particle's momentum, and $P_l(\cos\theta_{ij})$ is the
$l$th Legendre polynomial of the angle $\theta_{ij}$ between particles
$i$ and $j$.
The $h^{so}_l$ terms contain information on the correlation between
the $B$ candidate direction and the direction of the rest of the
event. The odd $h^{oo}_l$ terms partially reconstruct the kinematics
of the other
$B$ in the event while the even terms quantify the sphericity of the
other side of the event.
We create a six-variable Fisher discriminant called the
Super Fox-Wolfram defined as
$$SFW = \sum_{l = 2,4}{\alpha_l\left({h^{so}_l \over h^{so}_0}\right)}
+ \sum_{l = 1-4}{\beta_l\left({h^{oo}_l \over h^{oo}_0}\right)},$$
where $\alpha_l$ and $\beta_l$ are the Fisher coefficients.

\begin{figure}[htbp]
\centerline{\epsfysize 2.5 truein\epsfbox{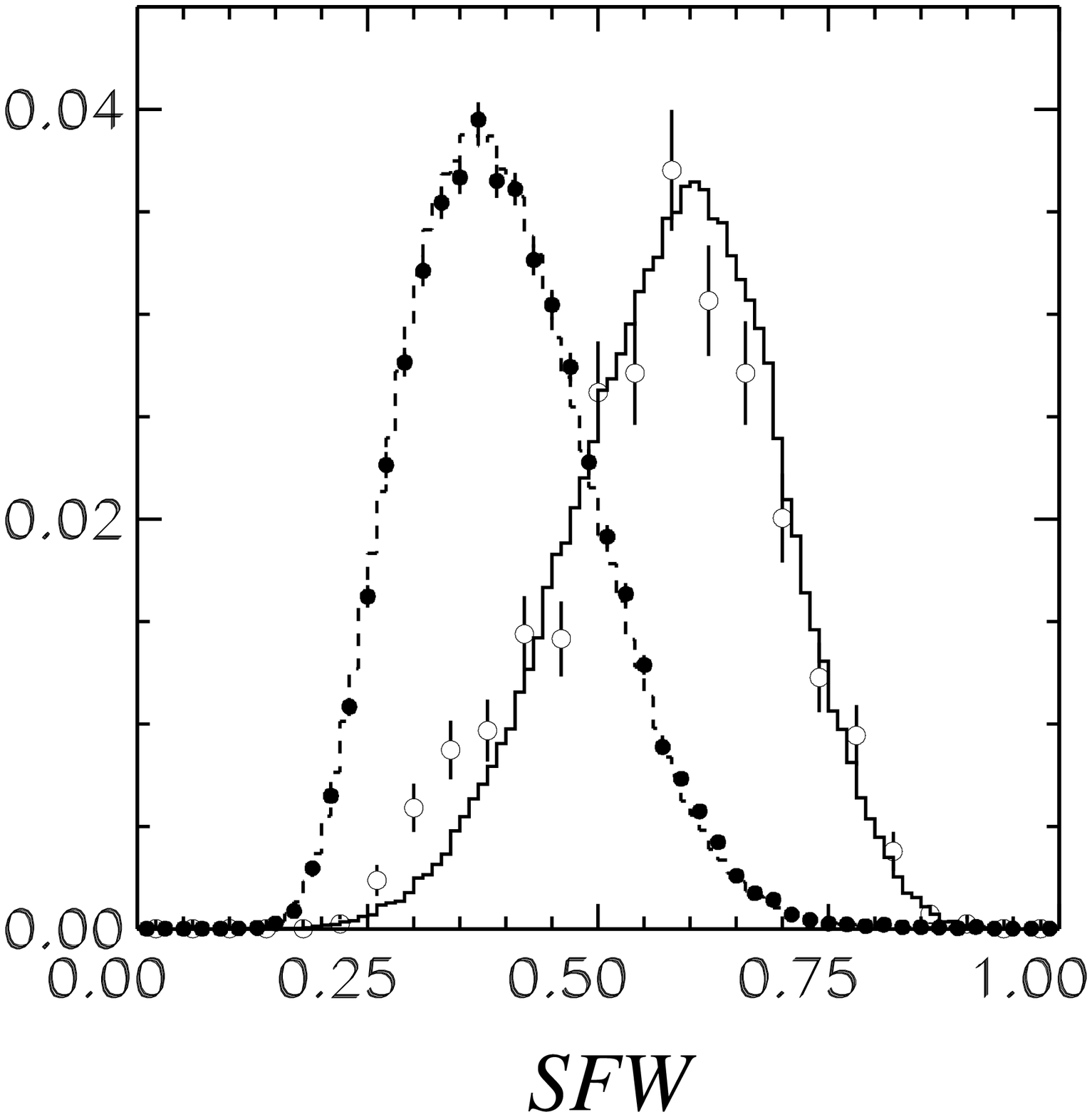}\epsfysize 2.5
truein\epsfbox{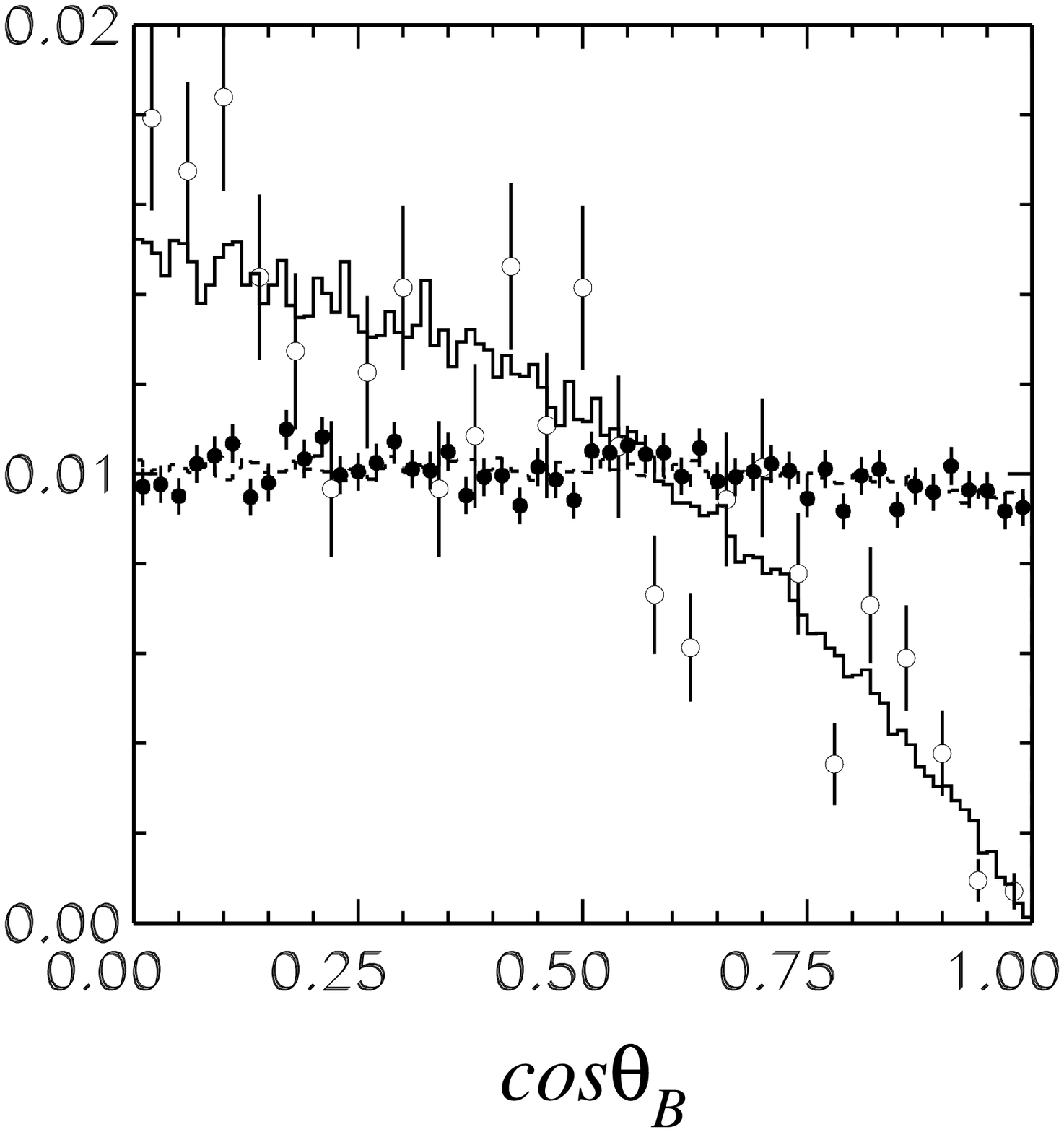}}\nopagebreak
\centerline{\epsfysize 2.5 truein\epsfbox{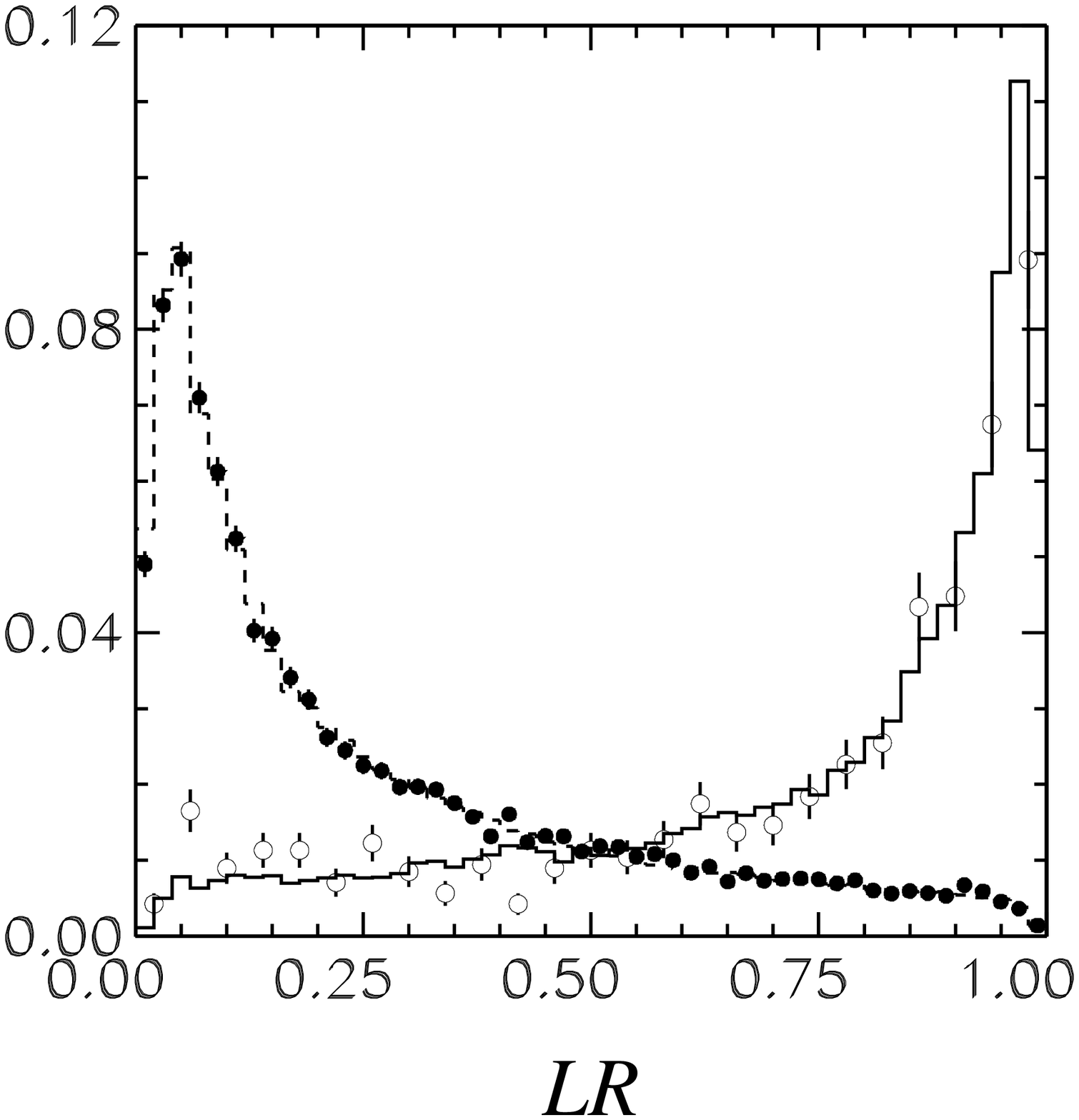}\epsfysize 2.5
truein\epsfbox{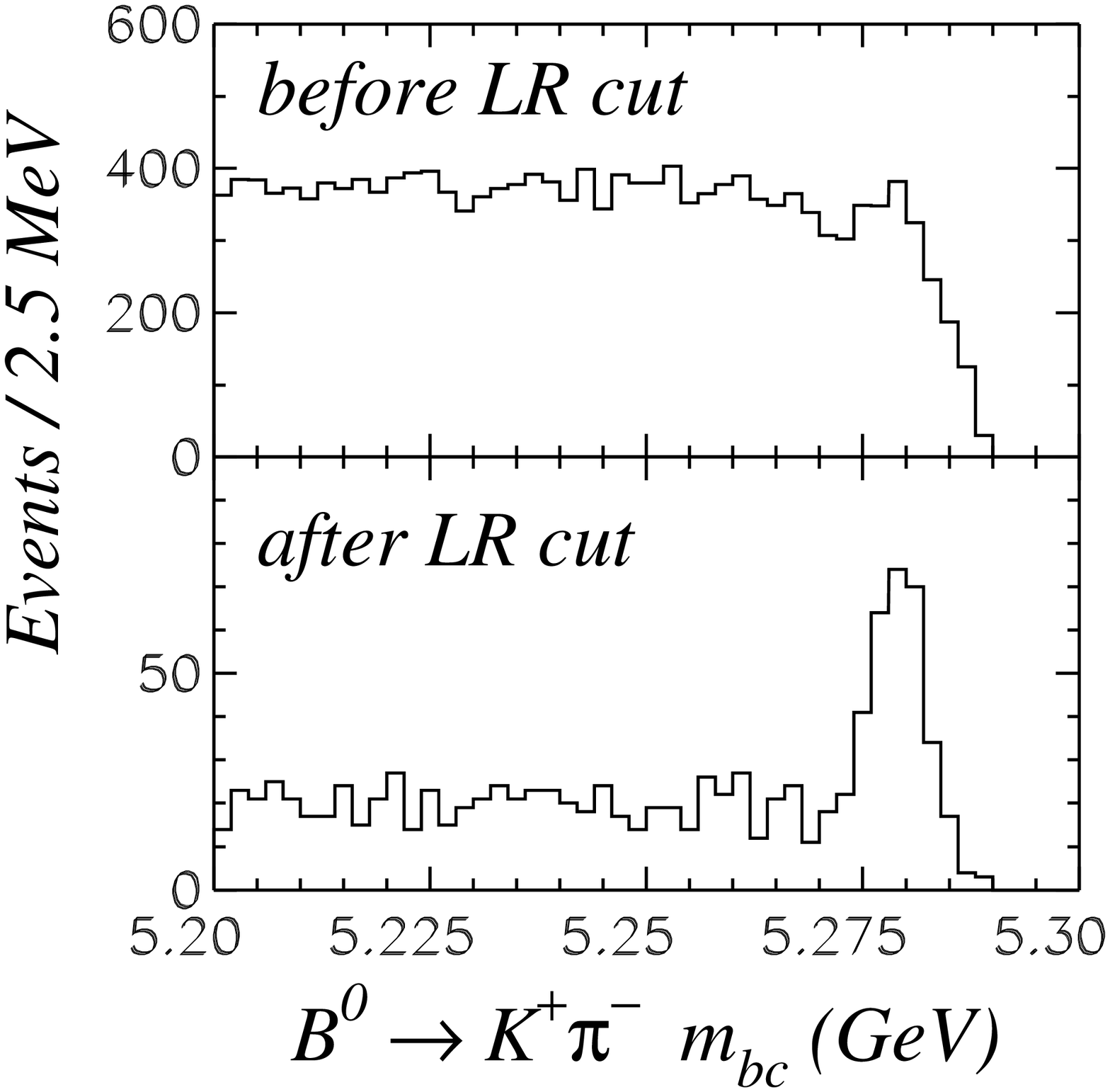} }
\caption{Continuum suppression variables: SFW (top left),
$\cos\theta_B$ (top right), and the combined likelihood ratio (bottom left). The solid curves are the signal
PDFs derived from MC.  The dashed curves are the continuum PDFs derived
from sideband data.  The open points are the $B^+\rightarrow
D^0\pi^+$, $D^0\rightarrow K^-\pi^+$ data sample.  The solid points
are off-resonance data.  The bottom right distribution is the beam
constrained mass distribution for the $B^0\rightarrow K^+\pi^-$ data
sample before and after requiring the likelihood ratio cut $LR>0.8$.}
\label{fig:qq}
\end{figure}

The $SFW$ variable is combined with the $B$ flight direction
with respect to the beam axis, $\cos\theta_B$, to form a single
likelihood
$${\cal L}_{B\bar{B}} = {\cal L}(SFW)_{B\bar{B}} \times
{\cal L}(\cos\theta_B)_{B\bar{B}}$$
 for signal and an equivalent
product for continuum, ${\cal L}_{q\bar{q}}$.  Continuum background is
suppressed by cutting on the likelihood ratio 
$$LR = { {\cal
L}_{B\bar{B}} \over {\cal L}_{B\bar{B}} + {\cal L}_{q\bar{q}}}.$$ 
These variables are shown in Fig.~\ref{fig:qq}.
The signal probability density functions (PDF) are derived from Monte
 Carlo (MC);
the continuum PDFs are taken from sideband data discussed below.
 The $SFW$ PDFs are modeled as the sum of a simple Gaussian and an asymmetric
 Gaussian~\cite{bib:asymmetric}
for both signal and continuum; the $\cos\theta_B$ PDF is modeled as a second-order polynomial for signal and is flat for continuum. 
We make separate requirements on $LR$ for each mode depending
on the expected background determined using sideband
data. As an example, Fig.~\ref{fig:qq} shows the  $B^0\rightarrow K^+\pi^-$ data
 sample before and after imposing the $LR>0.8$ requirement.
 
Table~\ref{tab:efficiency} lists the reconstruction, particle
identification, and continuum suppression efficiencies for each final state~\cite{bib:kskl}. 
The reconstruction and continuum suppression efficiencies are
determined using a GEANT-based MC~\cite{bib:geant}. 
The error in the reconstruction efficiencies are determined by
 embedding MC generated particles into hadronic event data
and comparing the efficiencies between the embedded events
and the default MC 
and also by measuring the relative yields of $D$ decays to various final states.
The charged track, $\pi^0$, and $K^0_S$ selection criteria efficiencies 
are tested by measuring the $D$ event yields before and after each cut is applied.
Further comparisons are made between kinematic distributions of
particles in sideband data (discussed below) and continuum MC events.  
Based on the results of
these studies, we assign a relative systematic error in the
reconstruction efficiencies of $2.5\%$ for
charged tracks, $6.3\%$ for $K^0_S$ mesons and $7.3\%$ for $\pi^0$ mesons.
The relative systematic error associated with the
continuum suppression cut is $4\%$ which is determined 
by taking the ratio of $B^+\rightarrow D^0\pi^+$ yields in data after
and before continuum suppression is applied and comparing to the MC efficiency.
The $B^0\rightarrow \pi^0\pi^0$ final state
includes an additional relative systematic error of $10\%$ to account for difficulties in
triggering and hadronic event selection for this mode.

\begin{figure}[htb]
\centerline{\epsfysize 2.5 truein\epsfbox{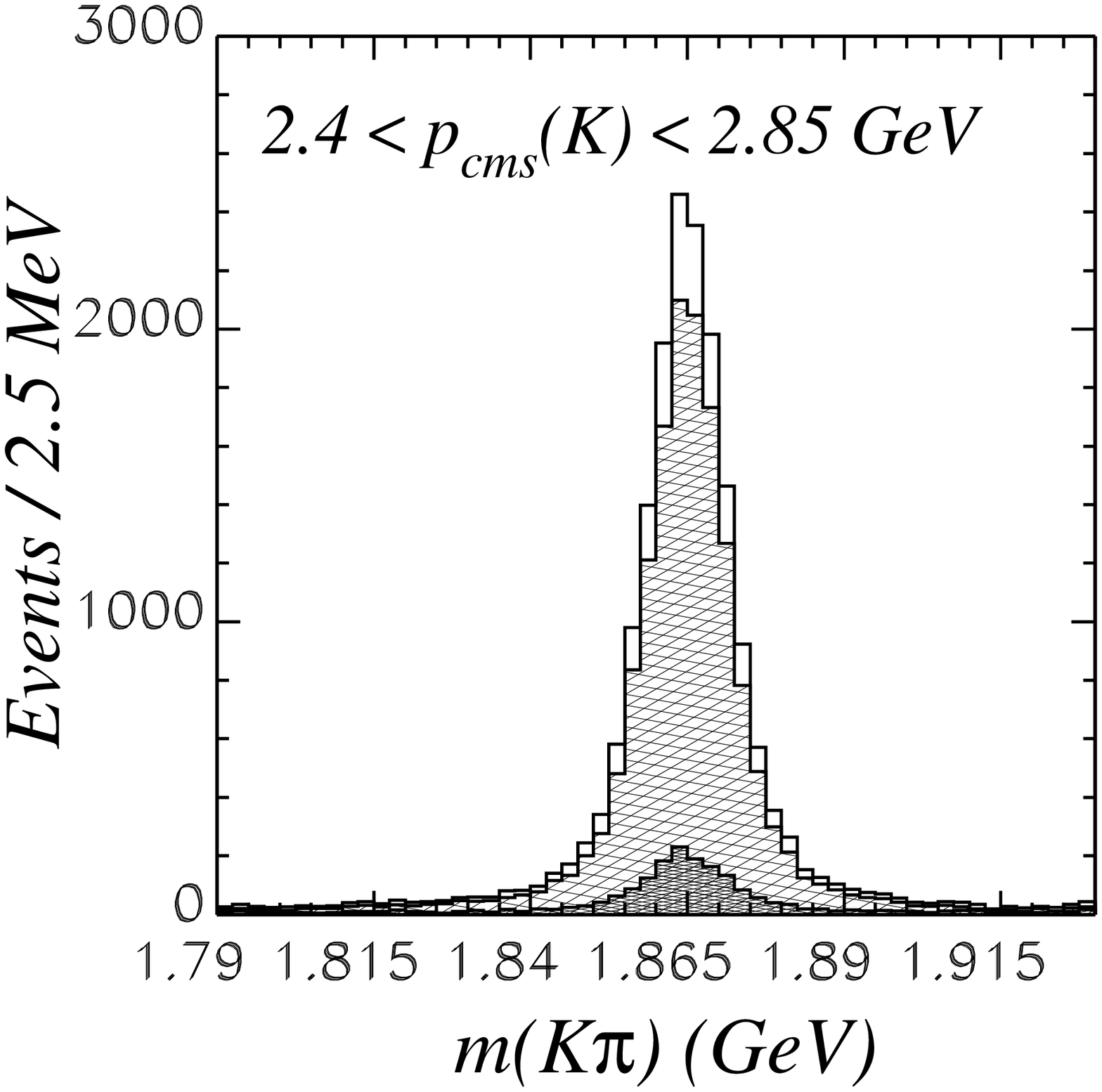}\epsfysize 2.5
truein\epsfbox{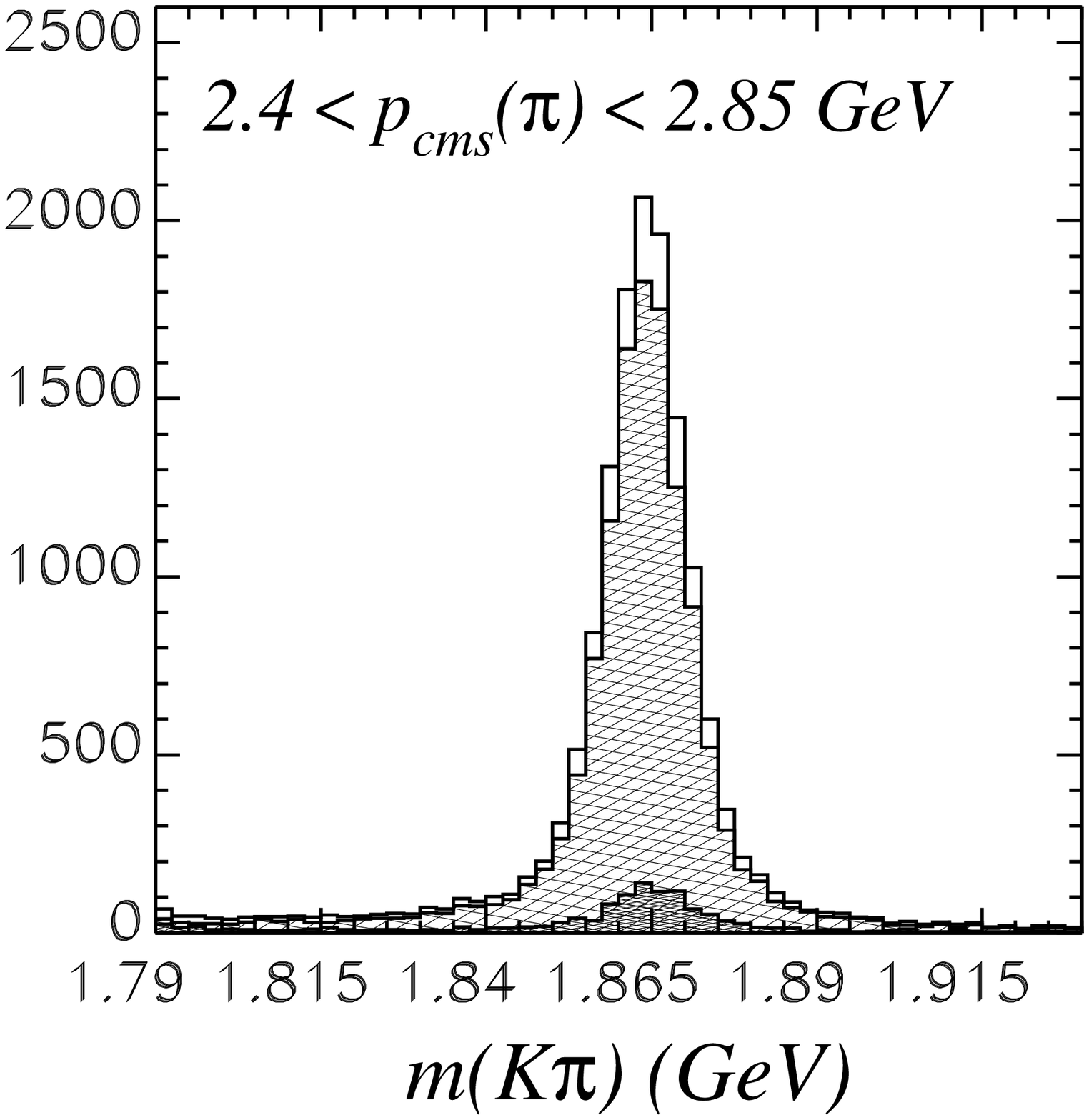}}
\centerline{\epsfysize 2.5 truein\epsfbox{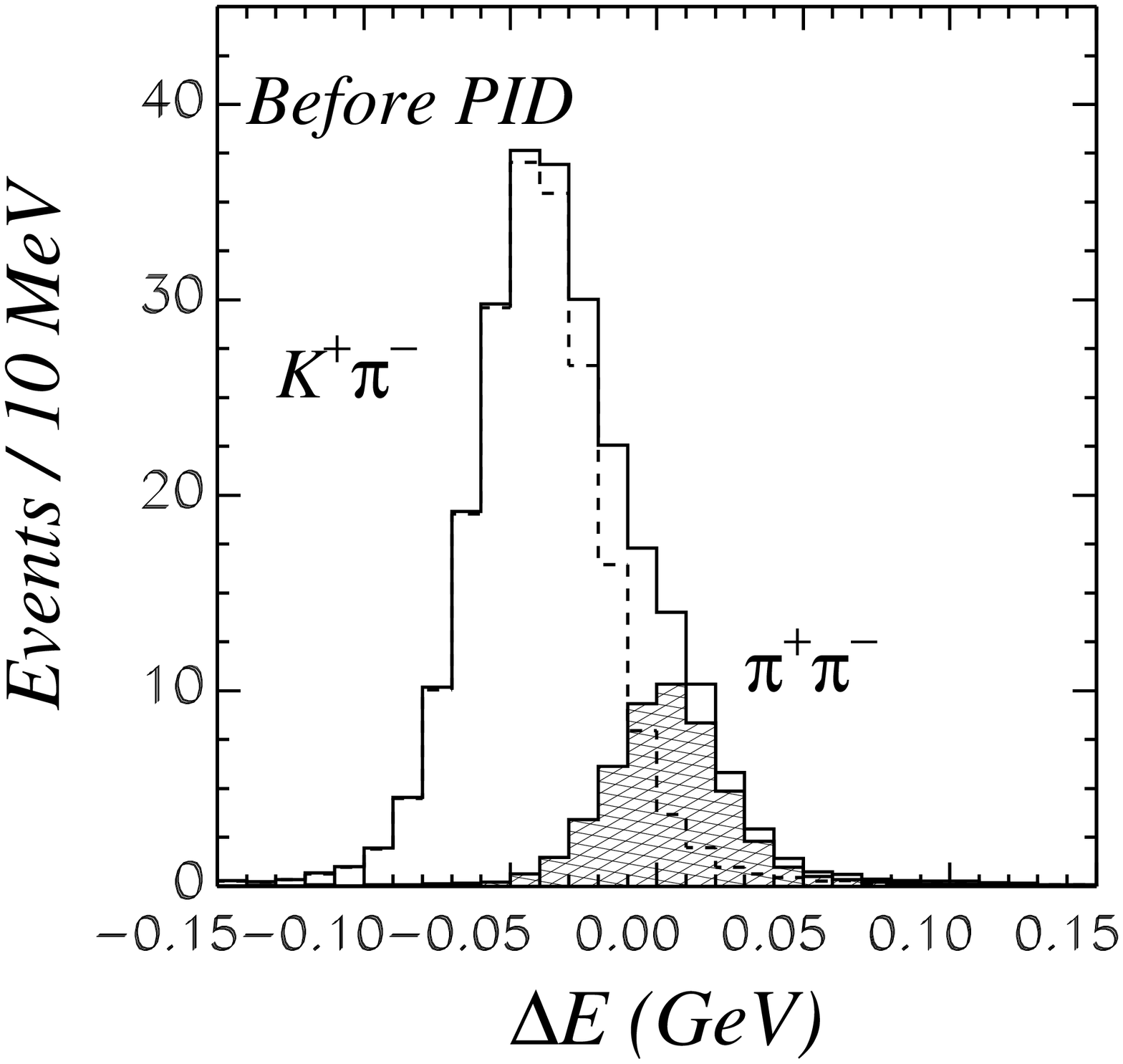}\epsfysize 2.5
truein\epsfbox{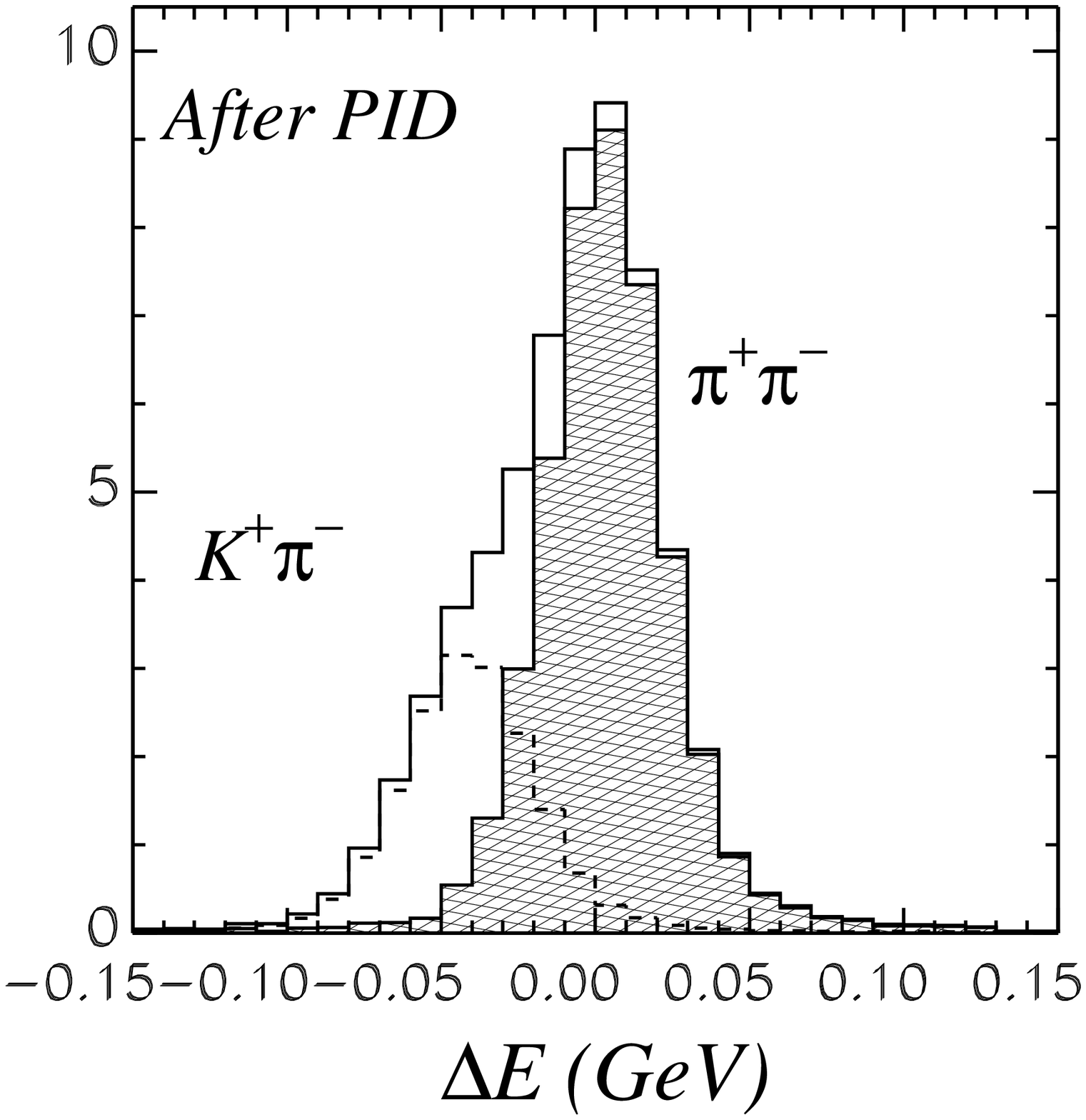}}\nopagebreak
\caption{The $D^{*+} \rightarrow D^0\pi^+$, $D^0
\rightarrow K^-\pi^+$ decay sample used to determine the PID
efficiencies and fake rates, shown in the top row. The distribution on
the left is the $D$ mass distribution where
the kaon daughter is required to have a similar momentum/$\cos\theta$
distribution as the two-body $B\rightarrow hh$ decay daughters.  The open histogram contains
all candidates.  The hatched histogram contains events where the kaon
passes PID cuts.  The solid histogram contains events where the kaon is
mis-identified as a pion.  The distribution on the right is the
corresponding figure for the pion daughter. The bottom row shows the combined
$B^0\rightarrow K^+\pi^-$ and $\pi^+\pi^-$ MC $\Delta E$ distribution
assuming a $4$:$1$ $K^+\pi^-$:$\pi^+\pi^-$ production ratio.  The
distribution on the left is before PID cuts are applied.  In the
distribution on the right, both tracks are required to be identified
as pions.  The solid histogram is the sum of $K^+\pi^-$ and
$\pi^+\pi^-$, the dashed histogram is the $K^+\pi^-$ component, and
the hatched histogram is the $\pi^+\pi^-$ component.}
\label{fig:showpid}
\end{figure}

A critical feature of the analysis is the measurement of the particle
identification efficiency and fake rate.  
These are determined using nearly pure samples of $K$ and $\pi$ mesons 
tagged using the
continuum $D^{*+}$ decay chain $D^{*+} \rightarrow D^0\pi^+$, $D^0
\rightarrow K^-\pi^+$. Figure~\ref{fig:showpid} shows the $K^-\pi^+$
invariant mass distributions before and after applying PID cuts.
For tracks in the
$B\rightarrow hh$ signal
region of $2.4$ GeV $<p <2.8$ 
GeV ($1.5$ GeV $<p_{ \rm lab}<4.5$  GeV) the $K$ efficiency and
fake rate are  $ \epsilon_K = 0.86$ and $f_K = 0.086$ (true $K$
fakes $\pi$);  the $\pi$ efficiency and fake rate are
$\epsilon_\pi = 0.88$ and $ f_\pi = 0.071$ (true $\pi$ fakes $K$).  The
relative systematic errors are $2\%$ in the efficiencies and $4\%$ in
the fake rates.  These errors are directly related to the sample purity.  Figure~\ref{fig:showpid} also shows
$B^0\rightarrow K^+\pi^-$ and $\pi^+\pi^-$ MC events before and after
applying pion identification cuts on both tracks.

\section{$B$ Reconstruction and Yield Extraction}

\begin{figure}[htb]
\centerline{\epsfysize 2.5 truein\epsfbox{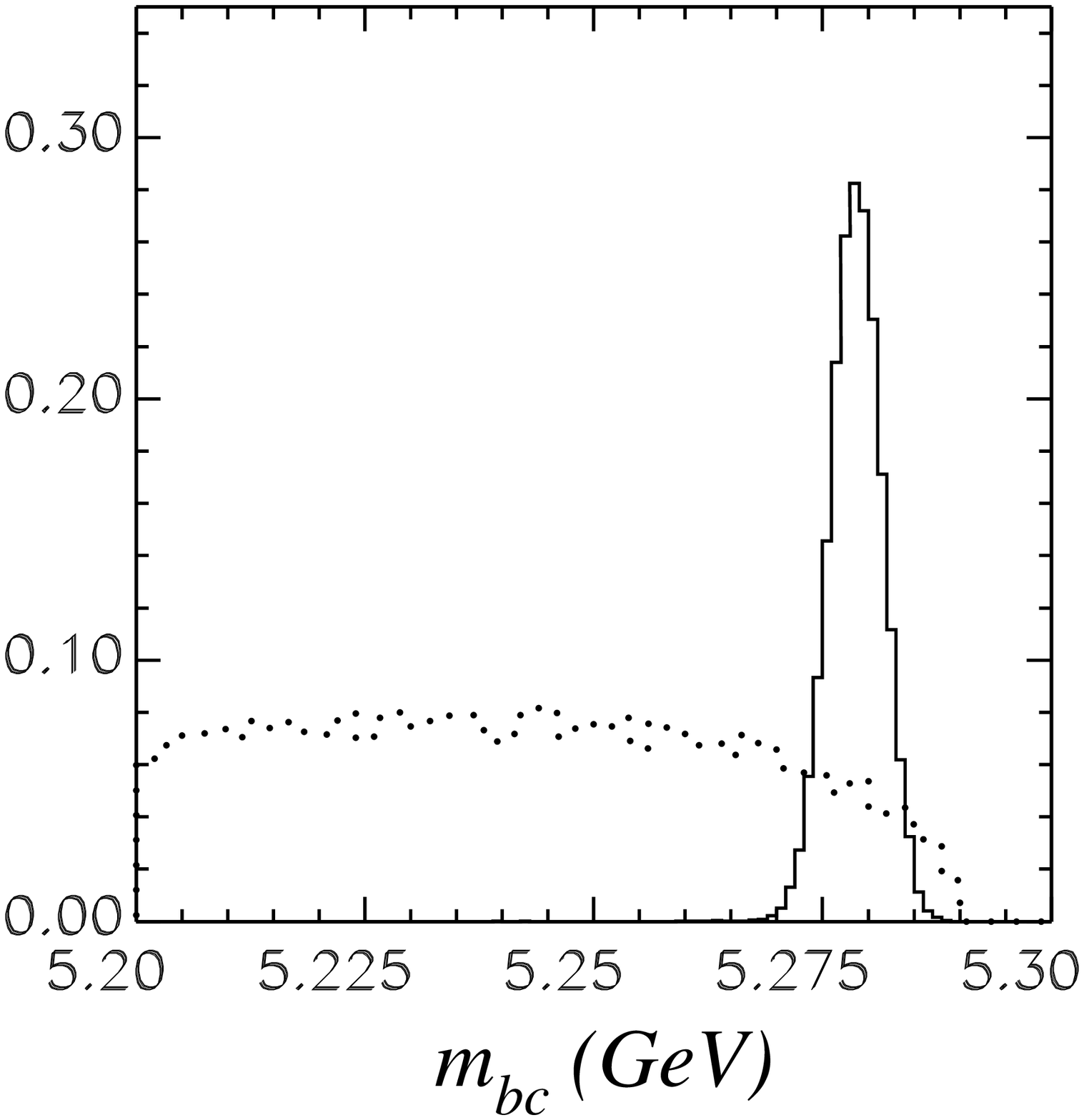}\epsfysize 2.5
truein\epsfbox{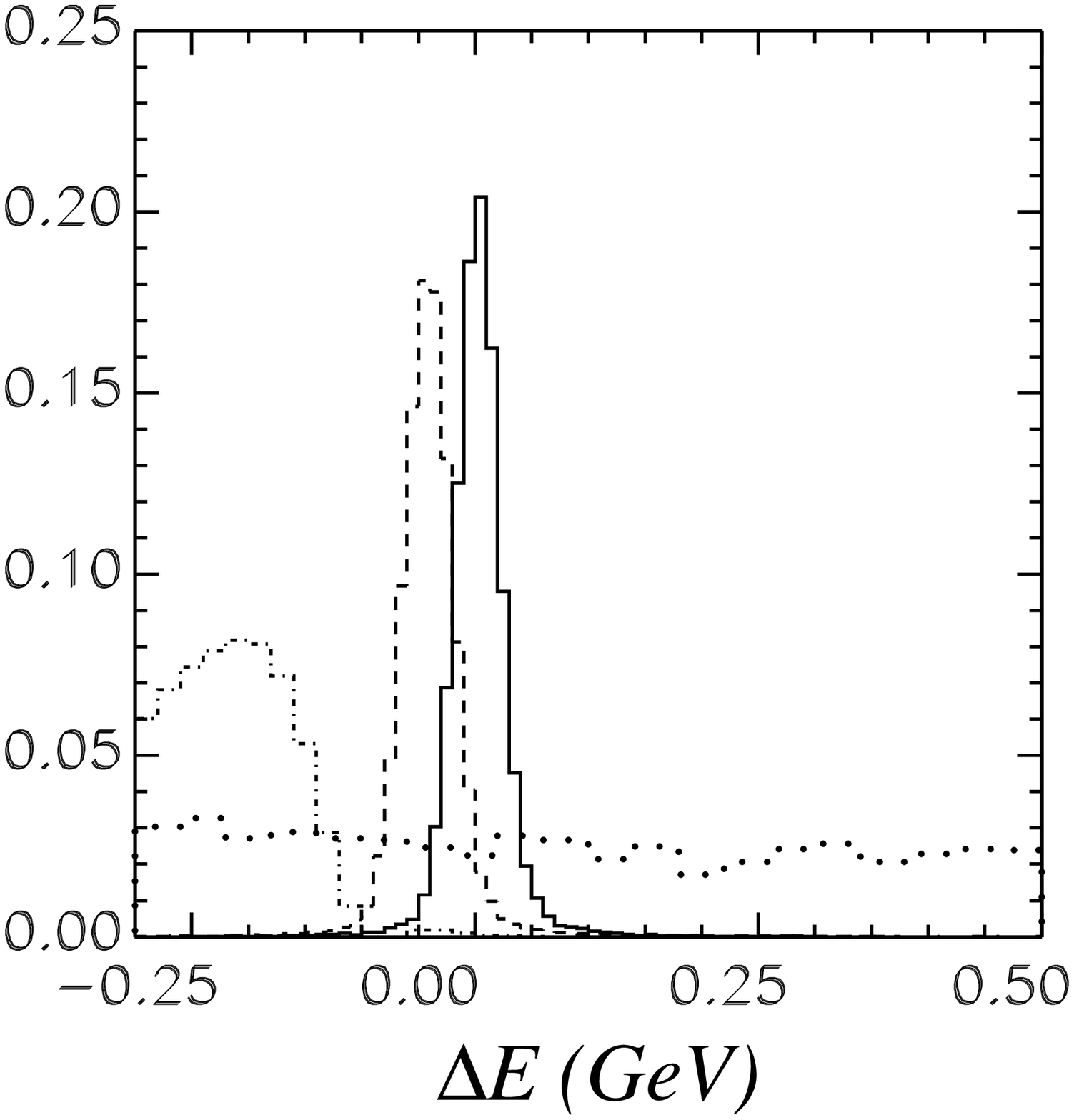}}
\centerline{\epsfysize 2.5 truein\epsfbox{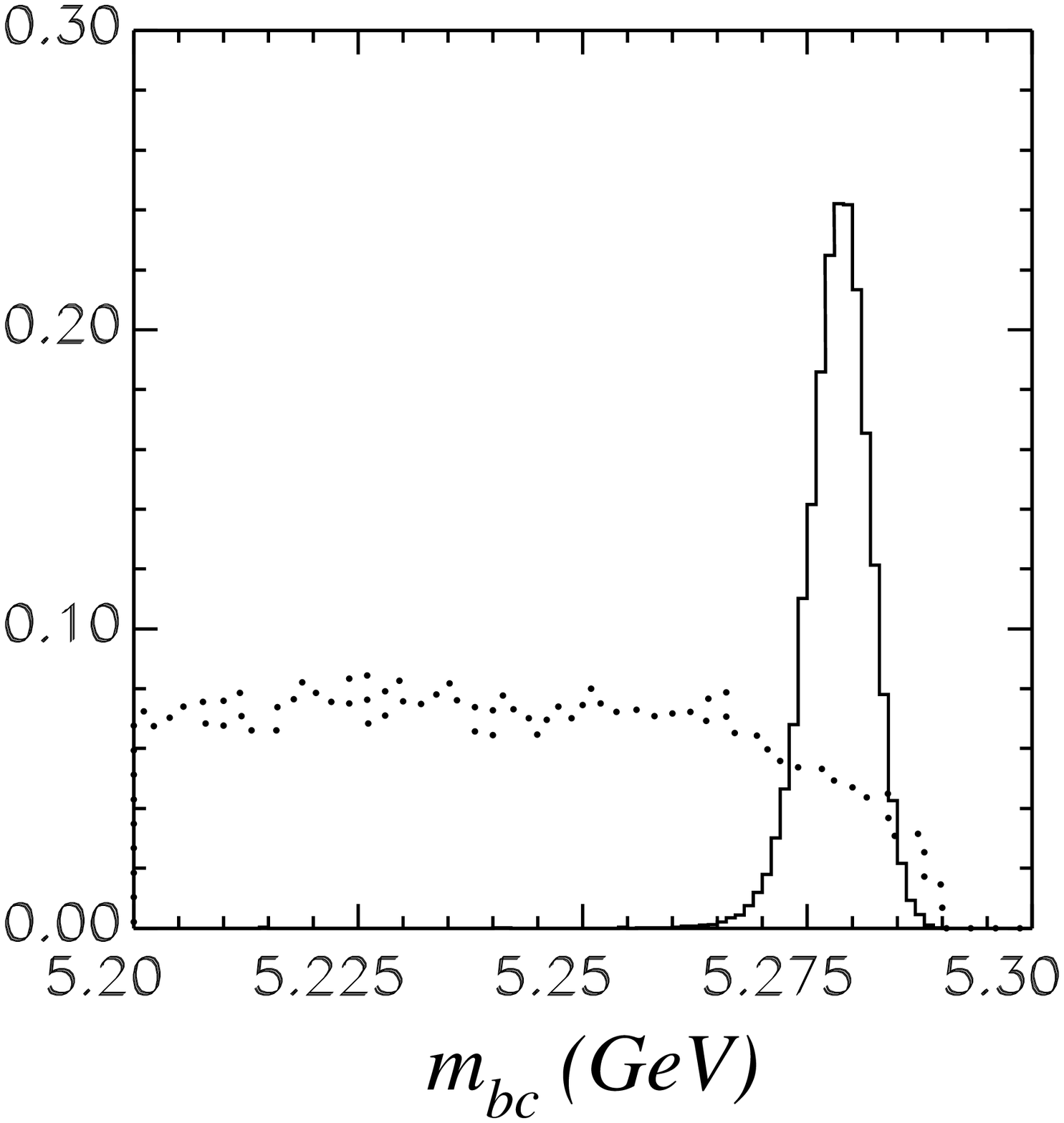}\epsfysize 2.5 truein\epsfbox{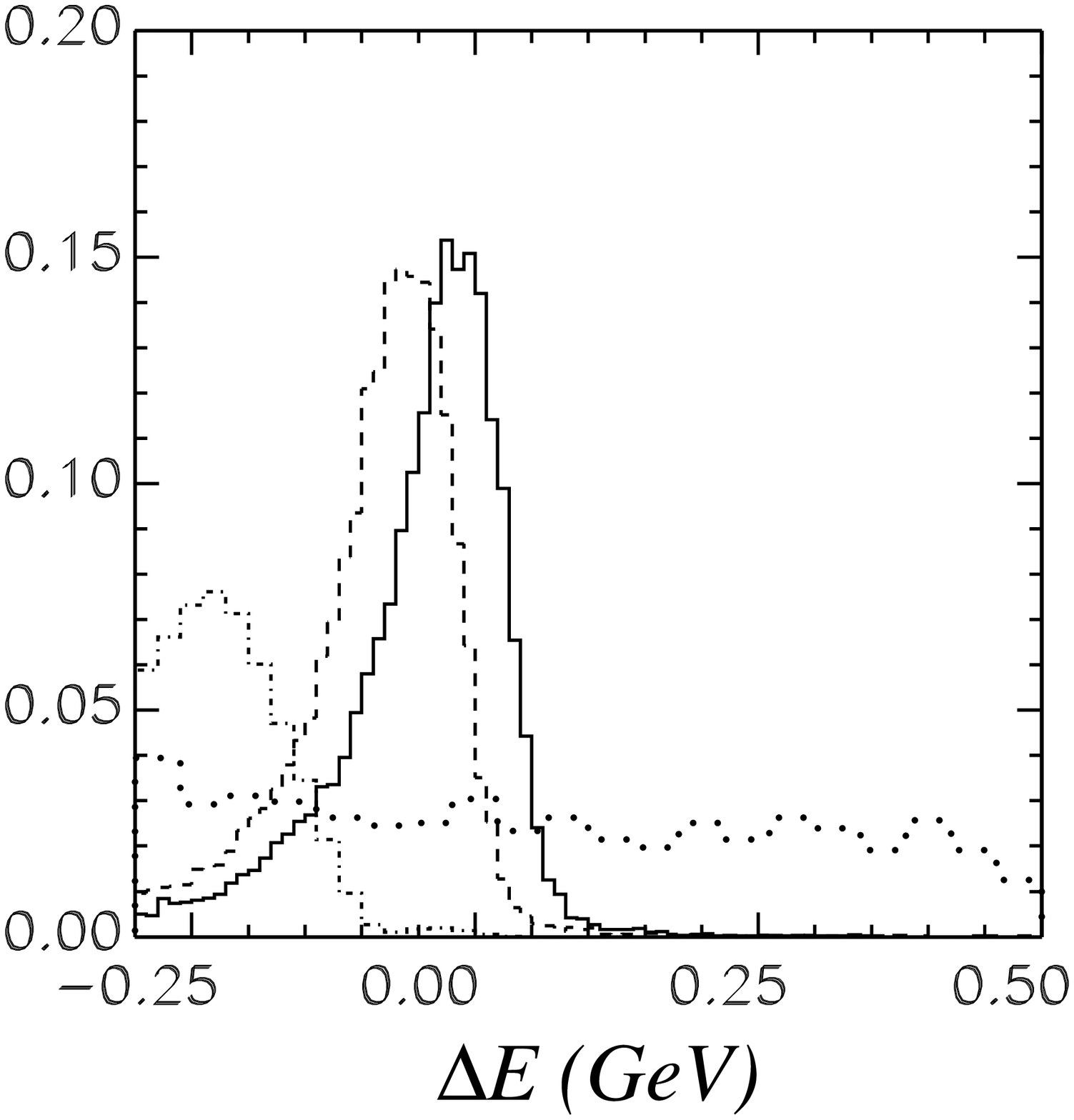}}
\caption{Kinematic reconstruction variables $m_{bc}$ (left) and
$\Delta E$ (right) for $B^0\rightarrow K^+\pi^-$ and $\pi^+\pi^-$ MC
(top) and $B^+\rightarrow K^+\pi^0$ and $\pi^+\pi^0$ MC (bottom).  The solid histograms are $B^{0(+)}\rightarrow
\pi^+\pi^{-(0)}$ MC.  The dotted histograms are off-resonance data.  The
dashed histograms are for $B^{0(+)}\rightarrow K^+\pi^{-(0)}$ MC
events, which are
indistinguishable in the $m_{bc}$ distribution but shifted by $-45$
MeV in $\Delta E$ due to pion mass assignment to the kaon track. The
dot-dashed histograms represent background from multi-body charmless $B$
meson decays.}
\label{fig:mbde}
\end{figure}

To reconstruct $B$ mesons we form two quantities: 
the energy difference, $\Delta
E = E_{B} - E_{\rm beam}$, and the beam constrained mass, $m_{bc} =
\sqrt{E_{\rm beam}^2 - p_B^2}$, where $E_{\rm beam} = \sqrt{s}/2 = 5.29$ GeV, 
and $E_{B}$ and $p_B$ are the reconstructed energy and momentum of the
$B$ candidate in the center of mass frame.  These are shown in Fig.~\ref{fig:mbde} for
$B^{0(+)}\rightarrow K^+\pi^{-(0)}$ and $\pi^+\pi^{-(0)}$ MC.  
Modes containing $\pi^0$s have a tail extending into the negative
$\Delta E$ region due to shower leakage out of the back of the
calorimeter and 
photon interactions with the material in front of the calorimeter.
We calculate the energy of final state
charged particles using a pion mass assumption.  This shifts $\Delta
E$ by $-45$ MeV for each charged kaon in the final state. 
The signal
yields are extracted by a binned maximum-likelihood fit 
to the $\Delta E$ distribution in the
region
$5.271 < m_{bc}< 5.289$ GeV (
 $m_{bc}>5.270$ GeV for modes containing $\pi^0$s)
  and $-300$ MeV $<\Delta E<500$ MeV. 
The yields are verified by fitting $m_{bc}$
in the $\Delta E$ signal region. A sideband region of $5.2$
GeV $< m_{bc} < 5.26$ GeV is used to study the continuum
background in the $\Delta E$ distribution, while a sideband of $150$
MeV $<\Delta E< 500$ MeV is used to study the continuum background in
the $m_{bc}$ distribution.

The $\Delta E$ fits include four components: signal,
crossfeed from other mis-identified $B\rightarrow hh$ decays, 
continuum background, and
backgrounds from multi-body and radiative charmless $B$ decays. These
are shown in Fig.~\ref{fig:mbde}.
The crossfeed component is shifted from the signal component by $45$
MeV as described above.
 The charmless $B$ decay background
is dominated by events where the $B$ meson decays to an $hh\pi$ final
state such as $\rho\pi$ or $f_0(980)K$ where one pion is not
reconstructed.  This shifts the charmless $B$ background by
 at least the mass \newpage \noindent  of the missing pion.  
We expect no backgrounds from $b\rightarrow c$ decays based on a large
MC sample.

For charged particle final
states, the $\Delta E$ signal is modeled with a Gaussian.  
For modes containing
$\pi^0$s, 
the signal is modeled as the sum of a primary Gaussian and a secondary
asymmetric Gaussian. 
The mean positions of the two are equal and the $+\Delta E$ $\sigma$
of the asymmetric Gaussian is constrained to equal the $\sigma$ of the
primary Gaussian.  The crossfeed component has an equal shape,
shifted by $45$ MeV for each mis-identified particle.

\begin{figure}[htbp]
\centerline{\epsfysize 2.5
truein\epsfbox{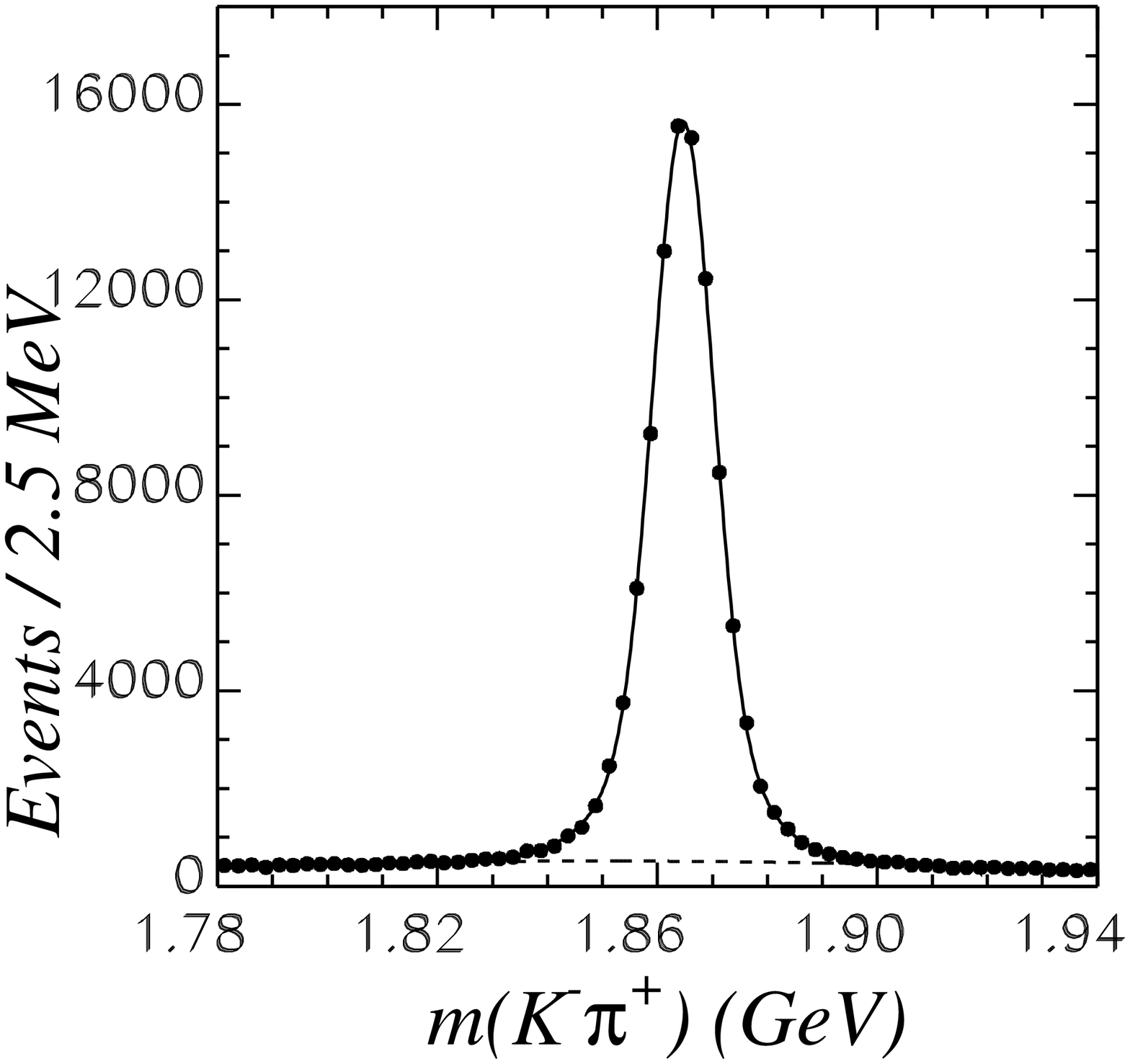}\epsfysize 2.5
truein\epsfbox{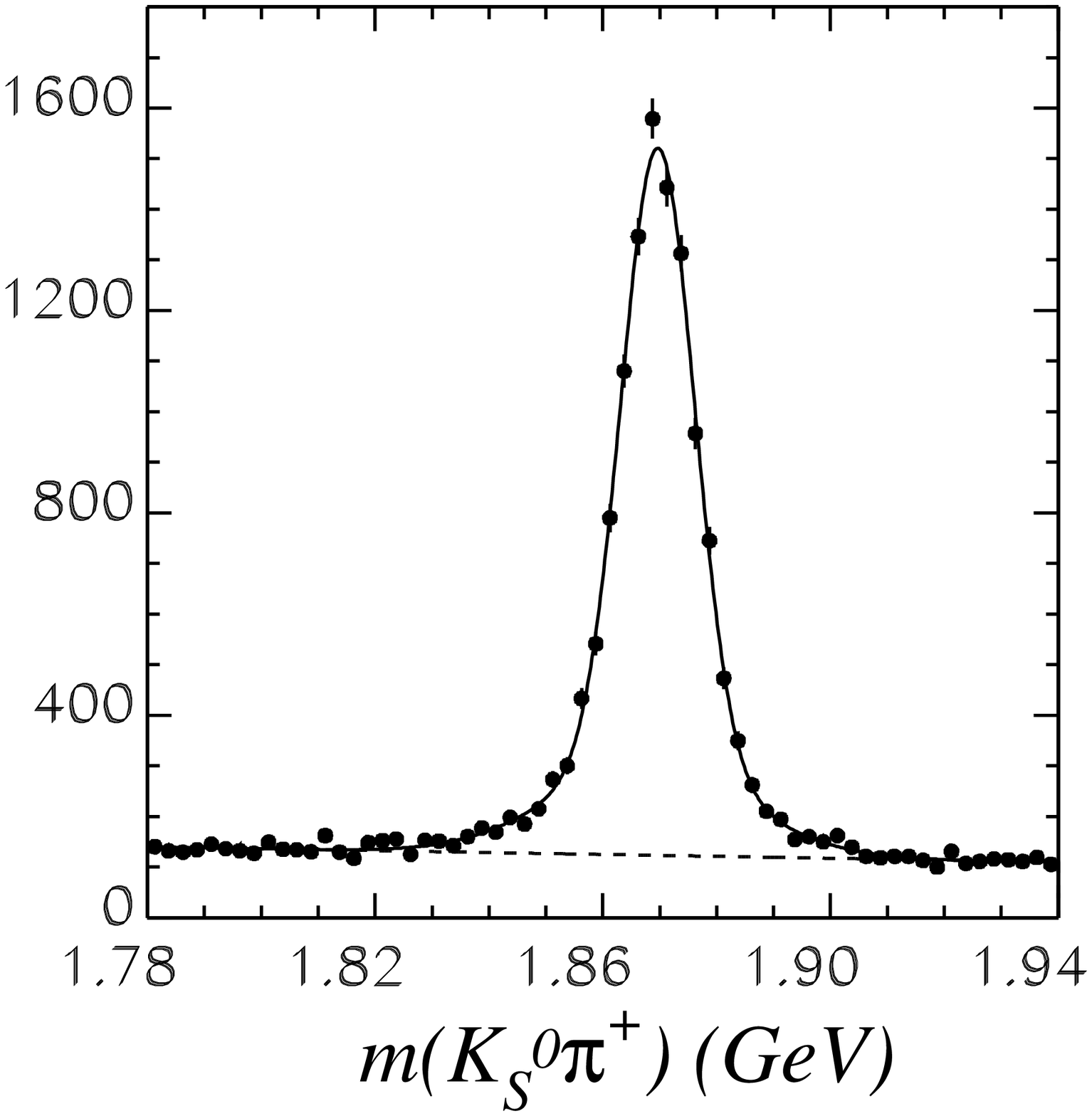}}
\centerline{\epsfysize 2.5 truein\epsfbox{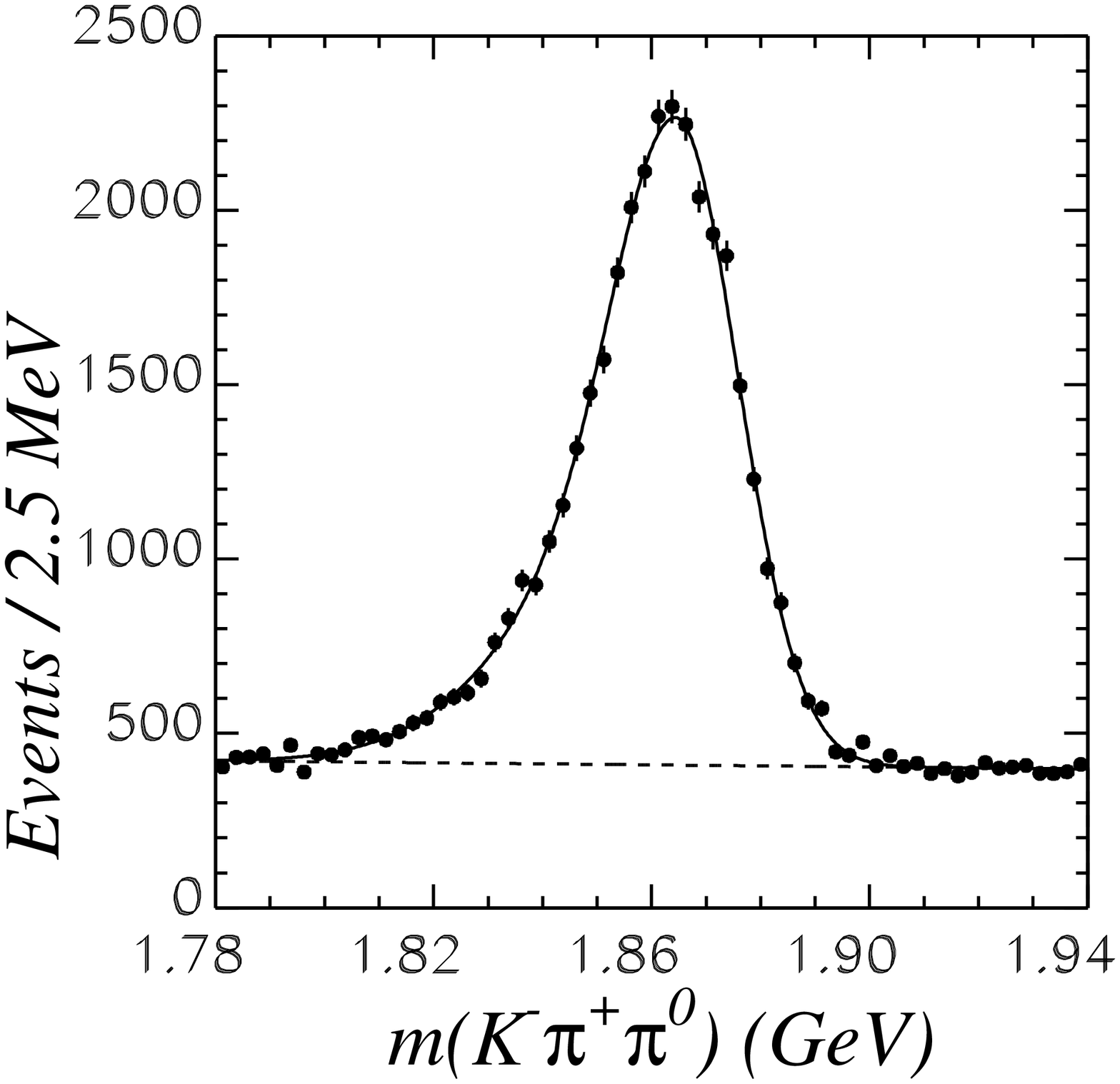}}
\caption{Mass spectra for inclusive $D^0\rightarrow
K^-\pi^+$ (top left), $D^+\rightarrow K^0_S\pi^+$ (top right), and
$D^0\rightarrow K^-\pi^+\pi^0$ (bottom) used to determine $\Delta E$
fit parameters.  In each case, momentum cuts
are placed on the $D$ daughter particles to simulate the momentum of $B$
daughter particles.}
\label{fig:ddata}
 \end{figure}

The widths of the $\Delta E$ signal distributions are determined using
inclusive high momentum
$D^0\rightarrow K^-\pi^+$, $K^-\pi^+\pi^0$, and $D^+\rightarrow
K^0_S\pi^+$ decays after requiring
the $D$ daughter particles to have a momentum range similar to
$B\rightarrow hh$ candidate particles.  These distributions are shown
in Fig.~\ref{fig:ddata}.  Comparisons between the $D$
mass widths in MC and data are used to scale the $B\rightarrow
hh$ $\Delta E$ MC widths.  This procedure is also used to
determine the ratio of primary to secondary Gaussians for modes
containing $\pi^0$s.

\begin{figure}[htbp]
\centerline{\epsfysize 2.5 truein\epsfbox{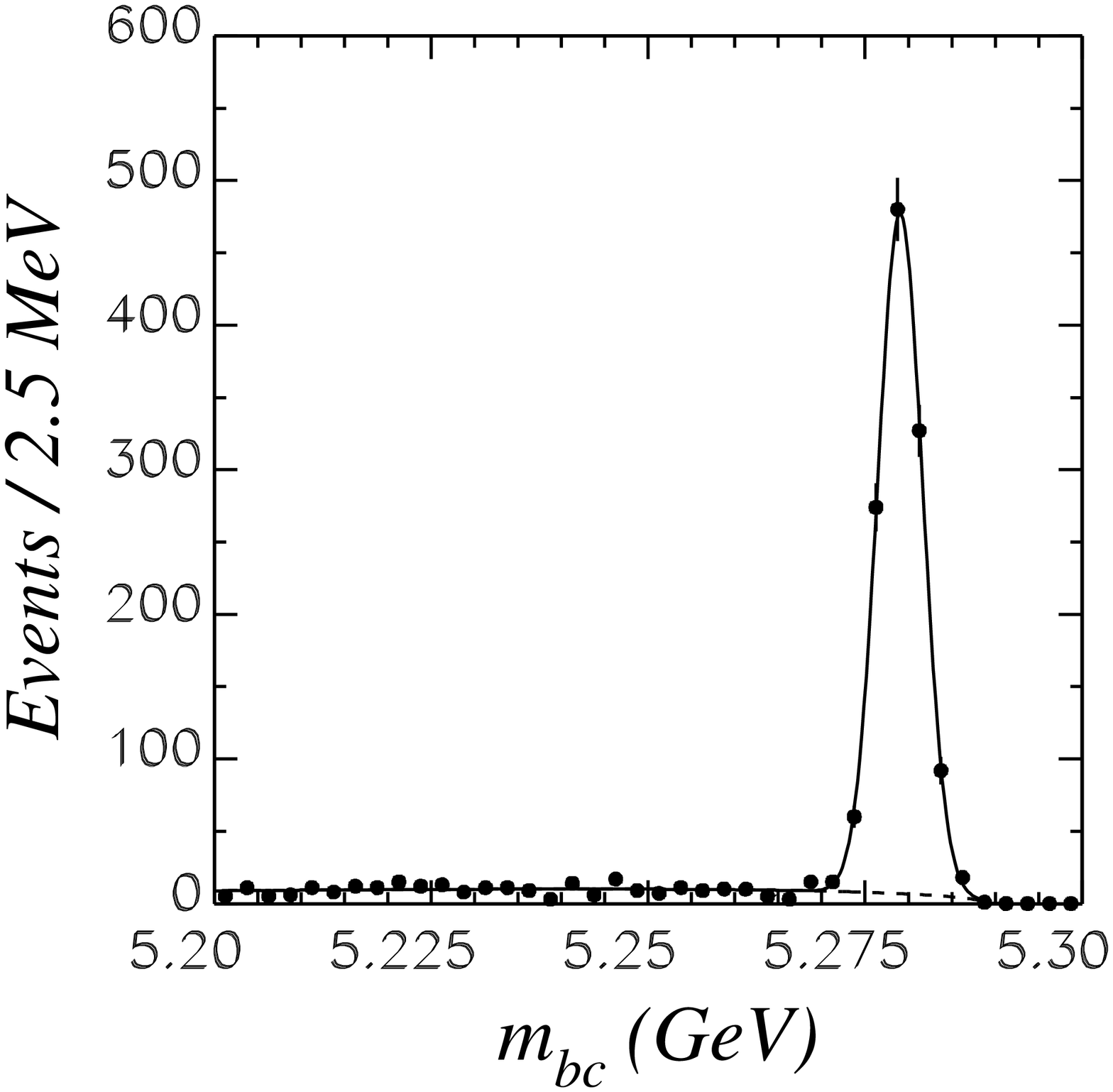}\epsfysize 2.5 truein\epsfbox{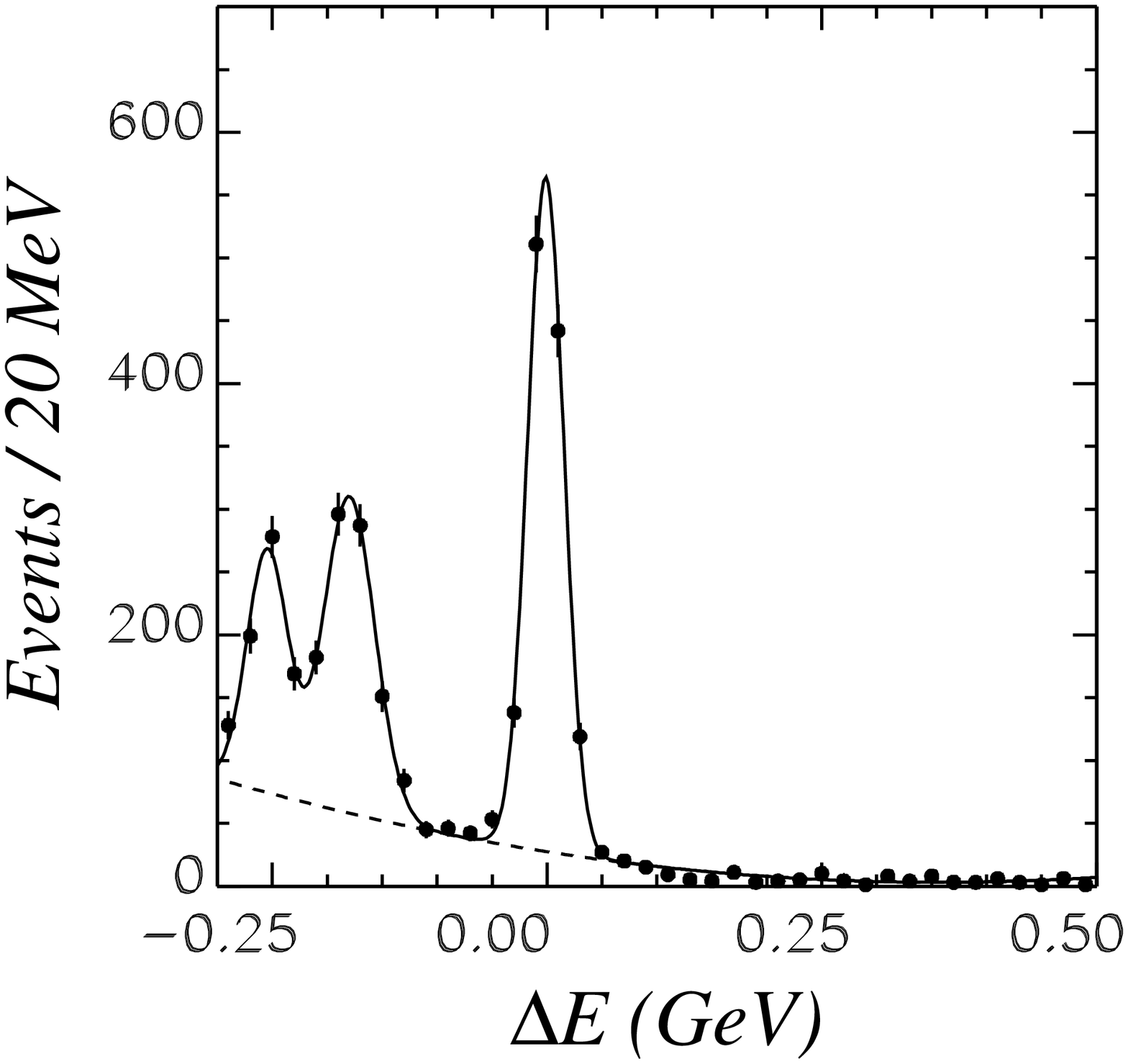}}
\caption{The $B^+\rightarrow D^0\pi^+$, $D^0\rightarrow K^-\pi^+$ data sample used to determine
$m_{bc}$ (left) and $\Delta E$ (right)
fit parameters. The $\Delta E$ distribution contains backgrounds from
$B^0\rightarrow D^0\pi^+\pi^-$.}
\label{fig:dpidata}
 \end{figure}

The peak positions of the $\Delta E$ signal Gaussians are a function
of the beam energy and the momentum scale.  The beam energy
is determined using the peak position of the $m_{bc}$ distribution for
the $B^+\rightarrow
D^{0}\pi^+$, $D^0\rightarrow K^-\pi^+$ data sample shown in
Fig.~\ref{fig:dpidata}.  The momentum scale is determined using the
peak positions of the inclusive $D$ mass spectra discussed above as
well as the $\Delta E$ distribution for the $B^+\rightarrow D^0\pi^+$
data sample also shown in Fig.~\ref{fig:dpidata}.

\begin{figure}[htbp]
\centerline{\epsfysize 4 truein\epsfbox{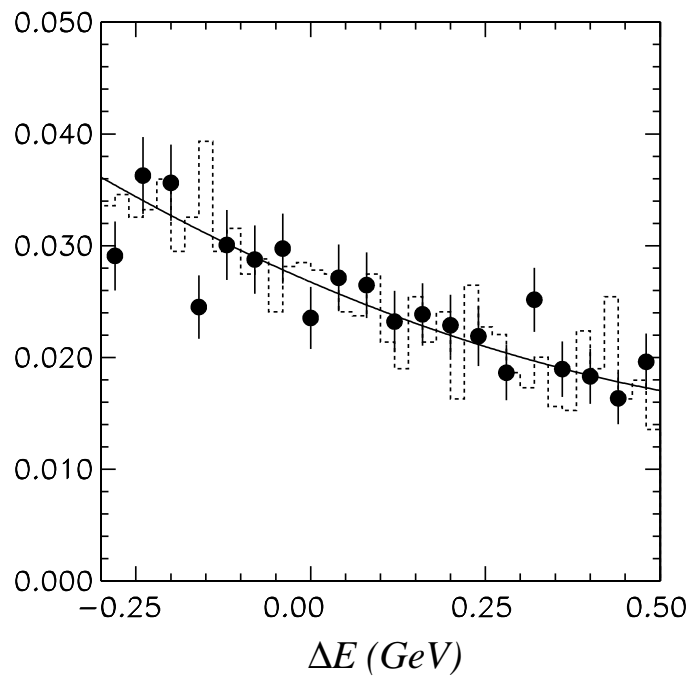}}
\caption{The $\Delta E$ distribution for
$B^+\rightarrow K^0_S\pi^+$ candidates in $m_{bc}$ sideband data
(dashed-histogram) compared to the same distribution in the $m_{bc}$
signal window for off-resonance data (black circles).  The solid curve
is a second-order polynomial used to parameterize the shape.}
\label{fig:sidebandtest}
 \end{figure}

The continuum background is modeled with a second-order polynomial with coefficients
determined from sideband data.  Figure~\ref{fig:sidebandtest}
demonstrates the validity of this method by comparing the
continuum $\Delta E$ background shape in on-resonance $m_{bc}$
sideband data to the shape in the $m_{bc}$ signal region
in the off-resonance data sample. Backgrounds from charmless $B$ decays
are modeled by a smoothed MC histogram.  

For all final states except $K^+\pi^0$ and $\pi^+\pi^0$, 
the normalizations of the four components are the only free parameters
in the fits. The significance of the signal yield above background is
determined by re-fitting the $\Delta E$ distribution without a signal
component and comparing the maximum likelihoods of the two fits.
Due to the large overlap of the signal and crossfeed components in
the $K^+\pi^0$ and $\pi^+\pi^0$ signals, we perform a simultaneous fit
to the $K^+\pi^0$ and $\pi^+\pi^0$ $\Delta E$ distributions
constraining the crossfeed to the expected values based on the PID fake
rates. 

The $m_{bc}$ distribution provides no discrimination among the
three $B$ decay components.  The sum of the three components is
modeled with the same functional form as the $\Delta E$ signal shapes
discussed above. We parameterize the continuum background
 with a function that behaves like phase space near the endpoint (the ARGUS
shape~\cite{bib:ARGUS}).

\section{Results}

\begin{figure}[htbp]
\centerline{\epsfysize 1.85 truein\epsfbox{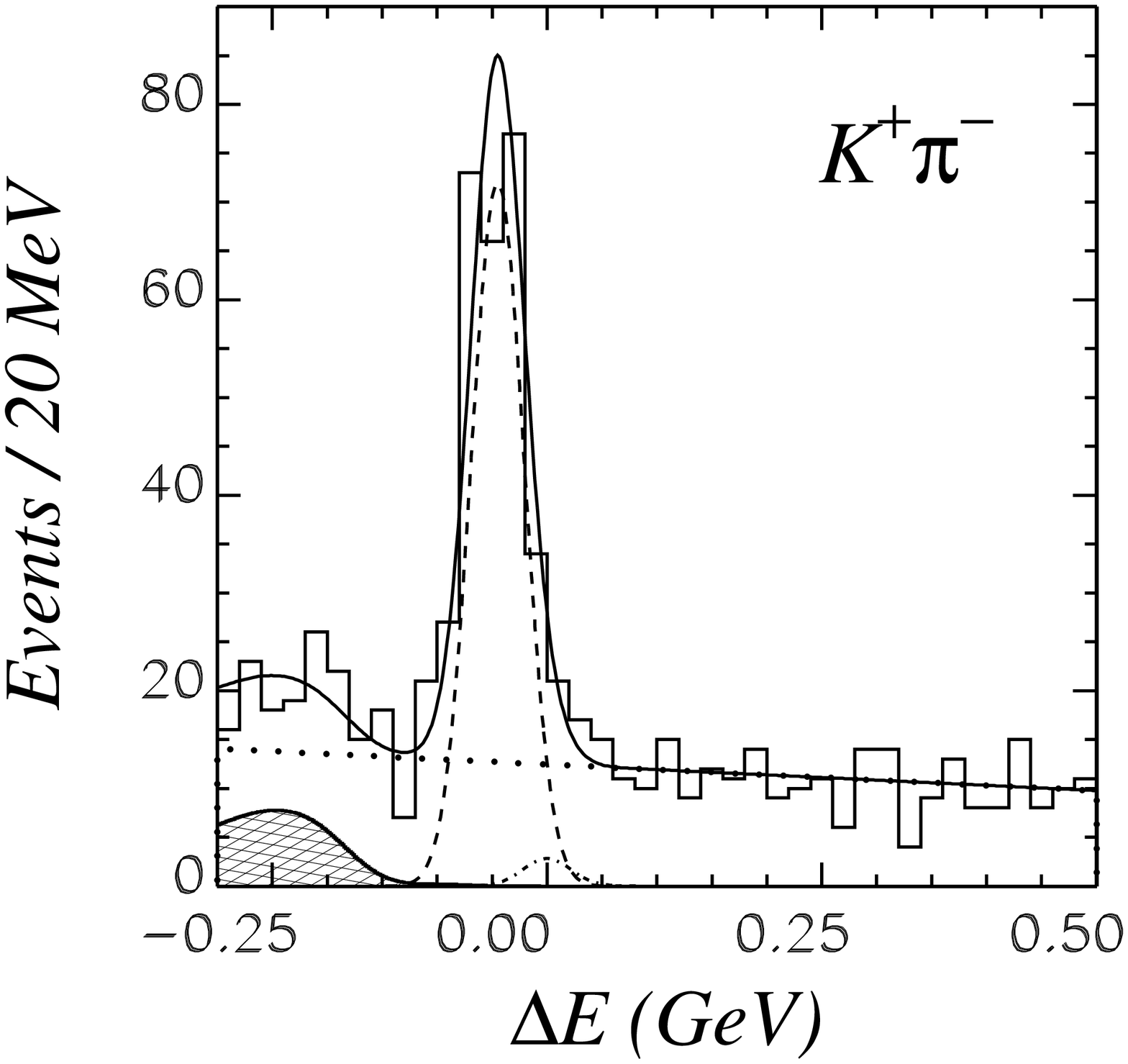}\epsfysize 1.85 truein\epsfbox{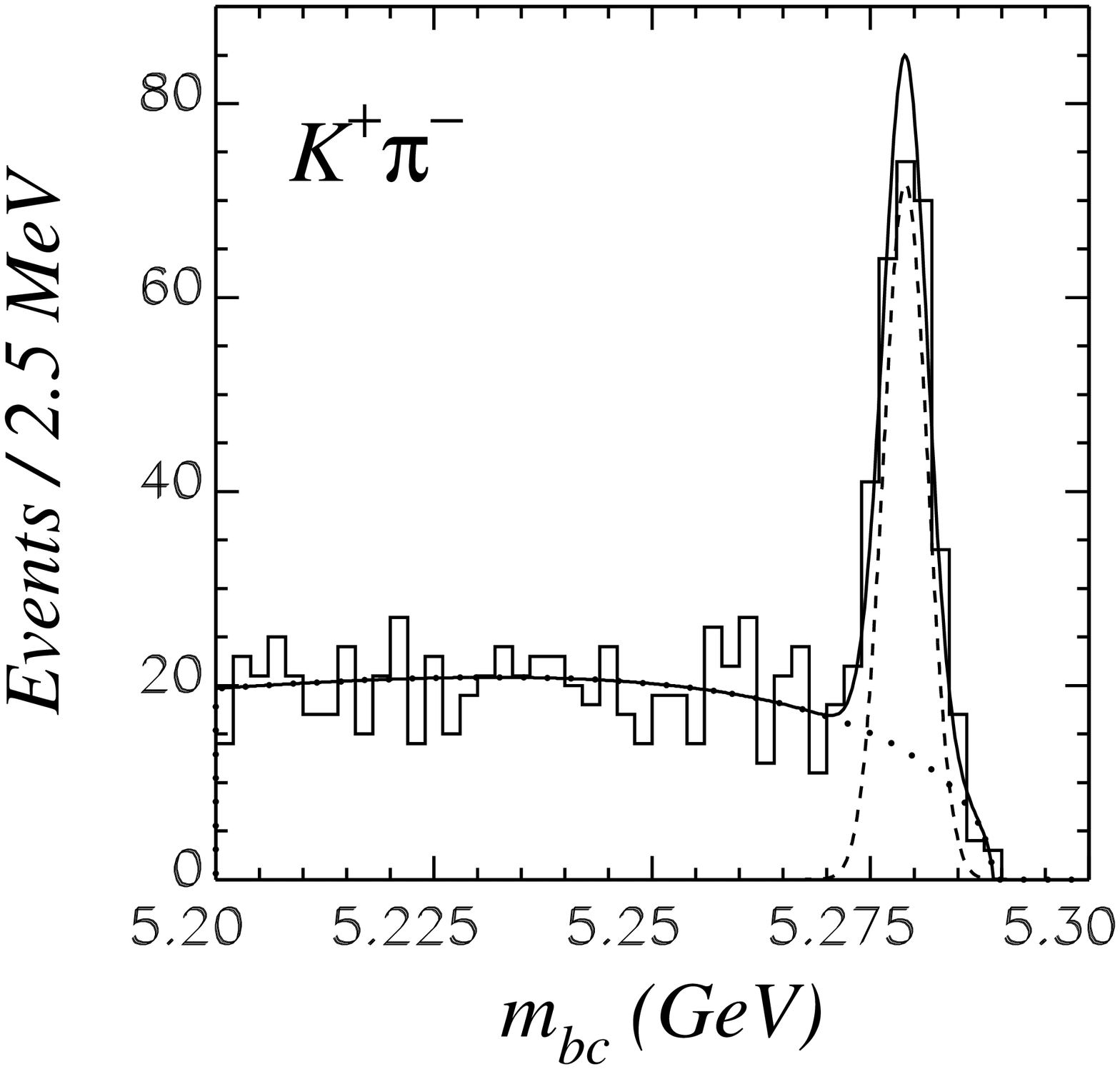} }\nopagebreak
\centerline{\epsfysize 1.85 truein\epsfbox{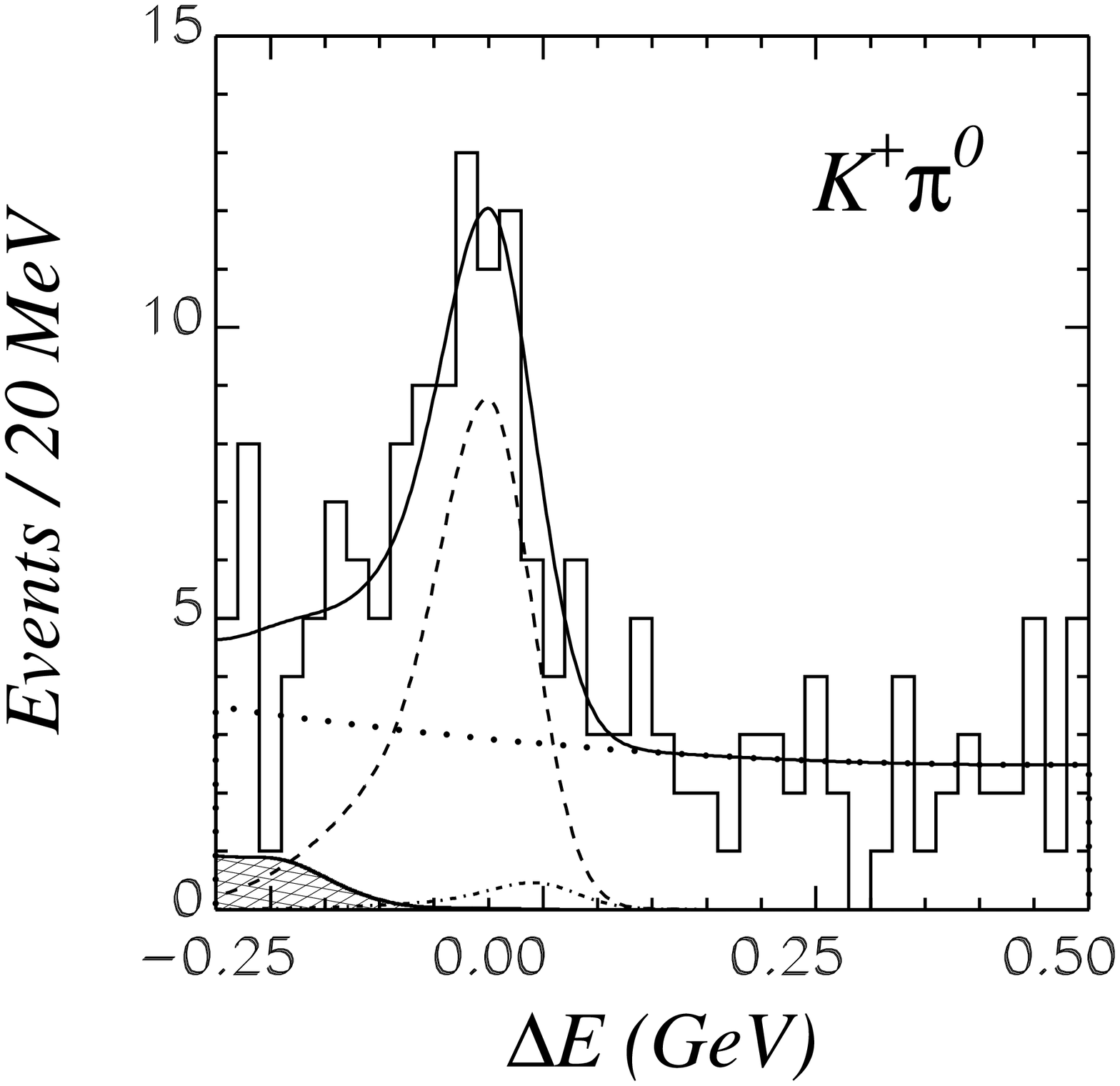}\epsfysize 1.85 truein\epsfbox{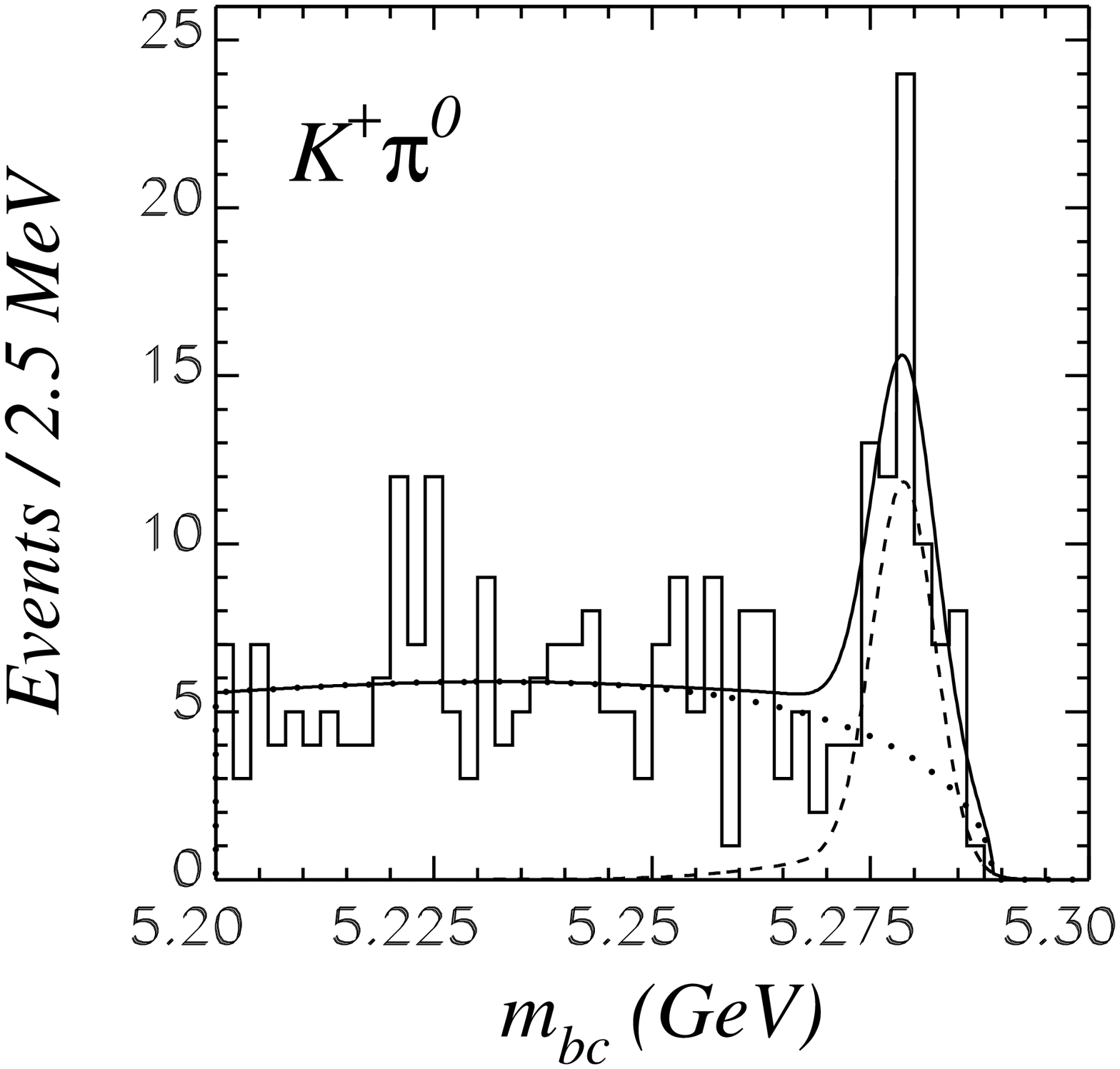} }\nopagebreak
\centerline{\epsfysize 1.85 truein\epsfbox{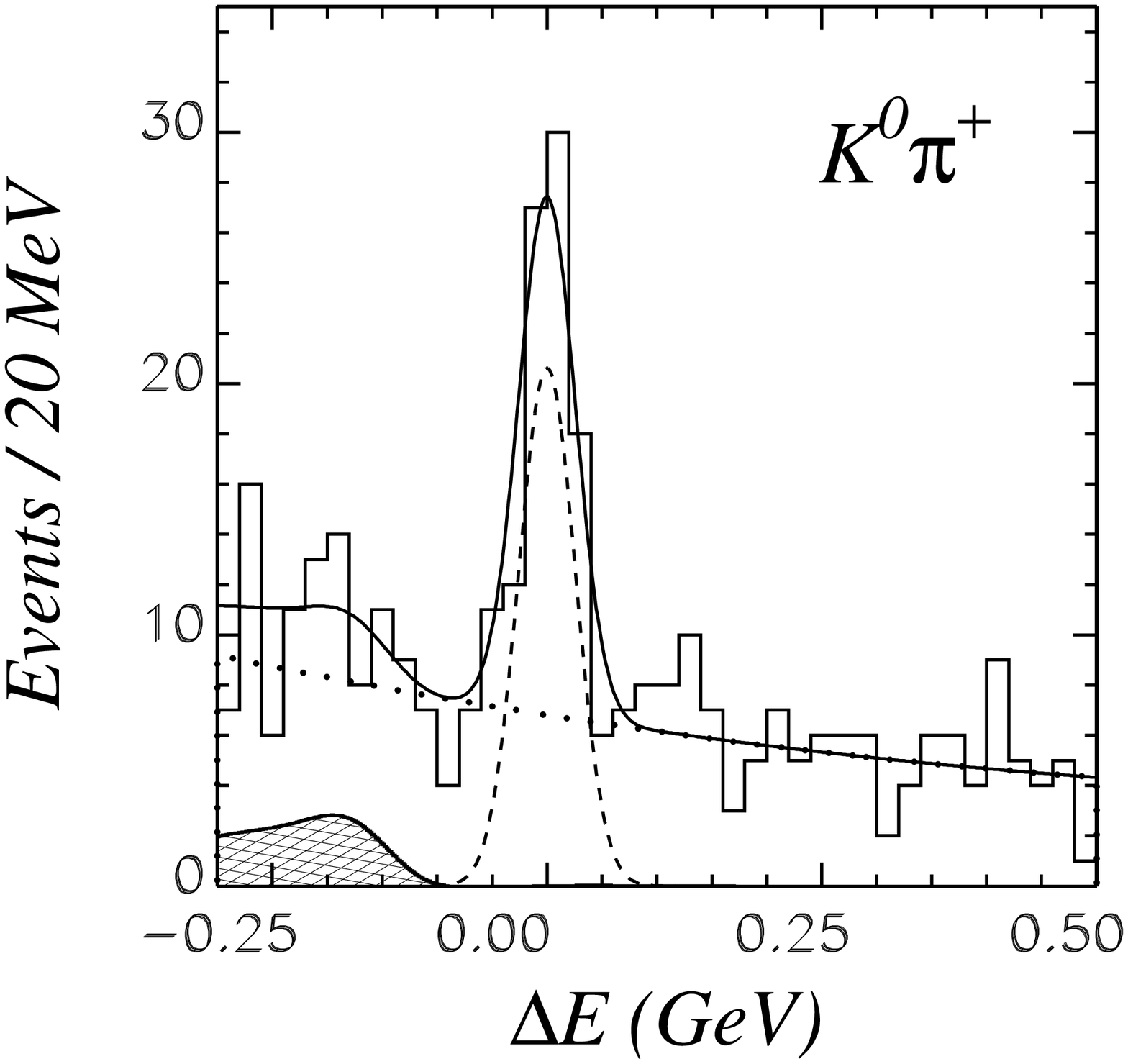}\epsfysize 1.85 truein\epsfbox{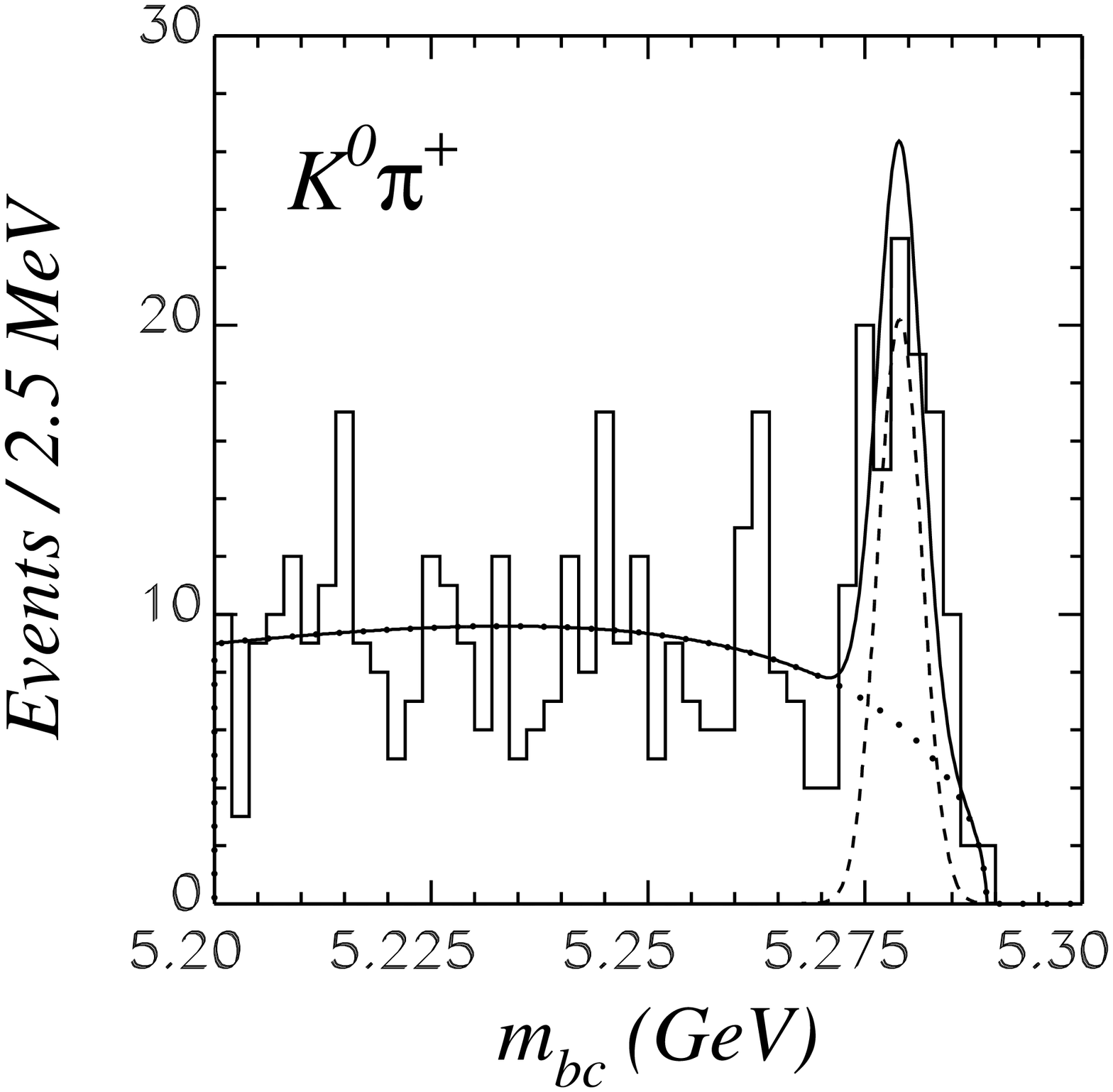} }\nopagebreak
\centerline{\epsfysize 1.85 truein\epsfbox{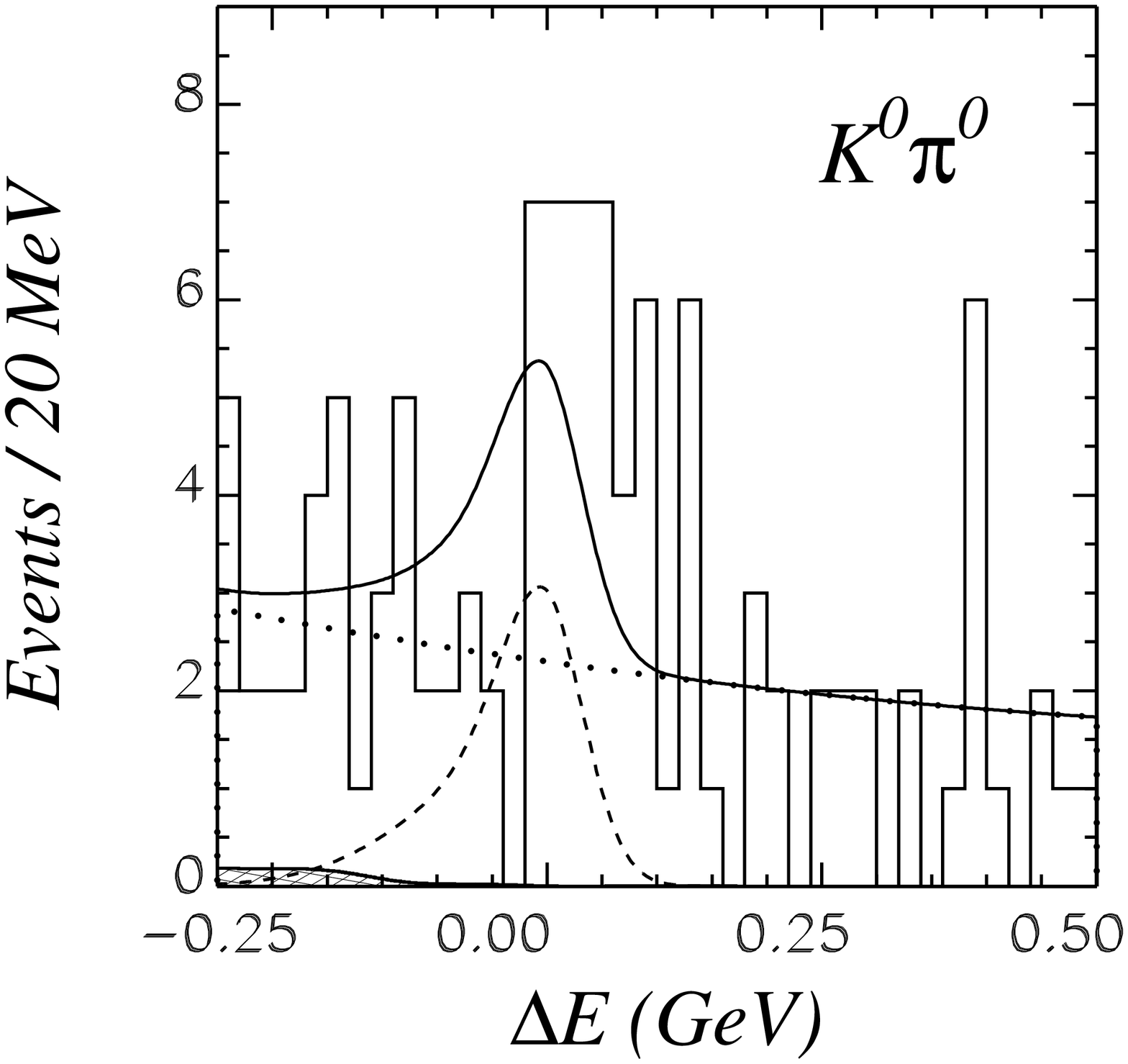}\epsfysize 1.85 truein\epsfbox{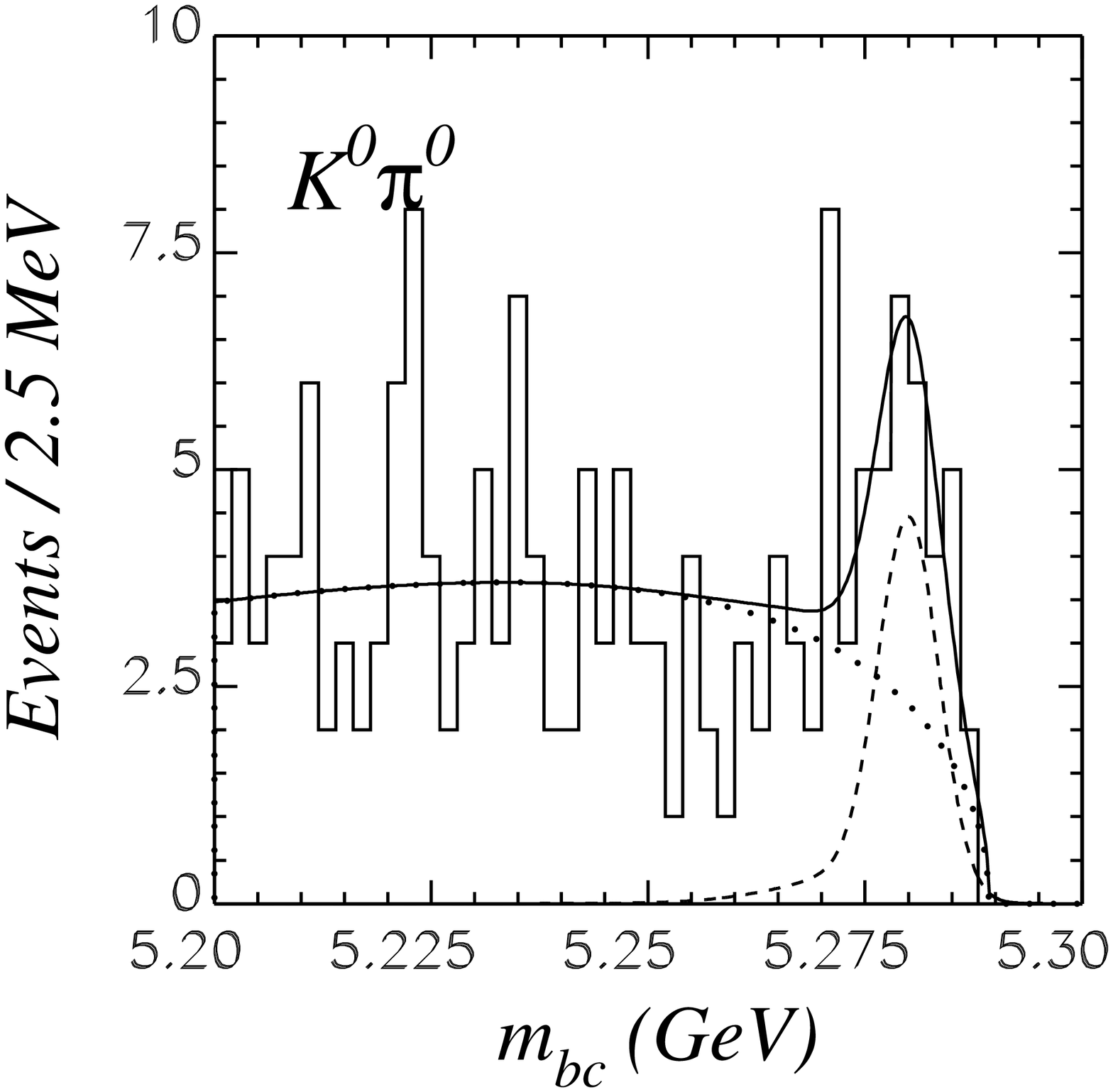} }\nopagebreak
\caption{The $\Delta E$ (left) and $m_{bc}$ (right) fits to the
$B\rightarrow K\pi$ event samples.  
The sum of the signal and
background functions is shown as a solid curve.  For the $\Delta E$
distributions, the dashed curve represents the signal component, the
dotted curve represents the continuum background, and the hatched
histogram represents the charmless $B$ background component.  For the
$K^+\pi^-$ and $K^+\pi^0$ distributions, the crossfeed components
from $\pi^+\pi^-$ and $\pi^+\pi^0$ are shown by dot-dashed curves
centered $45$ MeV above the signal components.  For the $m_{bc}$
distributions, the continuum background is represented by the dotted
curve while the sum of signal, charmless $B$ background, and crossfeed
components is shown by the dashed curve.  
}
\nopagebreak
\label{fig:kpires}
\end{figure}

\begin{figure}[htbp]
\centerline{\epsfysize 2.0 truein\epsfbox{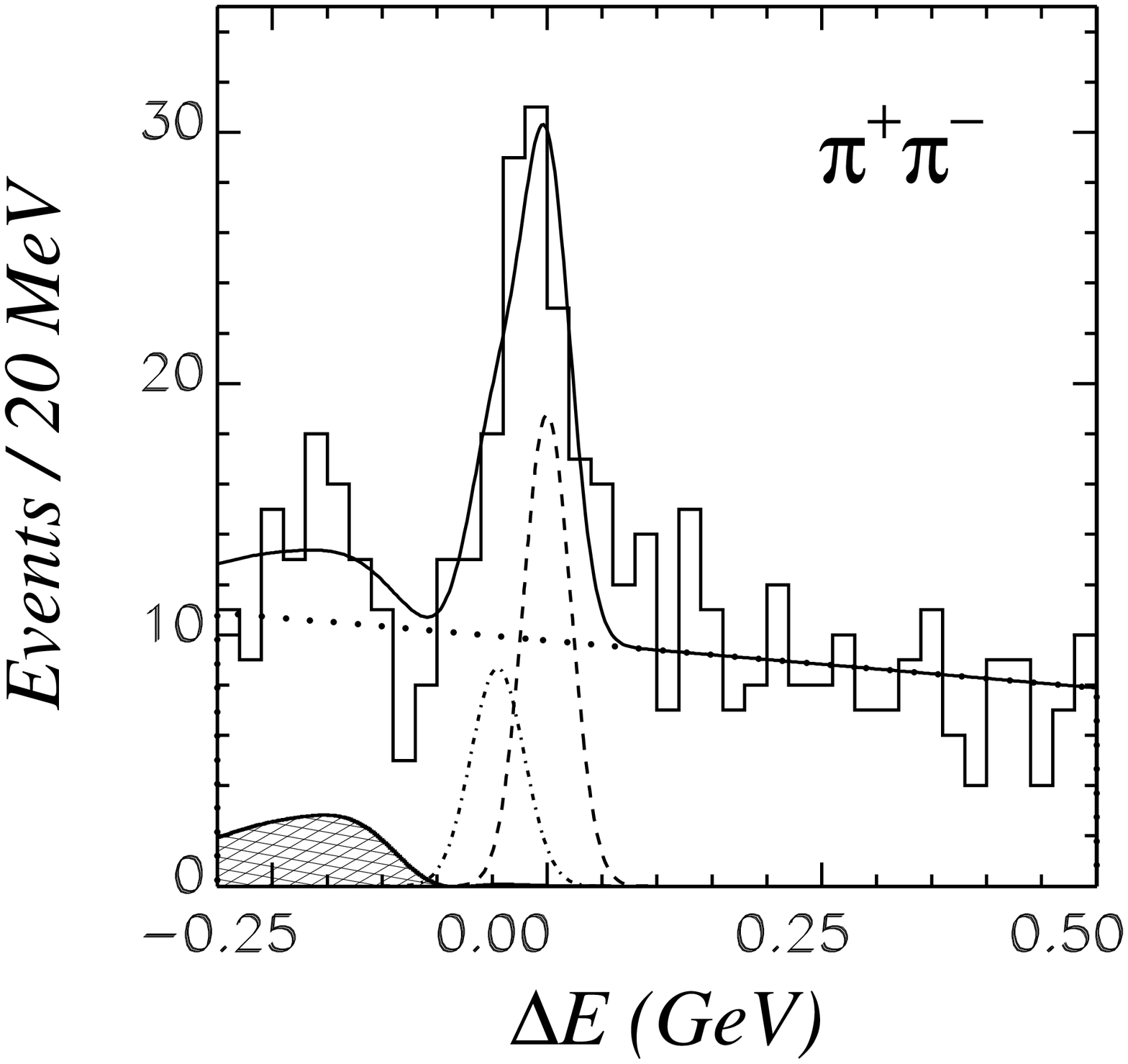}\epsfysize 2.0 truein\epsfbox{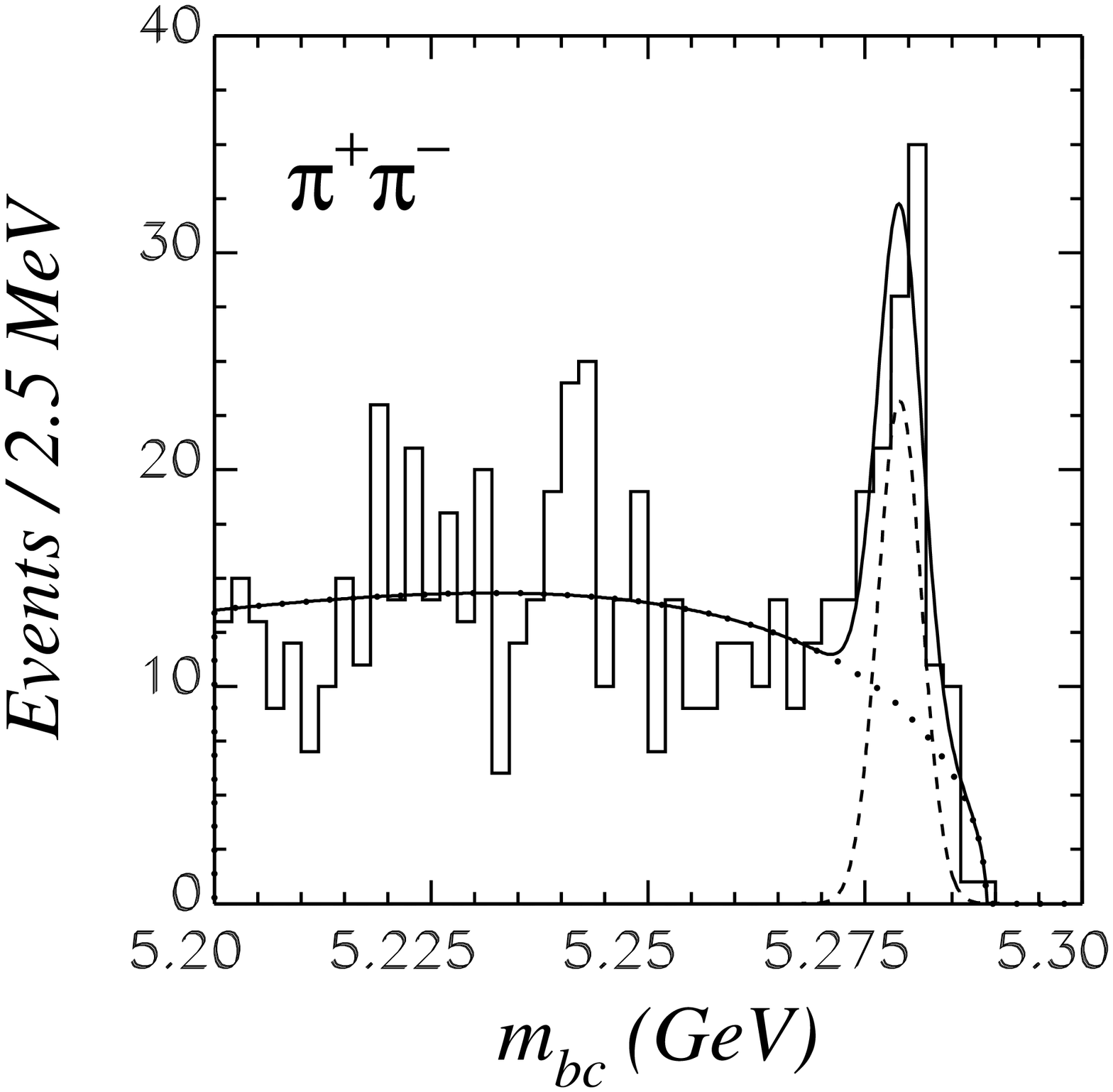} }
\centerline{\epsfysize 2.0 truein\epsfbox{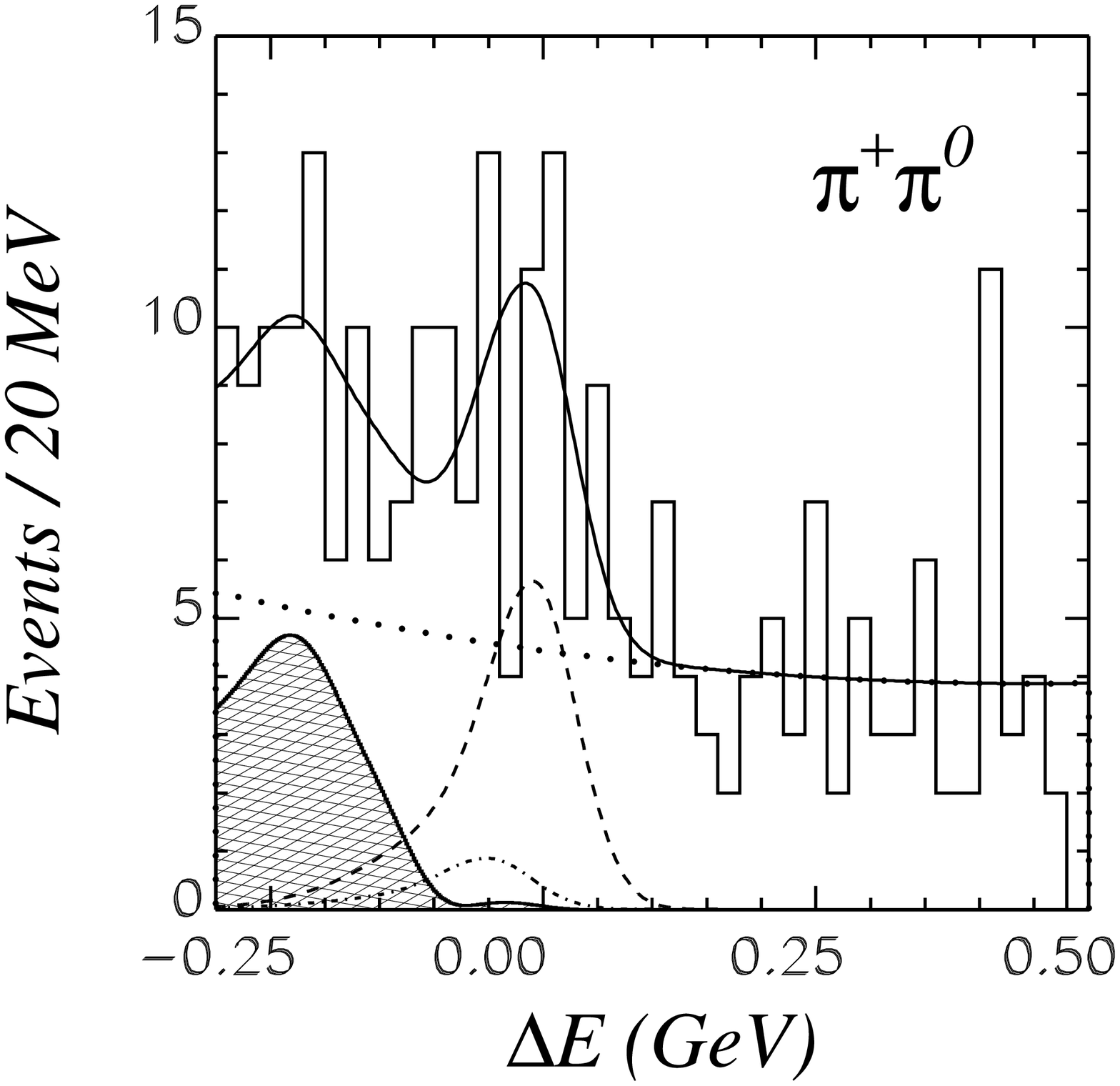}\epsfysize 2.0 truein\epsfbox{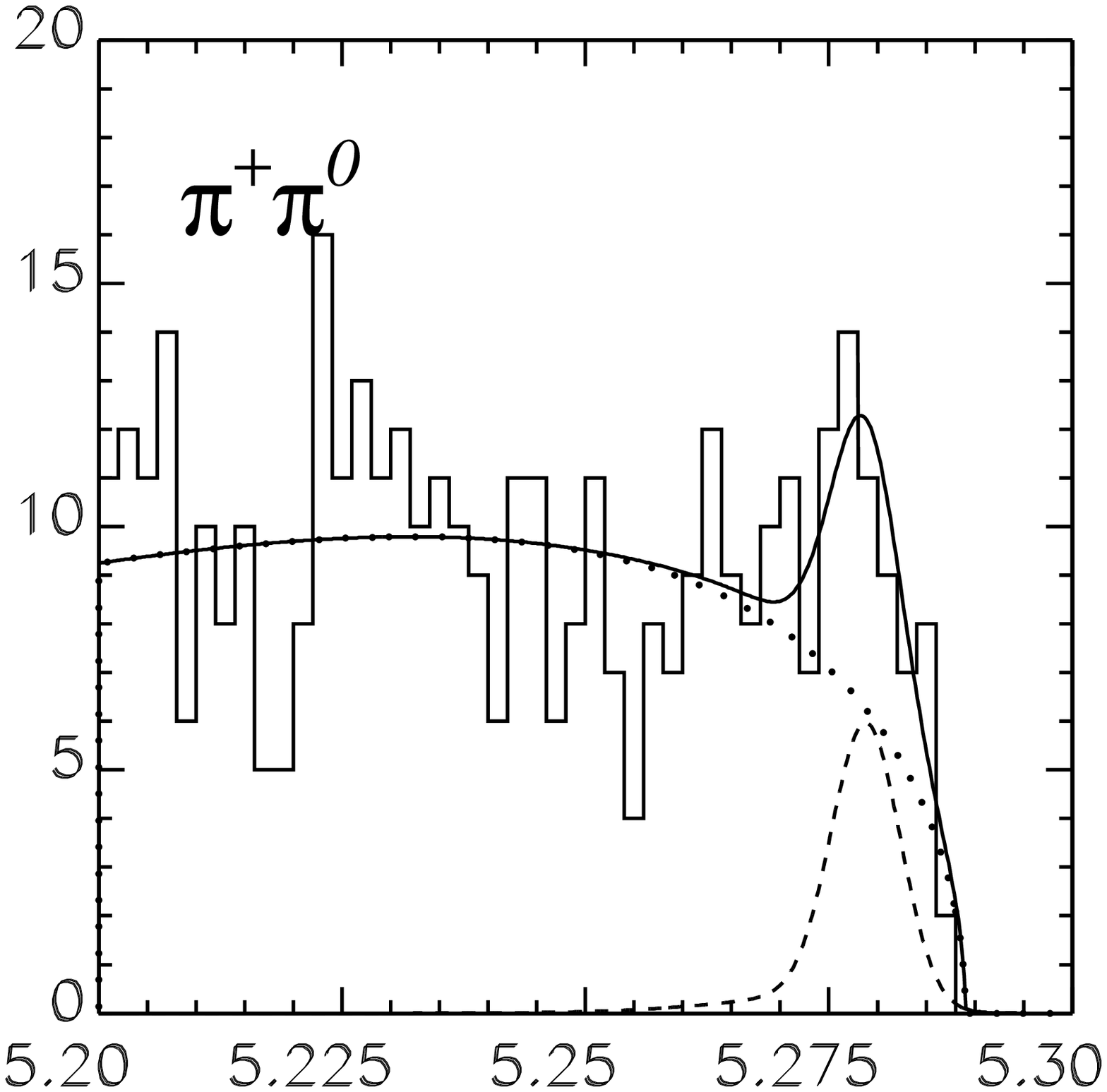} }
\centerline{\epsfysize 2.0 truein\epsfbox{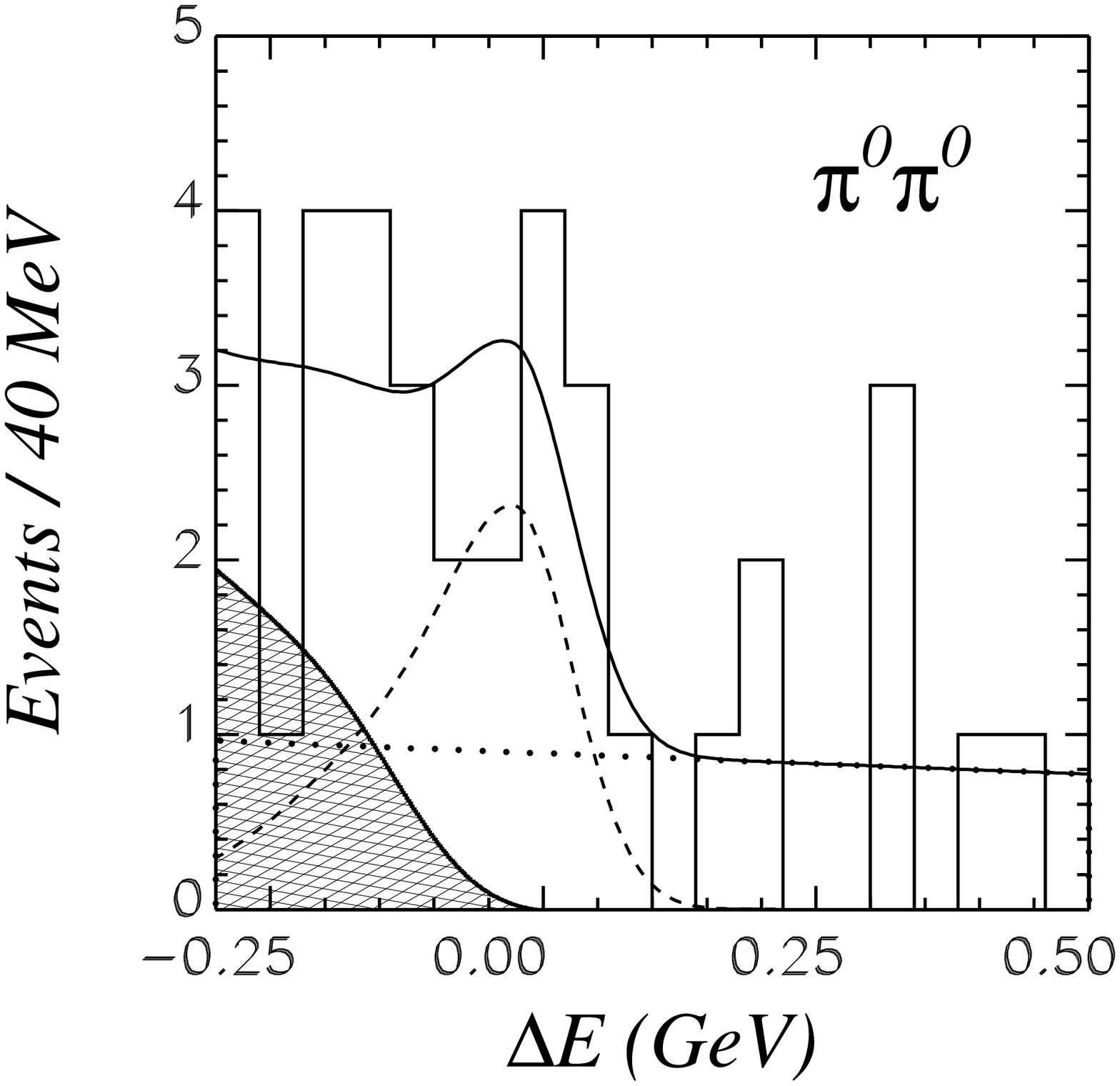}\epsfysize 2.0 truein\epsfbox{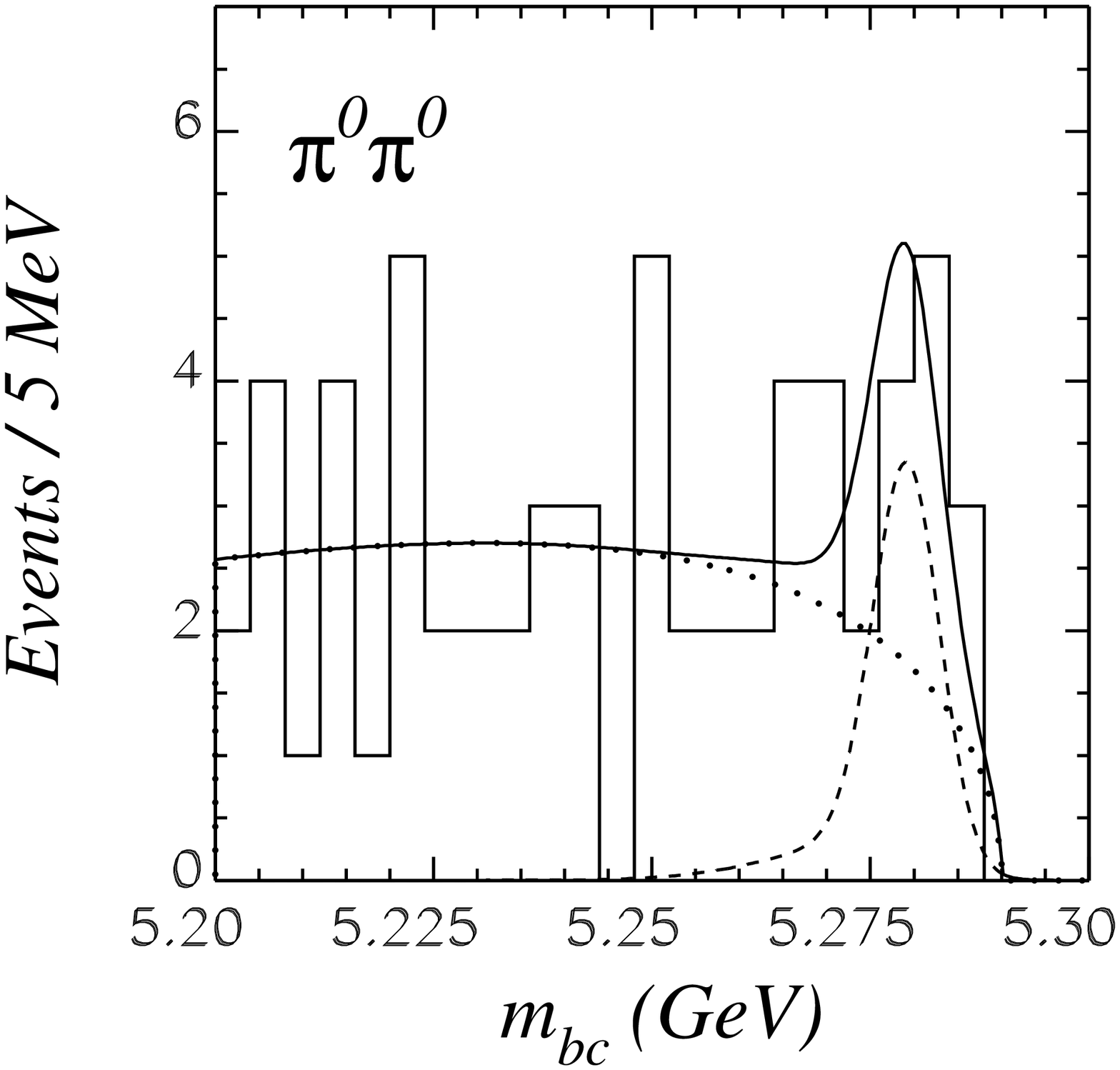} }
\caption{The $\Delta E$ (left) and $m_{bc}$ (right) fits to the
$B\rightarrow \pi\pi$ event samples. The sum of the signal and
background functions is shown as a solid curve.  For the $\Delta E$
distributions, the dashed curve represents the signal component, the
dotted curve represents the continuum background, and the hatched
histogram represents the charmless $B$ background component.  For the
$\pi^+\pi^-$ and $\pi^+\pi^0$ distributions, the crossfeed components
from $K^+\pi^-$ and $K^+\pi^0$ are shown by dot-dashed curves
centered $45$ MeV below the signal components.  For the $m_{bc}$
distributions, the continuum background is represented by the dotted
curve while the sum of signal, charmless $B$ background, and crossfeed
components is shown by the dashed curve.}
\label{fig:pipires}
\end{figure}

\begin{figure}[htbp]
\centerline{\epsfysize 2.0
truein\epsfbox{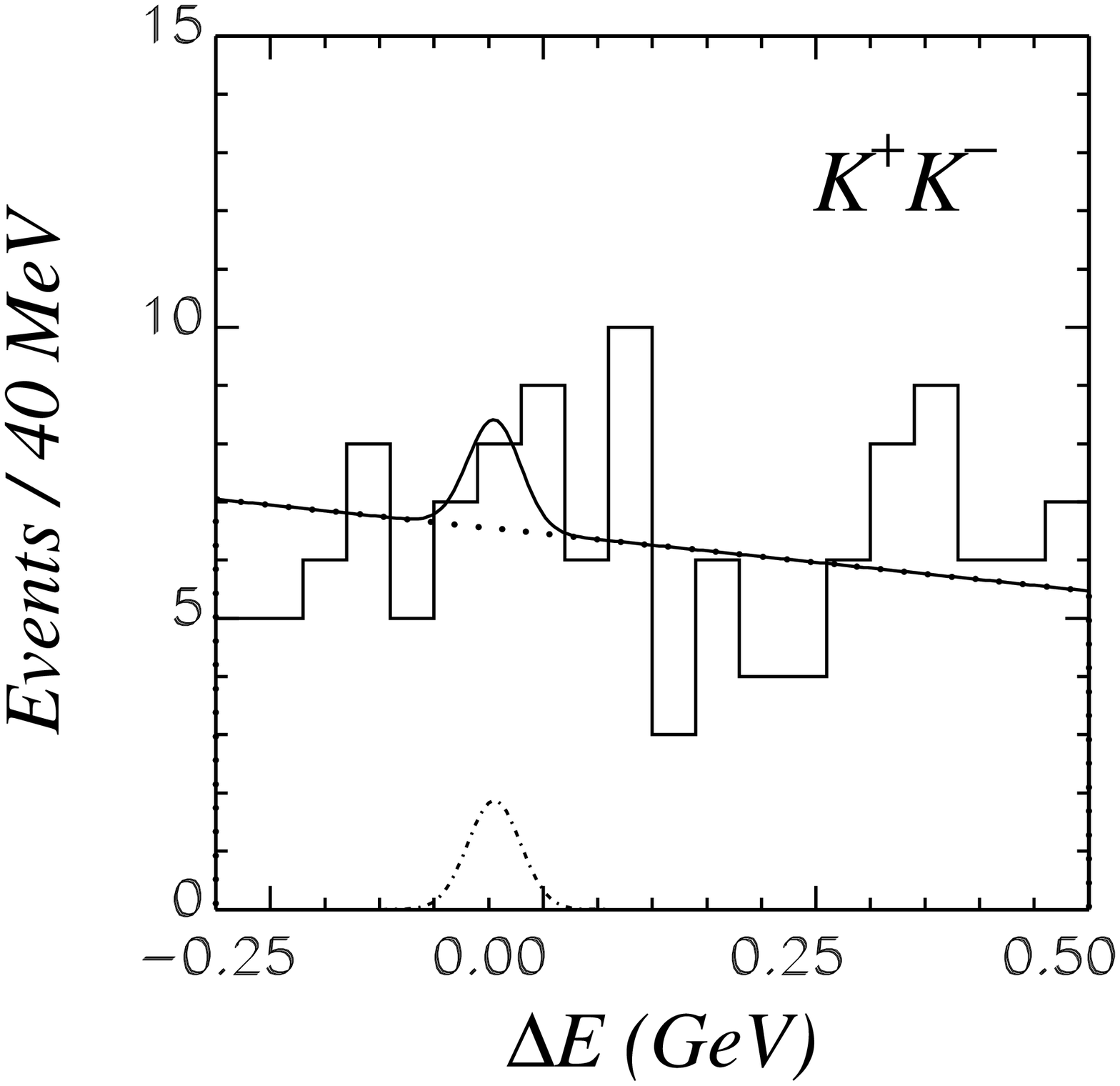}\epsfysize 2.0
truein\epsfbox{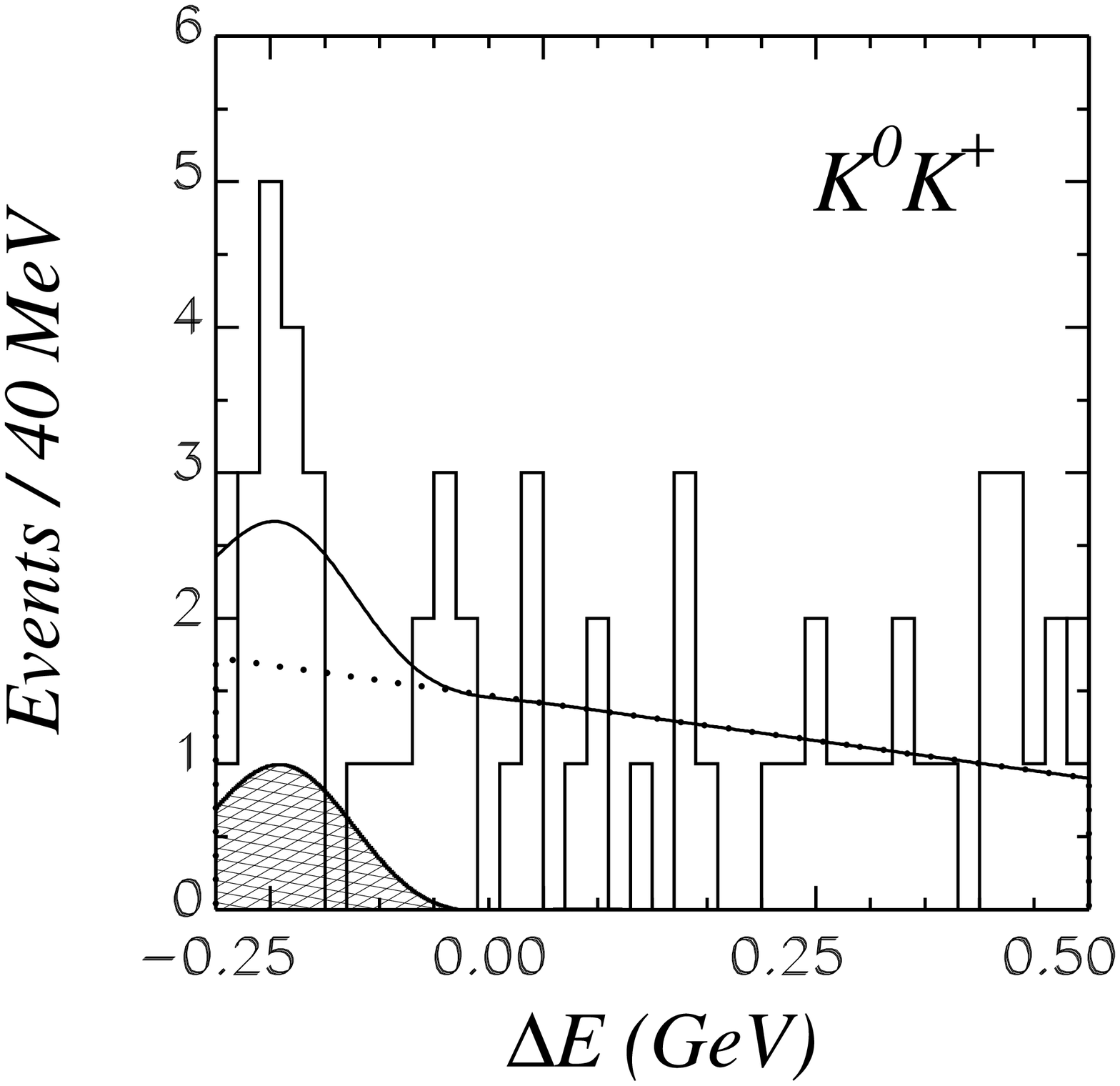}}
\centerline{ \epsfysize 2.0 truein\epsfbox{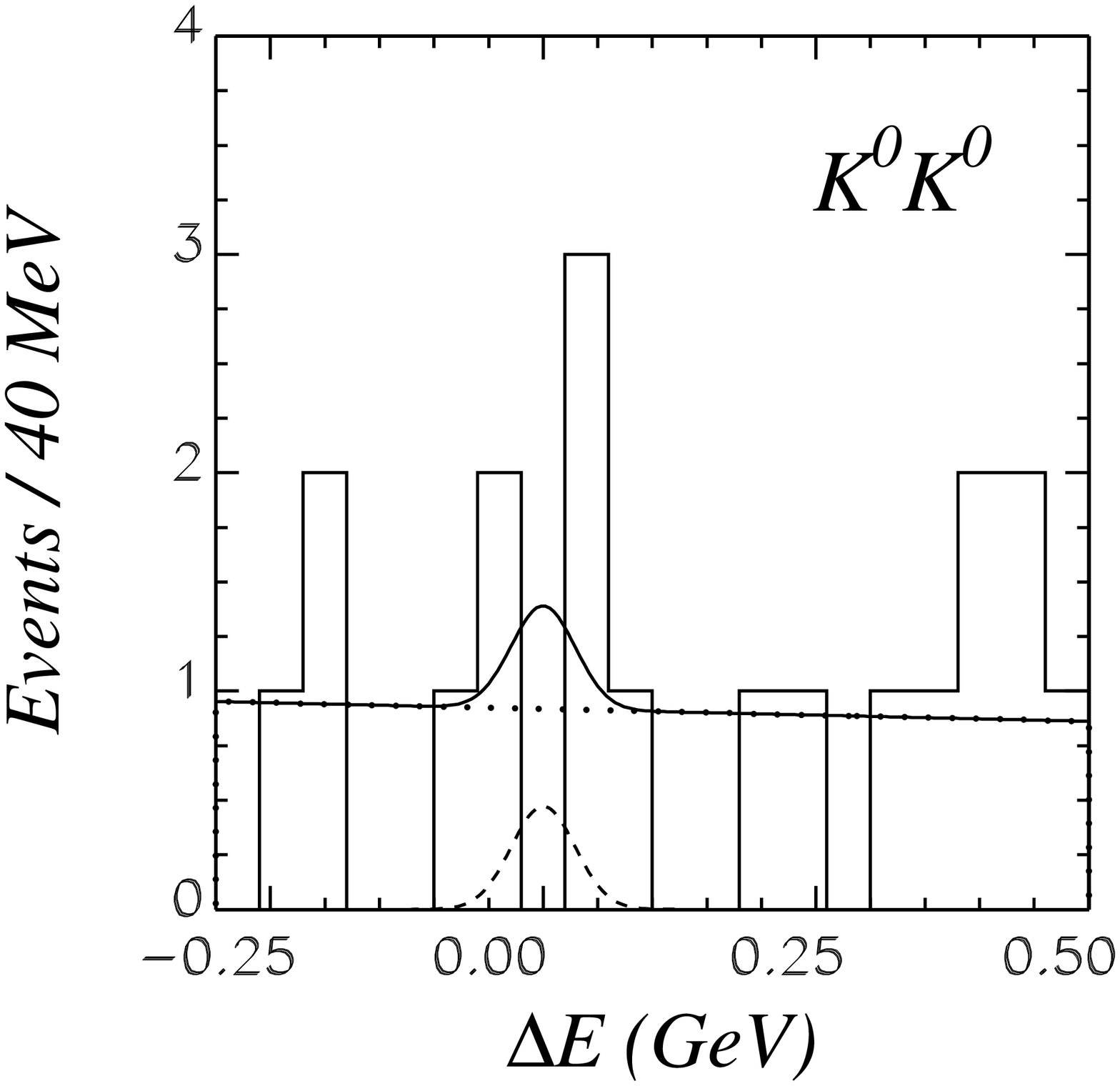} }
\caption{The $\Delta E$ fits to the
$B\rightarrow K\bar{K}$ event samples. The sum of the signal and
background functions is shown as a solid curve, the dotted curve
represents the continuum background. In the $B^0\rightarrow K^+K^-$
distribution, the dot dashed curve represents the $K^+\pi^-$ crossfeed. In the $B^+\rightarrow K^0_SK^+$
distribution, the hatched histogram represents the charmless $B$
background. In the $B^0\rightarrow K^0_SK^0_S$
distribution, the dashed curve represents the signal component.  }
\label{fig:kkres}
\end{figure}

Figures~\ref{fig:kpires}, ~\ref{fig:pipires}, and ~\ref{fig:kkres}
show the $\Delta E$ and $m_{bc}$ distributions for the $K\pi$,
$\pi\pi$ and $K\bar{K}$ final states, respectively.  The $\Delta E$ signal
yields and the significance above background are listed in
Table~\ref{tab:bfresults}~\cite{bib:errors}.  Using these results and the efficiencies
listed in Table~\ref{tab:efficiency}, we derive the branching
fractions listed in Table~\ref{tab:bfresults} based on the data
sample of $31.7$ million $B\bar{B}$ events~\cite{bib:fpmfoo}.  
In all cases, the
$m_{bc}$ fits give consistent results.  The systematic error in the
fitting procedure is determined by varying the parameters of the
fitting functions within their errors and measuring the change in the
signal yield.  The deviations
from the nominal yields are typically $1$ to $2$ events.
These deviations, along with the error in the efficiencies and
$N(B\bar{B})$ ($1\%$) are added in quadrature to give the systematic
error in the branching fractions.  For modes with significance below
$3\sigma$~\cite{bib:kspi0}, we report $90\%$ confidence
level upper limits~\cite{bib:likelihood} calculated with the
efficiency and $N(B\bar{B})$ reduced by their
systematic errors.

For the $K^+\pi^0/\pi^+\pi^0$ simultaneous fit, we re-fit the distributions after
removing the constraints.  The central values of the signal yields
differ by at most $2.3$ events.  These deviations are also included in
the systematic error.

The effects of backgrounds from charmless $B$ decayscan be demonstrated by
re-fitting the $\Delta E$ distributions in the region $\Delta E > -130$
MeV without a charmless $B$ background
component.  For modes with
$\pi^0$s in the final state, the yields deviate by as much as $12\%$,
clearly indicating the need for these components in the fit. 
There is almost no deviation in the yields of final states that do not
include a $\pi^0$.
In each fit, the measured charmless $B$ background yield agrees with
the expected values based on independent measurements of these
modes. However, the errors on the fitted yields and the expected
values are both large.

In Table~\ref{tab:partial}, we list ratios of partial widths among the
$K\pi$ and $\pi\pi$ modes~\cite{bib:fpmfoo}.  
The correlations between the numerator and
denominator are included in the systematic error calculation.  The
systematic error includes a $2.5\%$ fractional error from the ratio of charged to neutral $B$
meson lifetimes where we have used $\tau^+/\tau^0 = 1.091 \pm 0.027$~\cite{bib:belletime} to
convert the ratio of branching fractions to the ratio of partial widths.

\begin{figure}[htbp]
\centerline{\epsfysize 2.0 truein\epsfbox{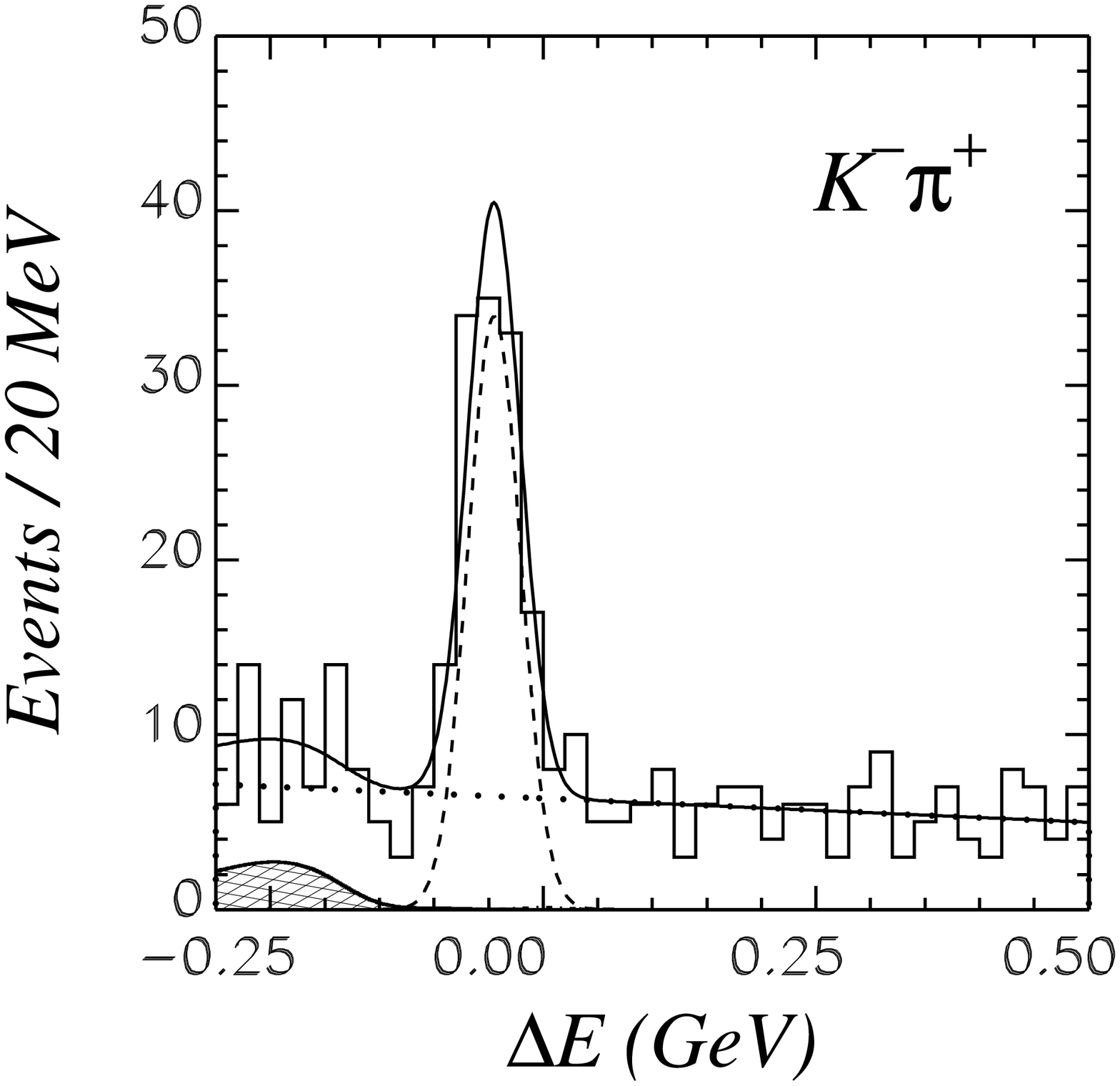}\epsfysize 2.0 truein\epsfbox{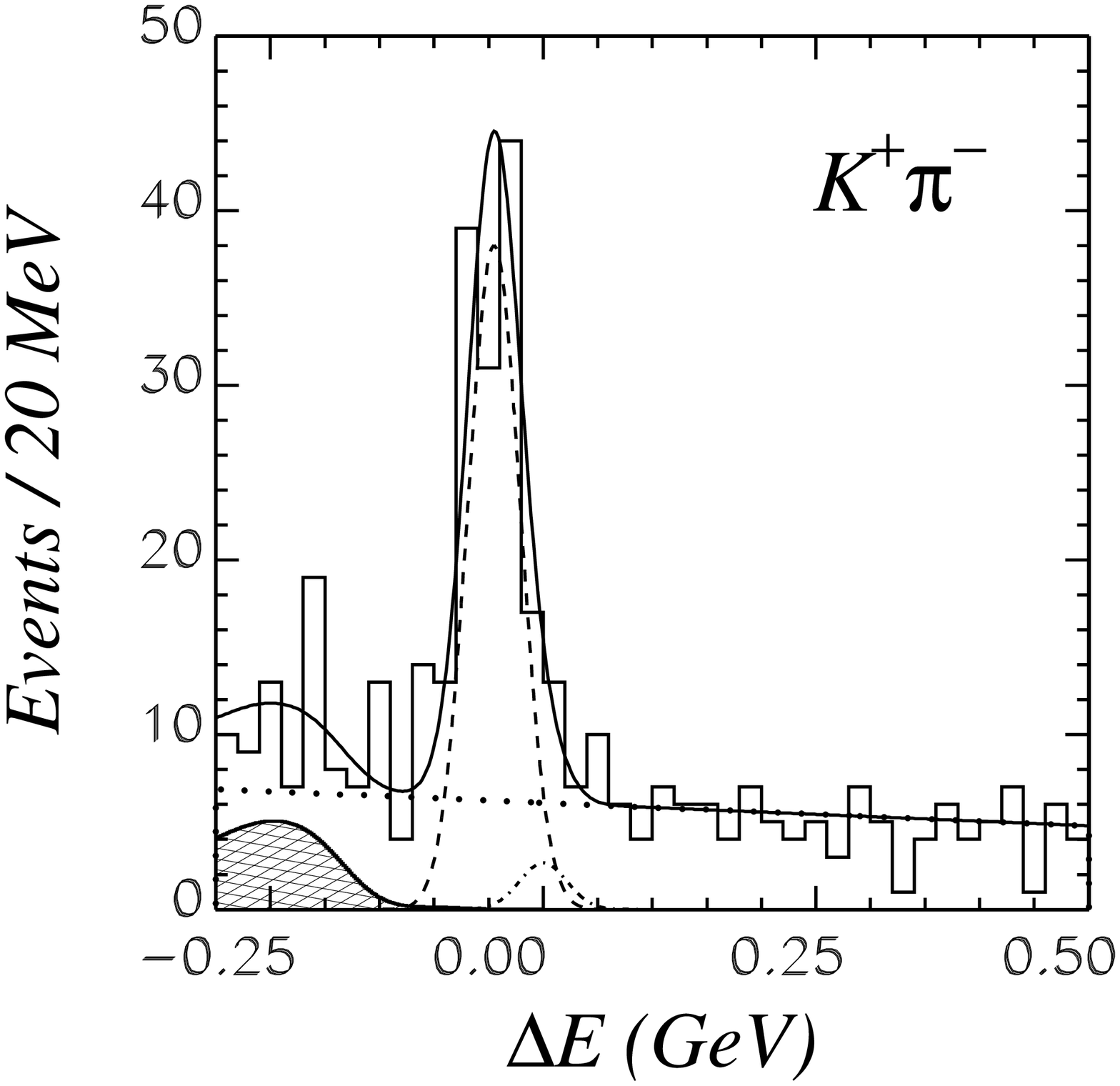} }
\centerline{\epsfysize 2.0 truein\epsfbox{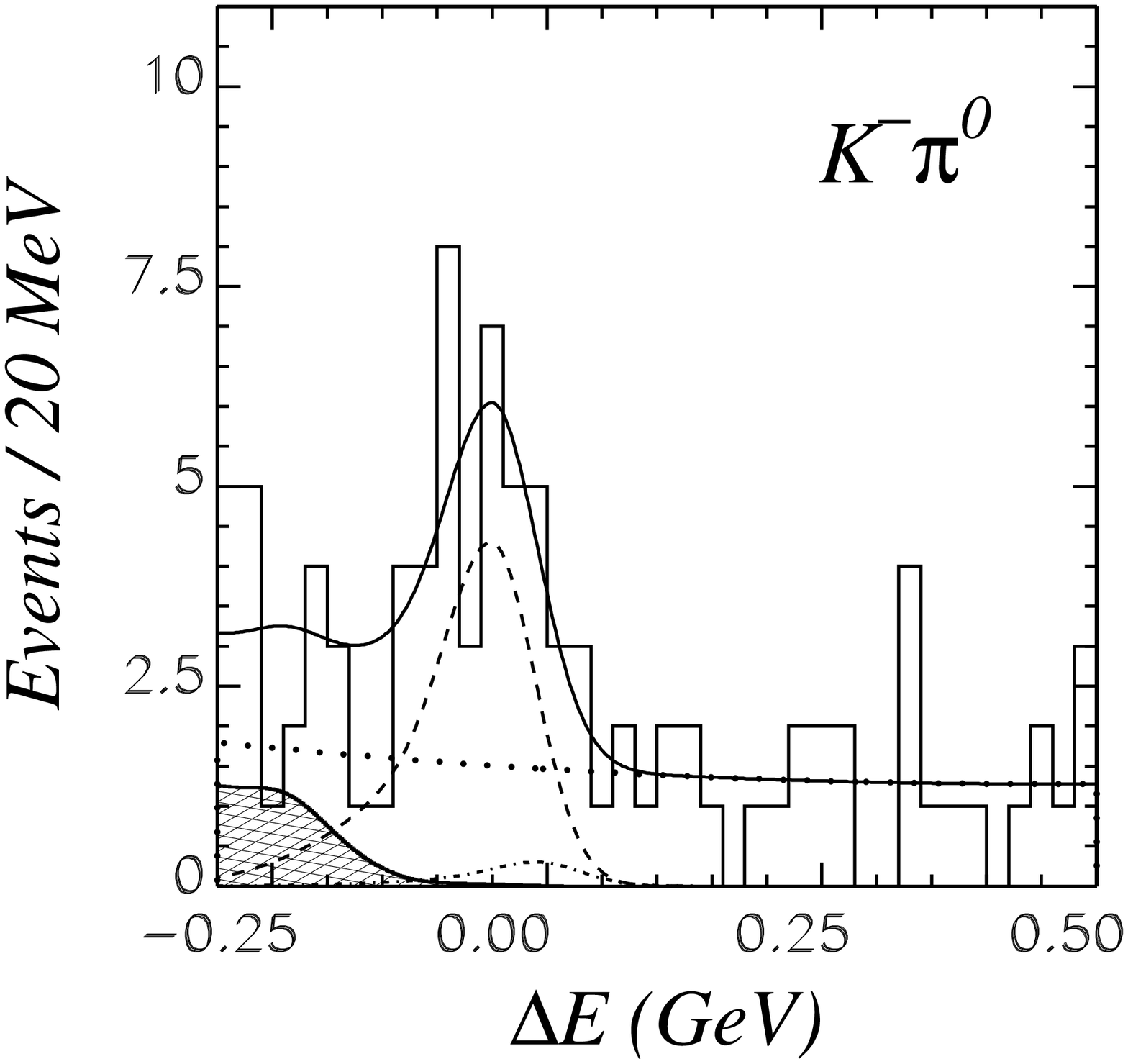}\epsfysize 2.0 truein\epsfbox{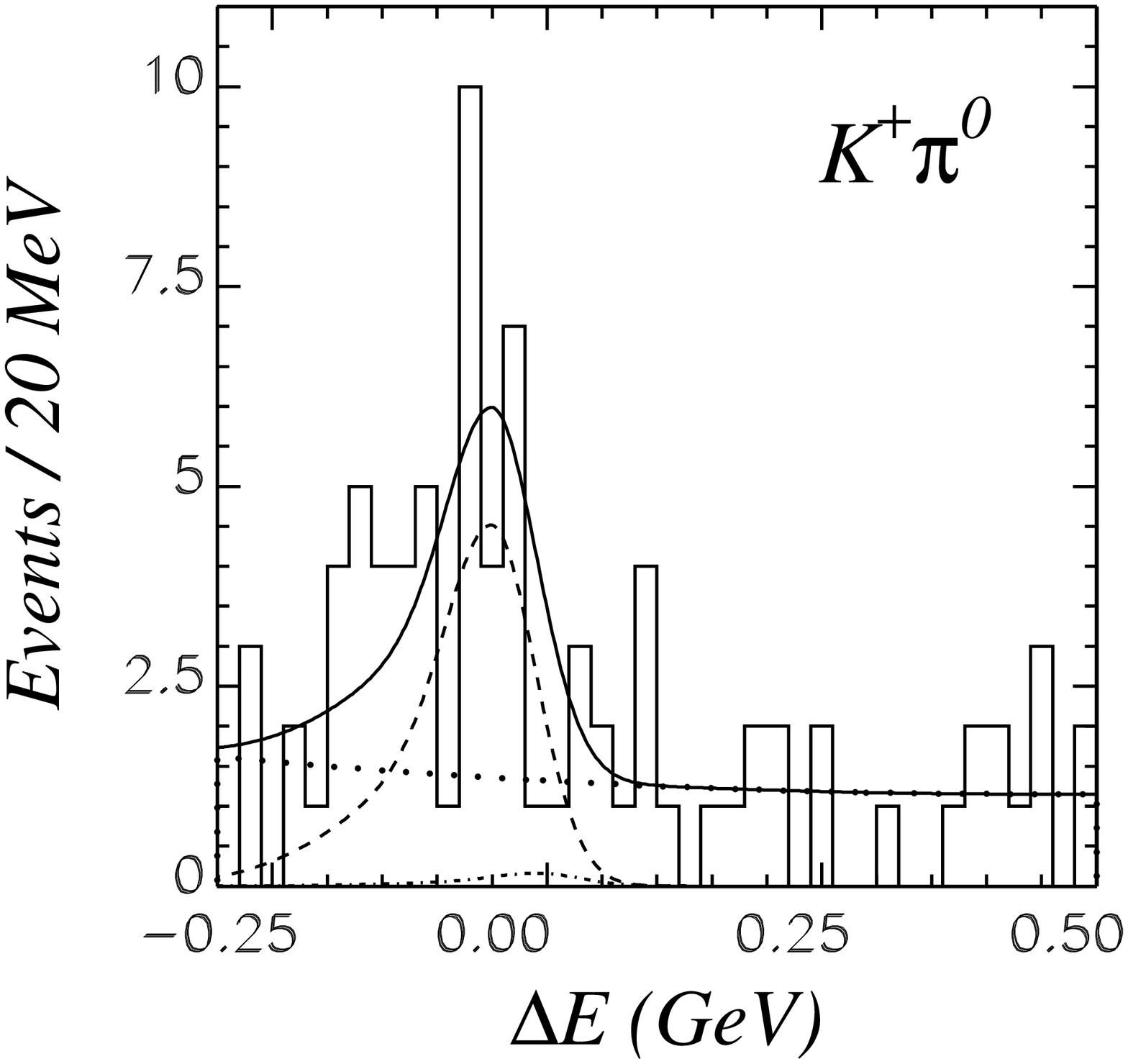} }
\centerline{\epsfysize 2.0 truein\epsfbox{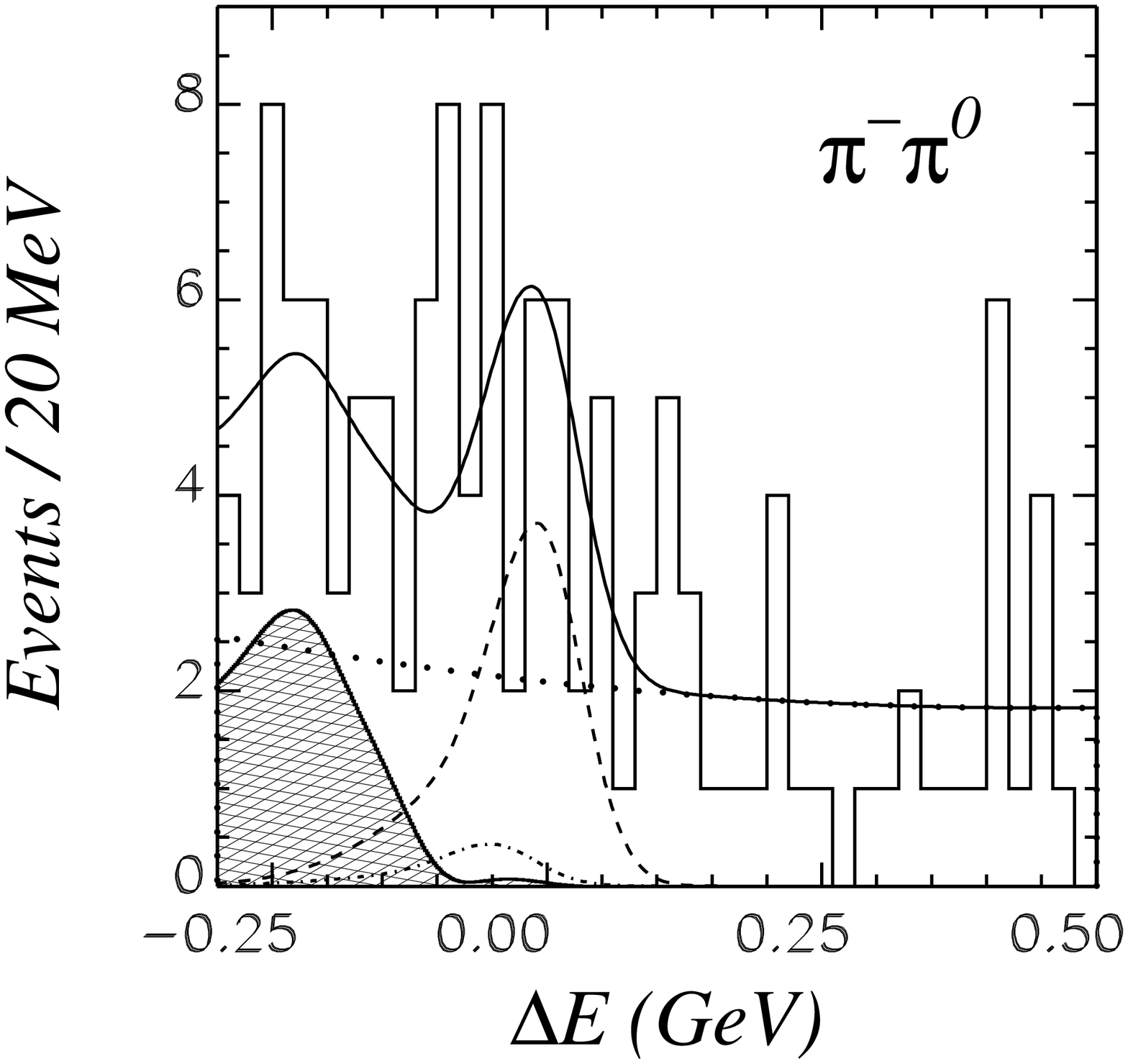}\epsfysize 2.0 truein\epsfbox{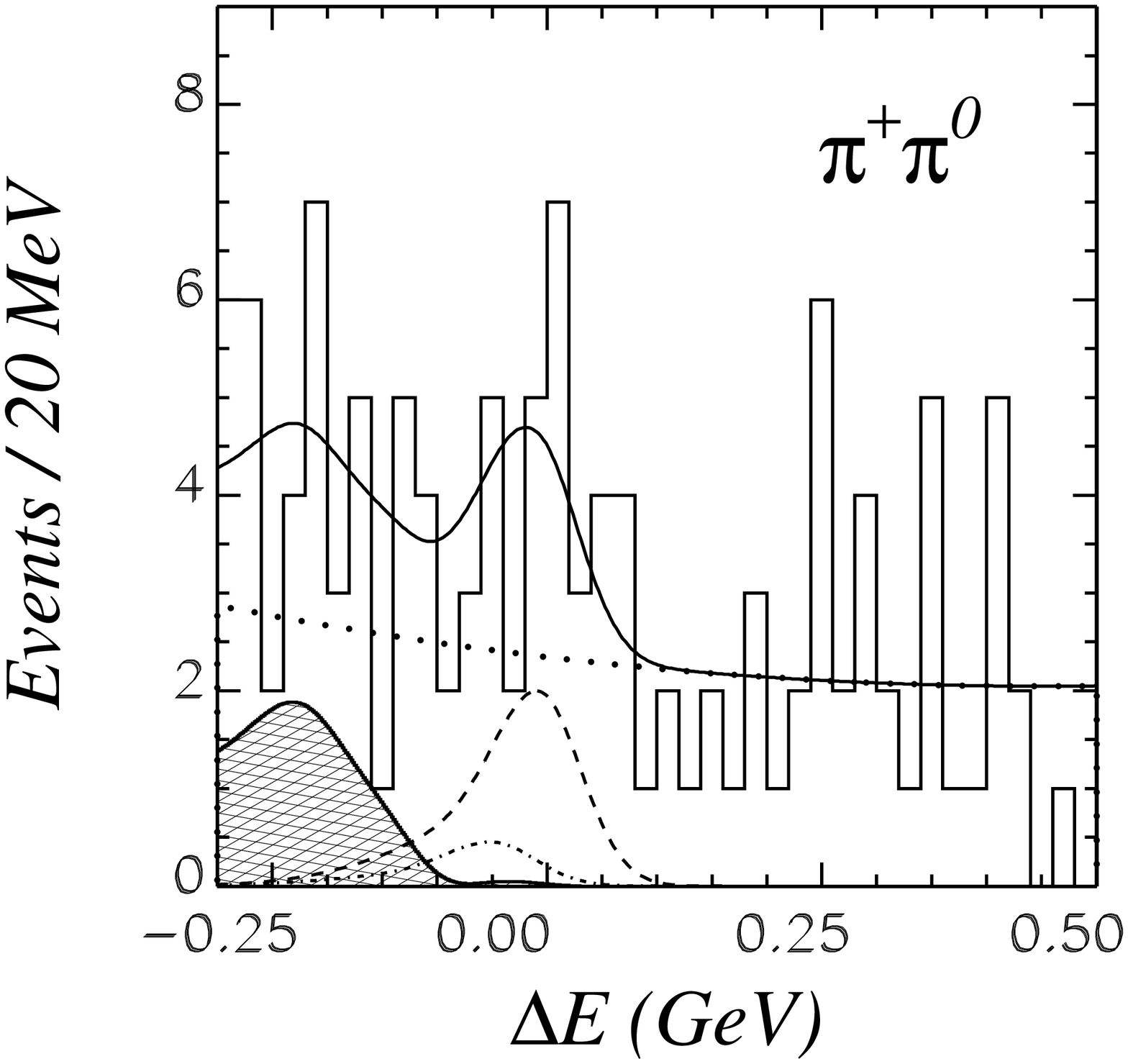} }
\caption{The $\Delta E$ distributions for $\bar{B}$ (left) and $B$
(right)  candidate decays to flavor-specific final states.   
The sum of the signal and
background functions is shown as a solid curve. 
The dashed curve represents the signal component, the
dotted curve represents the continuum background, and the hatched
histogram represents the charmless $B$ background component.   
The crossfeed components
are shown by dot-dashed curves
centered $45$ MeV from the signal components.}
\label{fig:acp}
\end{figure}

\begin{figure}[htbp]
\centerline{\epsfysize 2.0 truein\epsfbox{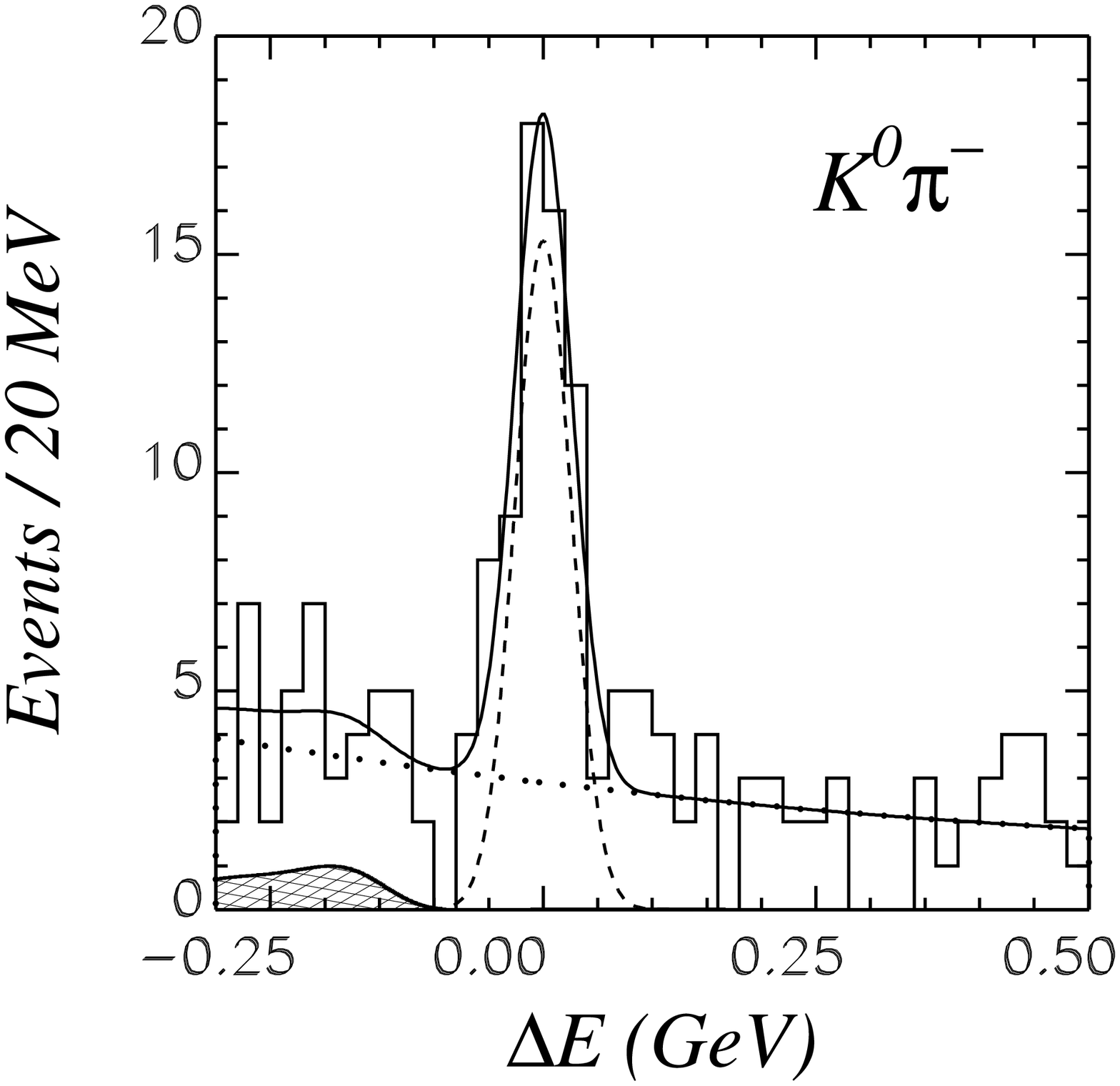}\epsfysize 2.0 truein\epsfbox{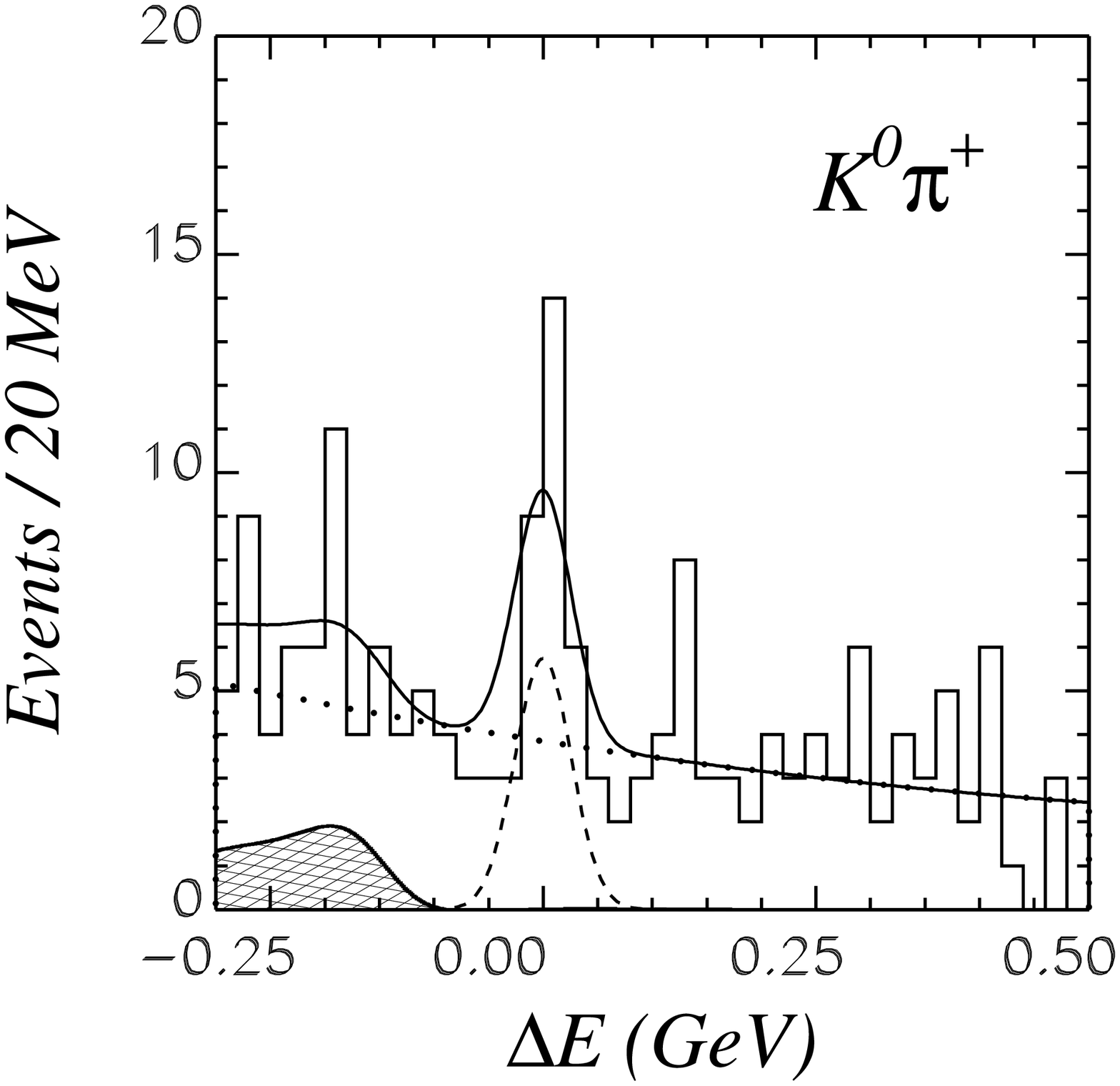} }
\centerline{\epsfysize 2.0 truein\epsfbox{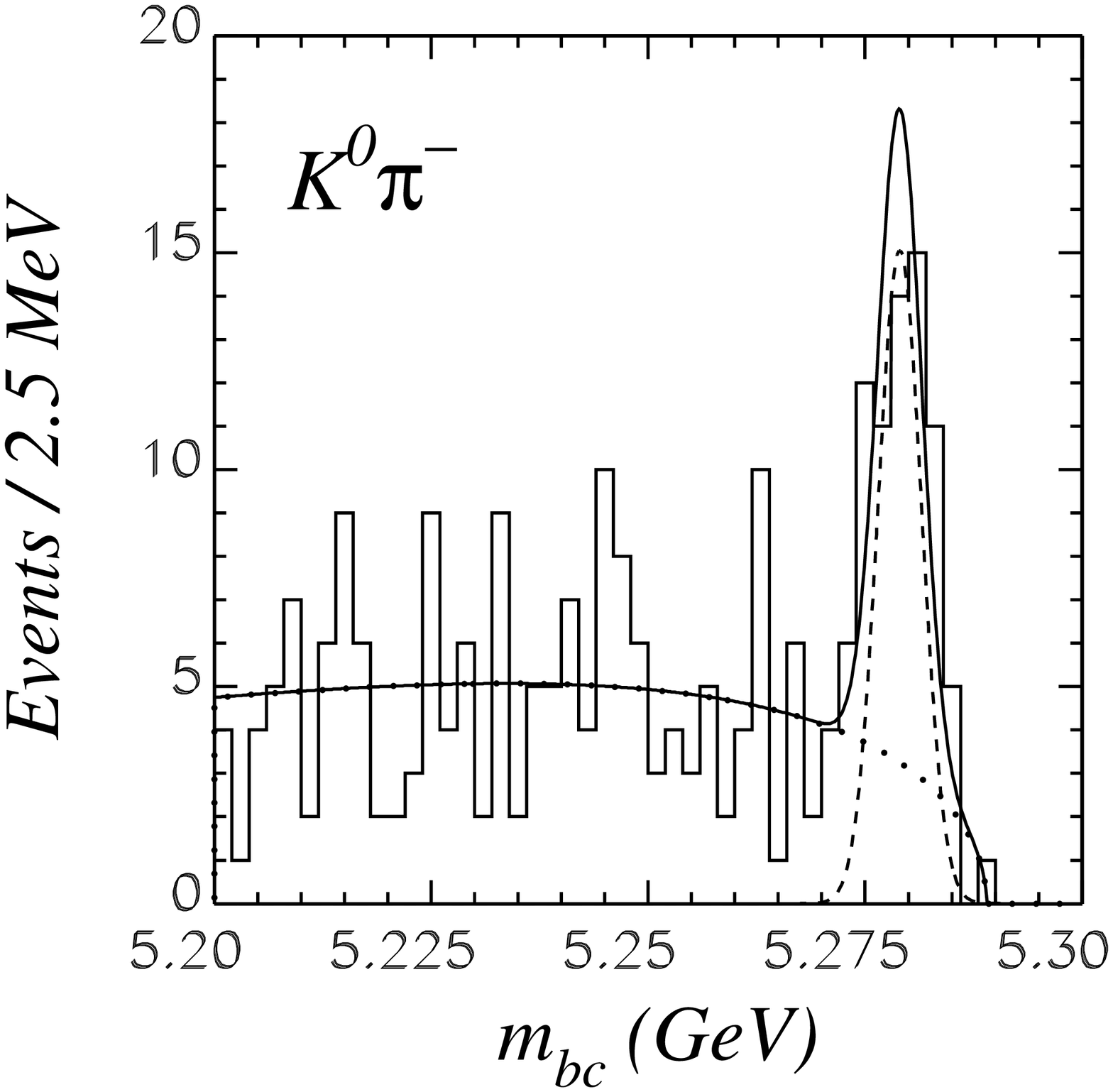}\epsfysize 2.0 truein\epsfbox{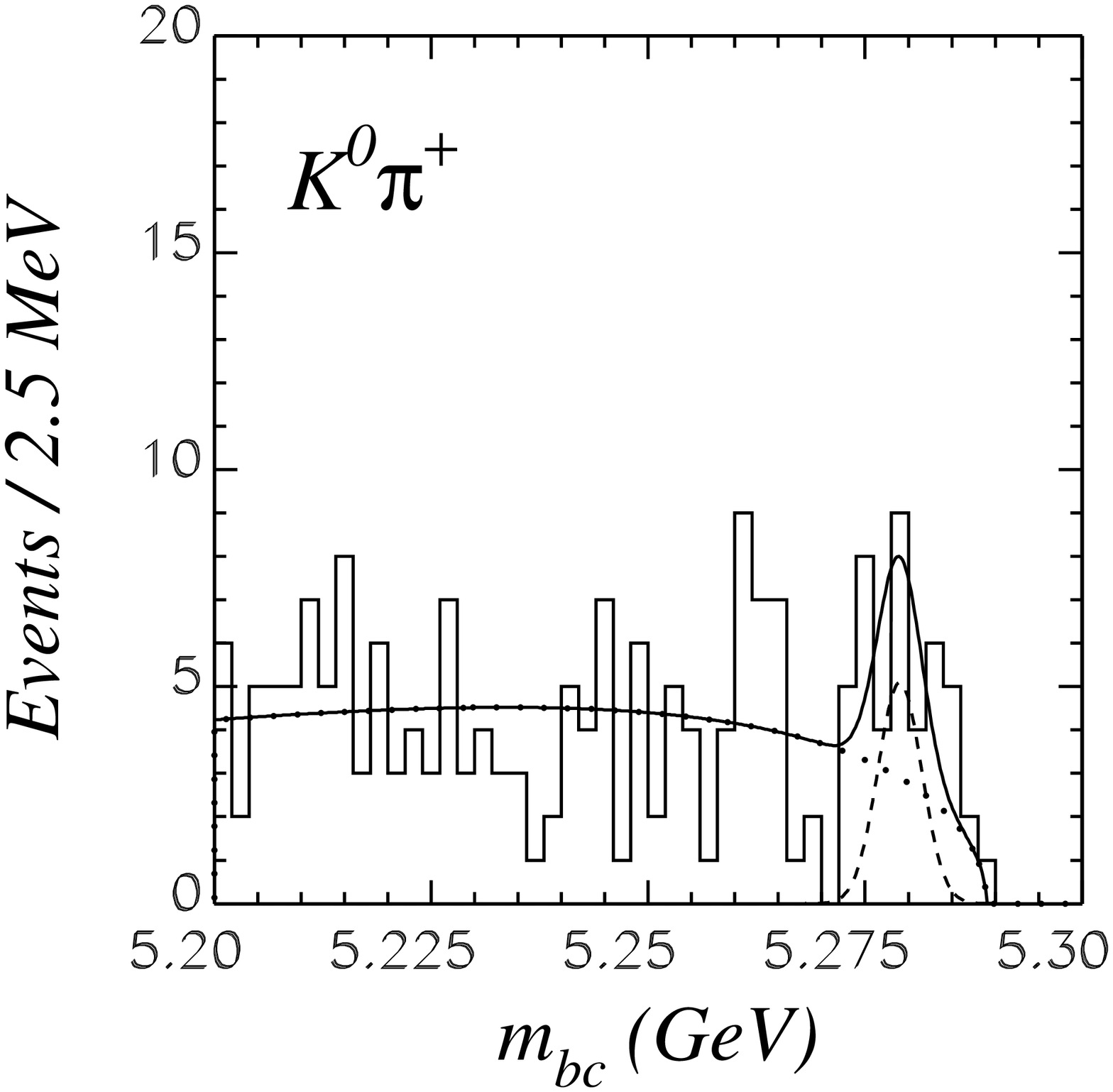} }
\caption{Distributions for $B^\mp \rightarrow
K^0_S \pi^\mp$.  The top row contains $\Delta E$ distribution for $\bar{B}$ (left) and $B$
(right) candidates.  The corresponding $m_{bc}$ distributions are
shown in the bottom row.
The sum of the signal and
background functions is shown as a solid curve. 
The dashed curve represents the signal component, the
dotted curve represents the continuum background, and the hatched
histogram represents the charmless $B$ background component.   
}
\label{fig:kspiacp}
\end{figure}

We measure partial-rate asymmetries in all measured flavor-specific modes.  
The $\Delta E$ distributions are shown separately for ${\bar B}$ and
$B$ modes for the $K^\mp\pi^\pm$, $K^\pm\pi^0$, and $\pi^\mp\pi^0$
final states in Fig.~\ref{fig:acp}. Figure~\ref{fig:kspiacp} shows
the corresponding distributions for the $K^0_S\pi^\mp$ final states.  The fitting results and 
partial-rate asymmetries are listed in Table~\ref{tab:acpres}.  Here,
the $90\%$ confidence intervals assume Gaussian statistics and are
expanded linearly by the systematic error.  The systematic errors are dominated
by fitting systematics but also include a $1\%$ contribution, added in
quadrature, to account for possible detector-based asymmetries, as discussed below.
In the $K^\mp\pi^\pm$ final
states, the asymmetry is corrected by a dilution factor of $0.984\pm
0.001$, due to double misidentification of $K^+\pi^-$ as $K^-\pi^+$.

\begin{figure}[htbp]
\centerline{\epsfysize 2.0
truein\epsfbox{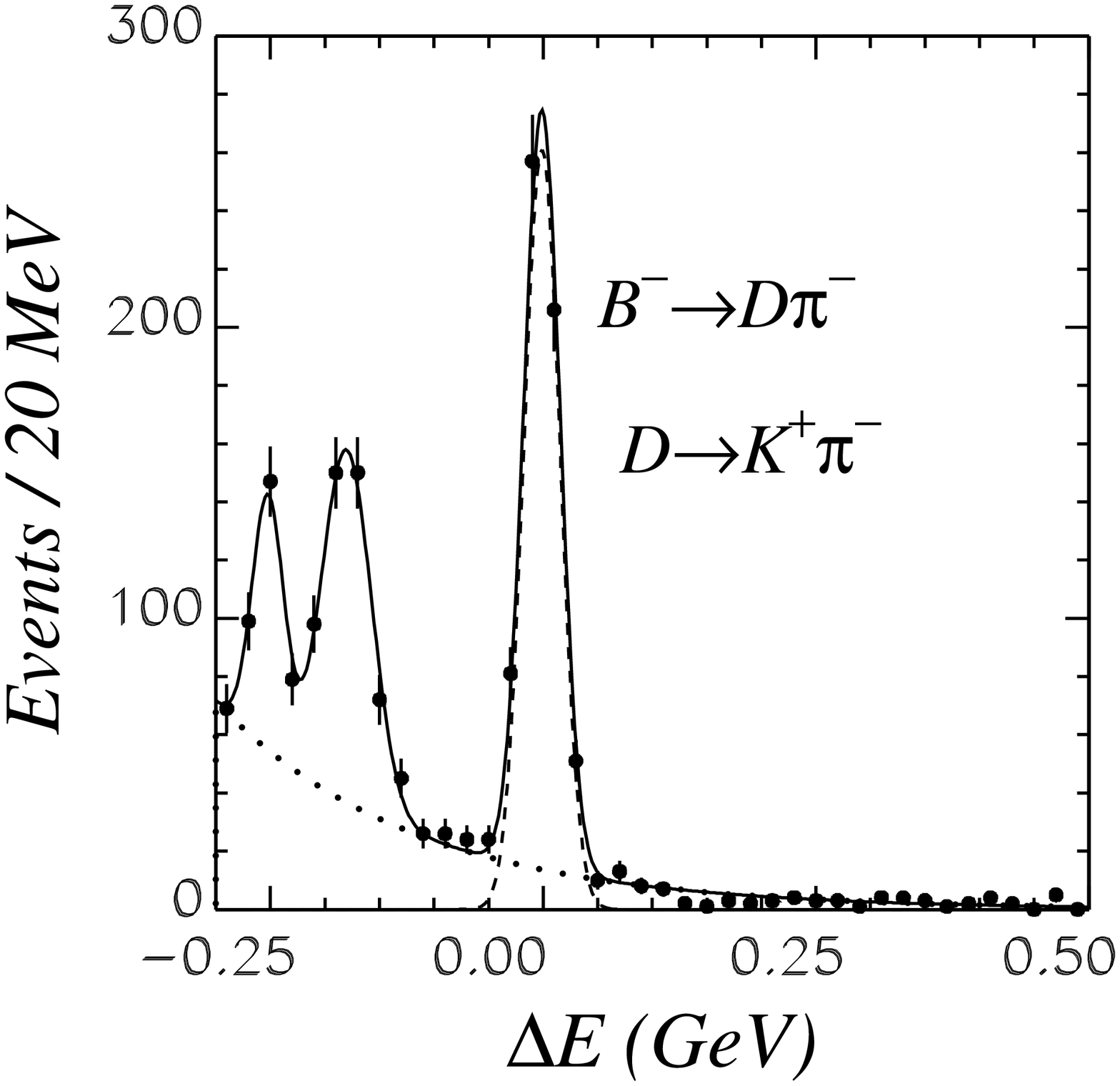}\epsfysize 2.0
truein\epsfbox{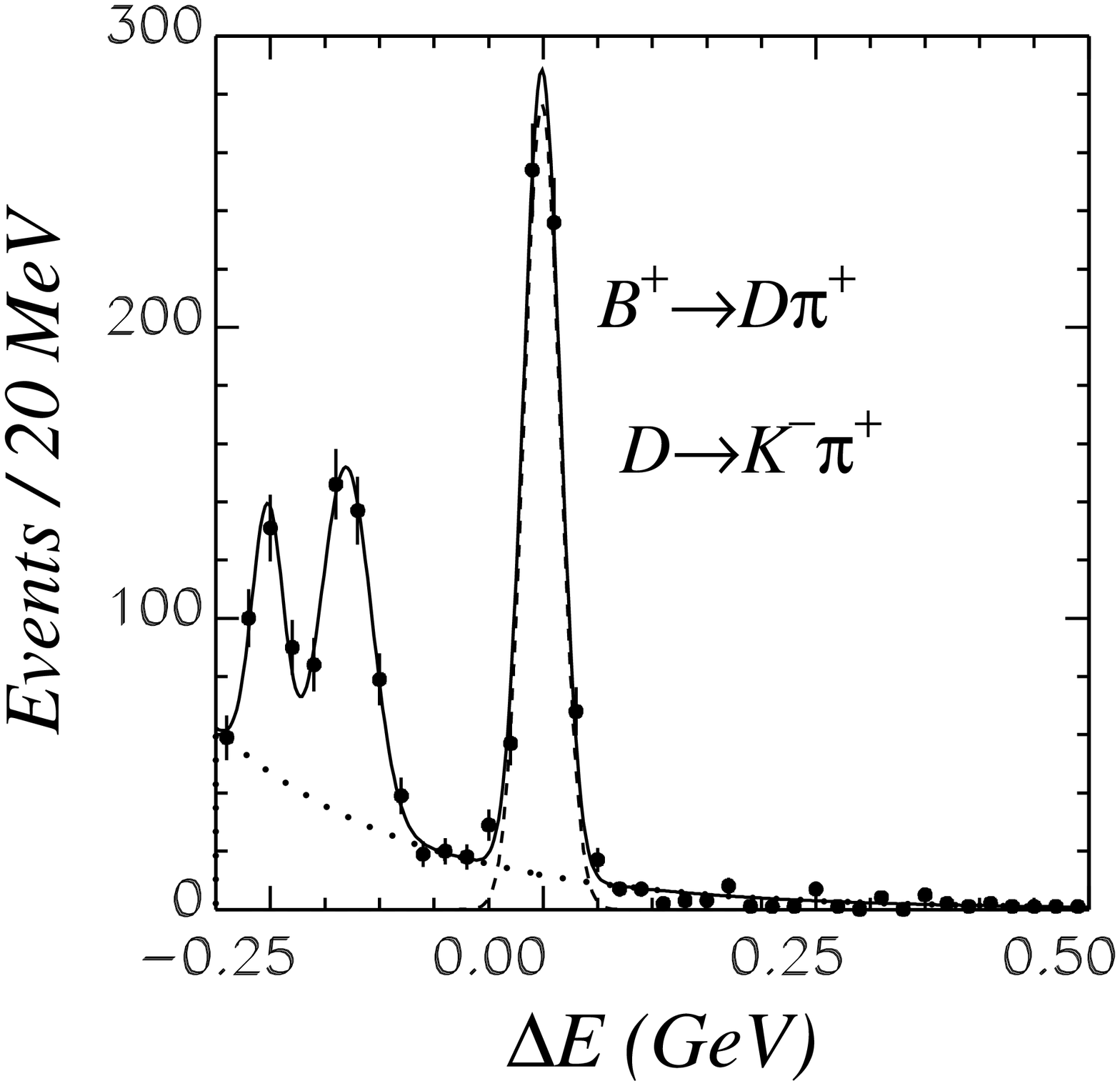} }
\centerline{\epsfysize 2.0 truein\epsfbox{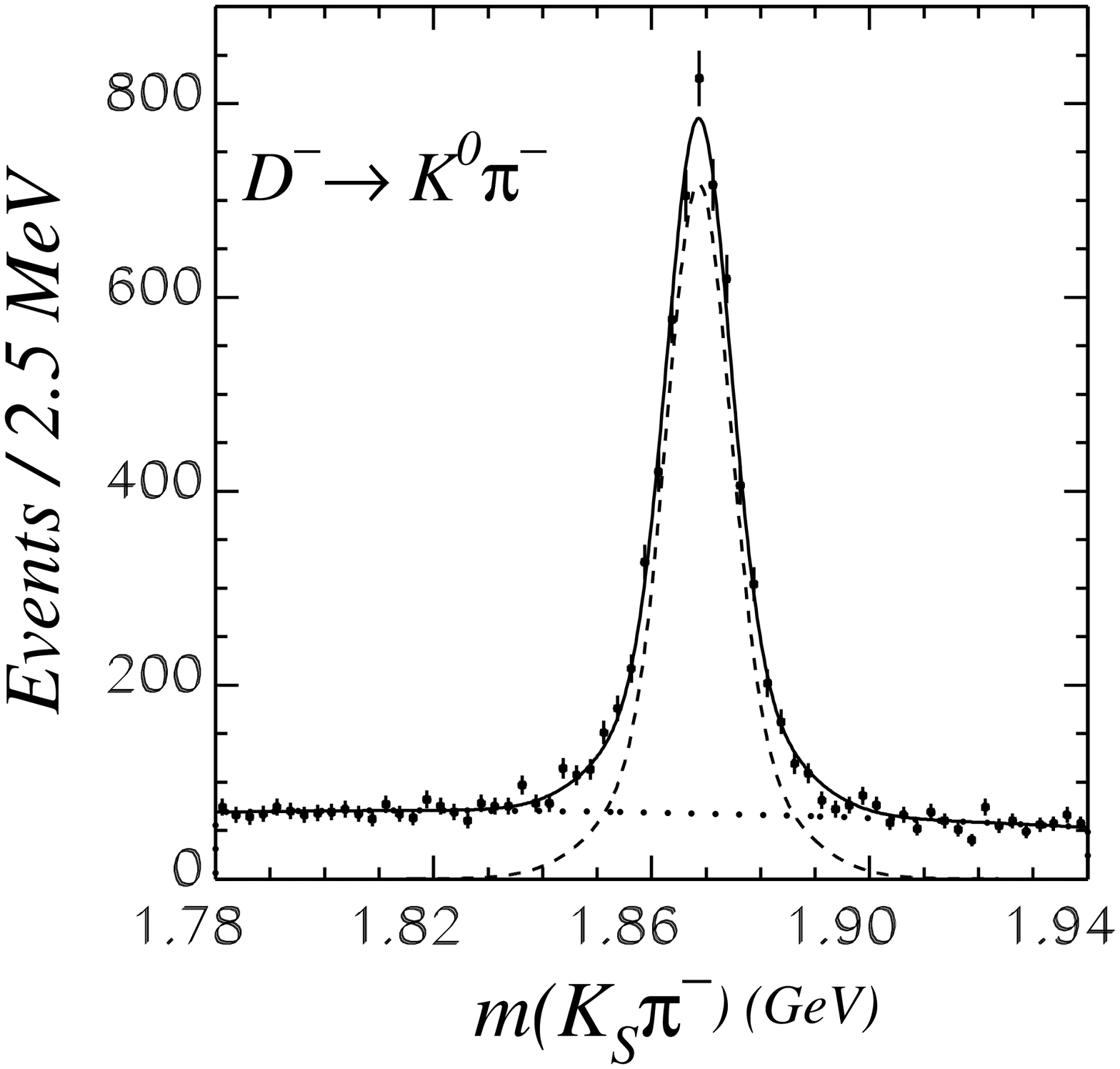}\epsfysize 2.0 truein\epsfbox{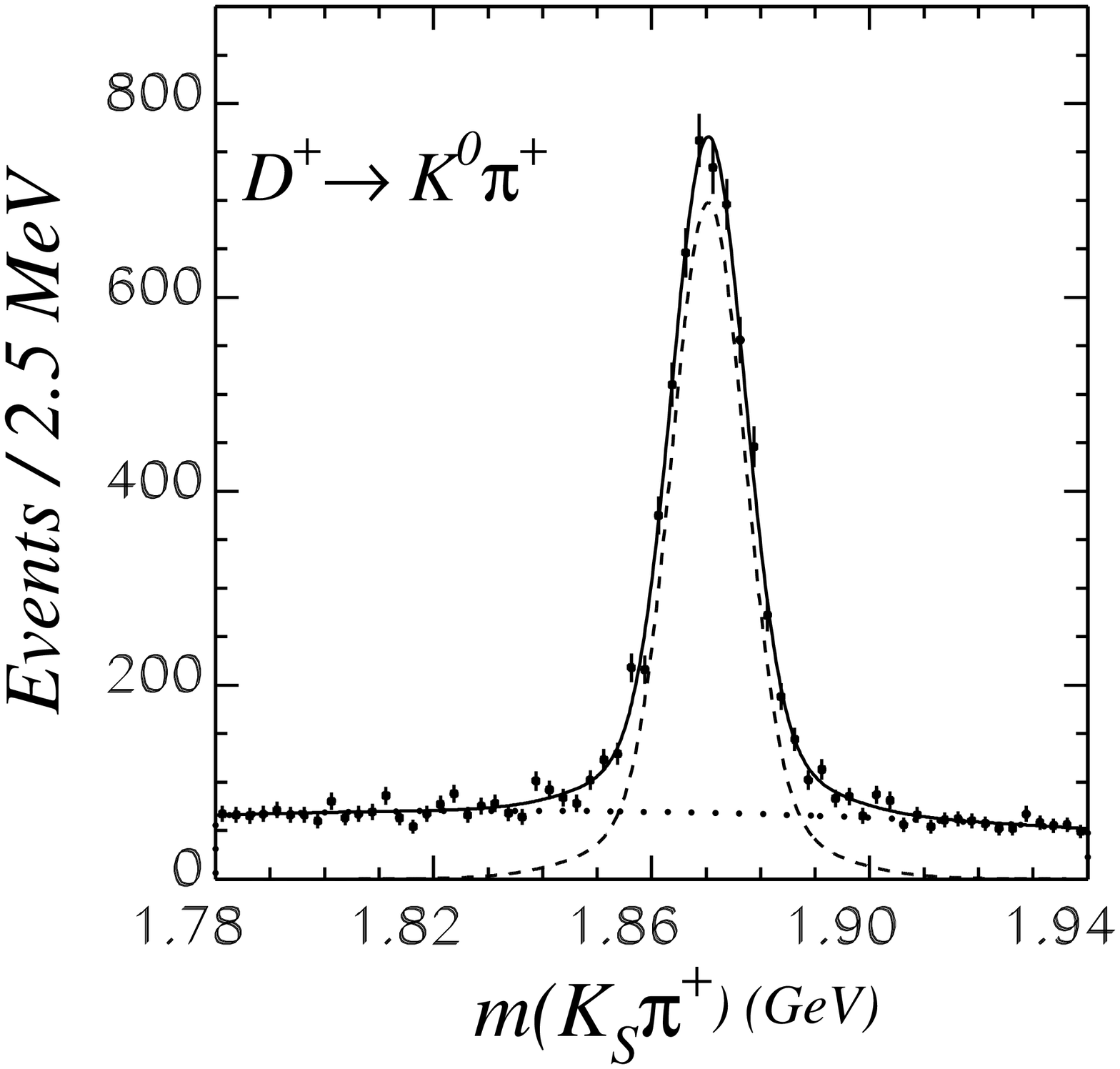} }
\caption{Control samples for the $B^\mp \rightarrow
K^0_S \pi^\mp$ ${\cal A}_{CP}$ measurement.  
The top row contains the $\Delta E$ distributions for $B^-\rightarrow
\bar{D}^0\pi^-$, $\bar{D}^0\rightarrow K^+\pi^-$ on the left and the
charge conjugate decay on the right after applying identical PID and
continuum suppression cuts as for the $K^0_S\pi^\mp$ analysis. 
The bottom row contains inclusive $D^-\rightarrow K^0_S\pi^-$ (left)
and $D^+\rightarrow K^0_S\pi^+$ (right) mass spectra after
making momentum cuts on the $D$ daughter particles to simulate the momentum of $B$
daughter particles.}
\label{fig:kspiacptest}
\end{figure}

Four samples are used to verify the symmetric performance of the Belle
detector for high momentum particles. An inclusive sample of tracks
in the two-body decay momentum bin $2.4$ GeV$ < p < 2.85$ GeV is used for
tracking efficiency tests before and after PID cuts are applied.  
Events in $m_{bc}$ sideband data further
test the reconstruction efficiency along with the continuum suppression
cut efficiency.  We also check the difference between inclusive high momentum $D^{0(+)}
\rightarrow K^-\pi^+$, $K^0_S\pi^+$, and $K^-\pi^+\pi^0$
decays and their charge conjugates that test for asymmetries in the detector
resolution. The entire reconstruction procedure is applied
to the $B\rightarrow D\pi^\mp$; 
$D \rightarrow K^\mp\pi^\pm$, $K^0_S\pi^\mp$, and
$K^\mp\pi^\pm\pi^0$ data samples.  
The results are listed in
Table~\ref{tab:acptest}.  The inclusive track sample yields an
asymmetry in the track reconstruction efficiency of $(N(h^-) - N(h^+))/(N(h^-) + N(h^+)) = (-3.6 \pm
0.3)\times 10^{-3}$.  Considering the statistical precision of the
current data set, we ignore this very small asymmetry.  We also see a
$1.8$ MeV shift between the inclusive $D^-\rightarrow K^0_S\pi^-$ and
$D^+\rightarrow K^0_S\pi^+$ mass peaks indicating an approximate
$0.1\%$ momentum scale difference between positive and negative
tracks.
This shift has been taken
into account when determining the error in the peak position of the
$\Delta E$ distributions for modes with odd numbers of tracks in the
final state.  Furthermore, as shown in the Table~\ref{tab:acptest}, the shift does not
result in an asymmetry between the efficiency to reconstruct the two
flavors.  
All other mass and width parameters are consistent within
the errors between the two flavors for the three $D$ decay channels. 
Figure~\ref{fig:kspiacptest} shows the samples most relevant to the $B^\mp
\rightarrow K^0_S\pi^\mp$ final states.
With the exception of the inclusive track sample, 
all are consistent with zero asymmetry within the error and we
conclude that detector based asymmetries are below $1\%$.

\section{Discussion and Conclusions}

We have presented measurements of the branching fractions of all
$B\rightarrow K\pi$ final states and the $B\rightarrow \pi^+\pi^-$ and
$\pi^+\pi^0$ final states.  We see no significant evidence for the decays
$B^0\rightarrow \pi^0\pi^0$ or $B\rightarrow K\bar{K}$ and set $90\%$
confidence level
upper limits on their branching fractions.  Furthermore, we see no
clear evidence for partial-rate asymmetries between the $\bar{B}$ and
$B$ decay amplitudes in these modes.  

The partial-rate asymmetry
between $B^-\rightarrow K^0\pi^-$ and $B^+\rightarrow K^0\pi^+$ of
$0.46 \pm 0.15 \pm 0.02$ has a non-zero significance of $2.9\sigma$.
Since this is below $3\sigma$, we defer
claiming evidence for an asymmetry until a larger data sample is collected.
 
The $\Gamma(\pi^+\pi^-)/\Gamma(K^+\pi^-)$ and
$\Gamma(\pi^+\pi^-)/2\Gamma(\pi^+\pi^0)$ partial width ratios are
significantly below
$1$.  This could be an indication of large destructive interference
between Tree and Penguin amplitudes for
the $B^0 \rightarrow \pi^+\pi^-$ decay although theoretical uncertainties
pertaining to the relative sizes of the interfering amplitudes  
are still large~\cite{bib:gronau,bib:beneke,bib:keum}.

Knowledge of the branching fraction for $B^0\rightarrow
\pi^0\pi^0$ is required for an isospin analysis of $B\rightarrow
\pi\pi$ decays and for the extraction of the CP violation parameter
$\sin 2\phi_2$ ~\cite{bib:gronau2}.  
With the present data set we set a limit of
 ${\cal B}(B^0\rightarrow \pi^0\pi^0) <6.4\times 10^{-6}$ 
at the $90\%$ confidence level.  If the excess of $\sim\!13$
events is indeed due to $B^0\rightarrow \pi^0\pi^0$
events, the
corresponding branching fraction would be ${\cal B}(B^0\rightarrow
\pi^0\pi^0) = (3.2\pm 1.5 \pm 0.7)\times 10^{-6}$.

\section{Acknowledgments}

We wish to thank the KEKB accelerator group for the excellent
operation of the KEKB accelerator.
We acknowledge support from the Ministry of Education,
Culture, Sports, Science, and Technology of Japan
and the Japan Society for the Promotion of Science;
the Australian Research Council
and the Australian Department of Industry, Science and Resources;
the National Science Foundation of China under contract No.~10175071;
the Department of Science and Technology of India;
the BK21 program of the Ministry of Education of Korea
and the CHEP SRC program of the Korea Science and Engineering Foundation;
the Polish State Committee for Scientific Research
under contract No.~2P03B 17017;
the Ministry of Science and Technology of the Russian Federation;
the Ministry of Education, Science and Sport of the Republic of Slovenia;
the National Science Council and the Ministry of Education of Taiwan;
and the U.S.\ Department of Energy.

\clearpage

\newpage

\newpage

\newpage

\newpage

\newpage

\begin{table}[htbp]
\caption{Efficiencies to reconstruct the $B\rightarrow hh$ modes.
Listed are the efficiencies for reconstruction, particle
identification, continuum suppression, the final combined efficiencies, and
products of efficiency times intermediate branching fraction.  }
\begin{tabular}{l c c c c c }
Mode & Rec. & PID & $q\bar{q}$ & Final & $\epsilon\times$ B.F. \\
\hline  
\hline
$K^+\pi^-$ & $0.73$ & $0.76$ & $0.55$ & $0.31$ & $0.31$ \\
$\pi^+\pi^-$ & $0.75$ & $0.77$ & $0.52$ & $0.30$ & $0.30$ \\
$K^+K^-$ & $0.71$ & $0.74$ & $0.38$ & $0.20$ & $0.20$ \\
\hline
$K^+\pi^0$ & $0.43$ & $0.86$ & $0.39$ & $0.14$ & $0.14$ \\
$\pi^+\pi^0$  & $0.46$ & $0.88$ & $0.39$ & $0.16$ & $0.16$ \\
\hline
$K^0\pi^+$ & $0.53$ & $0.88$ & $0.68$ & $0.32$ & $0.11$ \\ 
$K^0K^+$ & $0.51$ & $0.86$ & $0.38$ & $0.17$ & $0.06$ \\
\hline
$K^0\pi^0$ & $0.34$ & $1$ & $0.69$ & $0.23$ & $0.08$ \\
\hline
$\pi^0\pi^0$ & $0.31$ & $1$ & $0.41$ & $0.13$ & $0.13$ \\
\hline
$K^0\bar{K}^0$~\cite{bib:kskl} & $0.37$ & $1$ & $0.54$ & $0.20$ & $0.04$ \\
\end{tabular}
\label{tab:efficiency}
\end{table}

\begin{table}[htbp]
\caption{Signal yields, significance above background, and branching
fractions for $B\rightarrow hh$ modes assuming equal production
fractions for neutral and charged $B$ meson pairs.  We report $90\%$ confidence
level upper limits for the $B^0\rightarrow \pi^0\pi^0$ and $B\rightarrow K\bar{K}$ decays.}
\begin{tabular}{lccc}
 & Yield & Sig. & B.F. ($\times
10^{-5}$) \\
\hline
$K^+\pi^-$ & $217.6^{+18.6}_{-17.9} $ & $16.4 $ & $2.25 \pm 0.19 \pm 0.18$ \\
$K^+\pi^0$ & $58.5^{+11.3}_{-10.7} $ & $6.4 $ &
$1.30^{+0.25}_{-0.24} \pm 0.13$\\
$K^0\pi^+$ & $66.7^{+10.8}_{-10.1} $ & $7.6 $ & 
$1.94^{+0.31}_{-0.30} \pm 0.16$ \\
$K^0\pi^0$ & $19.8^{+8.3}_{-7.6} $ & $2.8 $ &
$0.80^{+0.33}_{-0.31}\pm 0.16$ \\
\hline
$\pi^+\pi^-$ & $51.0^{+ 11.6}_{-10.9} $ & $5.4 $ & $0.54 \pm 0.12 \pm 0.05$ \\
$\pi^+\pi^0$ & $36.7^{+11.5}_{-10.8} $ & $ 3.5$ & $0.74^{+0.23}_{-0.22} \pm 0.09$ \\
$\pi^0\pi^0$ & $12.5^{+6.2}_{-5.5} $ & $2.4 $ & $< 0.64 $ \\
\hline
$K^+K^-$ & $0^{+3.2}_{-0} $ & $ 0 $ & $< 0.09 $ \\
$K^0K^+$ & $0^{+2.0}_{-0} $ & $ 0 $ & $< 0.20 $ \\
$K^0\bar{K}^0$ & $0.9^{+2.9}_{-0.9} $ & $ 0 $ & $< 0.41 $ \\
\end{tabular}
\label{tab:bfresults}
\end{table}

\begin{table}[htbp]
\caption{Ratios of partial widths among the various $B\rightarrow
K\pi$ and $\pi\pi$ final states assuming equal production fractions
for neutral and charged $B$ meson pairs.  The ratios of branching fractions are
converted to ratios of partial widths using $\tau^+/\tau^0 = 1.091 \pm
0.027$ [34].}
\begin{tabular}{lc}
$\Gamma_2 /\Gamma_1$ & \\
\hline
$\pi^+\pi^-/K^+\pi^-$ & $0.24^{+0.06}_{-0.05} \pm 0.02$ \\
$2K^+\pi^0/K^0\pi^+$ & $1.34 \pm 0.33 \hskip 1mm^{+0.15}_{-0.14}$  \\
$(\tau^+/\tau^0)K^+\pi^-/K^0\pi^+$ & $1.27^{+0.22}_{-0.23} \pm 0.10$  \\
$K^+\pi^-/2K^0\pi^0$ & $1.41^{+0.56}_{-0.60} \hskip 1mm^{+0.28}_{-0.27} $  \\
$(\tau^+/\tau^0)\pi^+\pi^-/2\pi^+\pi^0$ & $0.40 \pm 0.15 \pm 0.05$  \\
$(\tau^+/\tau^0)\pi^0\pi^0/\pi^+\pi^0$ & $<0.83$ ($90\%$ C.L.)  \\
\end{tabular}
\label{tab:partial}
\end{table}

\begin{table}[htbp]
\caption{Partial-rate asymmetries.
 Listed are the number of signal events for each final state,
the ${\cal A}_{\rm CP}$ values with errors,  and their $90\%$ confidence
intervals, listed on the following line.   
 In the $K^\mp\pi^\pm$ final
states, the asymmetry is corrected for the dilution due to
double mis-identification.}
\begin{center}
\begin{tabular}{l c c c}
Mode & $N(\bar{B})$ & $N(B)$ & ${\cal A}_{\rm CP}$ ($90\%$ C.L.) \\
\hline
$K^\mp\pi^\pm$  & $102.8 \pm 12.6$ & $ 115.0 \pm 13.3$ & 
$ -0.06 \pm 0.09^{+0.01}_{-0.02}$ \\
& & & $-0.21$ : $0.09$ \\
$K^\mp\pi^0$  & $28.7 \pm 7.8 $ & $ 30.1\pm 7.7 $ &
$-0.02 \pm 0.19 \pm 0.02 $\\
& & & $-0.35 $ : $0.30 $
 \\
$K^0_S\pi^\mp$  & $49.5 \pm 8.4 $ & $18.6 \pm 6.3 $ &
$0.46 \pm 0.15 \pm 0.02$ \\
& & & $0.19 $ : $0.72$ 
\\
$\pi^\mp\pi^0$  & $ 24.2\pm 8.4 $ & $ 13.0\pm 7.3 $ & 
$0.30 \pm 0.30^{+0.06}_{-0.04} $\\
& & & $-0.23$ : $0.86 $ 
 \\
\end{tabular}
\end{center}
\label{tab:acpres}
\end{table}

\begin{table}[htbp]
\caption{Tests of detector based asymmetries.  The first row is based
on an inclusive track sample with $2.4$ GeV $< p < 2.85$ GeV in the
center of mass frame.  The following two rows are the asymmetries in
the same sample
after applying particle ID.  The fourth row is the asymmetry for inclusive $D$ meson
decays to high momentum $K^+\pi^-$, $K^+\pi^-\pi^0$, and $K^0_S\pi^+$
final states.  The fifth and sixth rows are the asymmetries in the $m_{bc}$
sideband and the $B^{+(0)}\rightarrow D^{0(-)}\pi^+$ data sample
before and after the $LR$ cut is applied.}
\begin{tabular}{lcc}
test sample &\multicolumn{2}{c}{ ${\cal A}_{\rm CP}$} \\
\hline
high $p$ tracks & \multicolumn{2}{c}{ $(-3.6\pm 0.3)\times 10^{-3}$} \\
with $K$ pid & \multicolumn{2}{c}{ $(-3.2\pm 0.5)\times 10^{-3}$}\\
with $\pi$ pid & \multicolumn{2}{c}{ $(-3.7\pm 0.3)\times 10^{-3}$} \\
\hline
high $p$ $D$ decays & \multicolumn{2}{c}{ $(-2\pm 3)\times 10^{-3}$} \\
\hline
& before $LR$ cut & after \\
\hline
$m_{bc}$ sideband & $(-0.07 \pm 0.17)\times 10^{-3} $& $0.01 \pm 0.05$ \\
$B\rightarrow D\pi^\mp$ & $ -0.045\pm 0.025 $& $-0.055 \pm 0.027$ \\
\end{tabular}
\label{tab:acptest}
\end{table}

\onecolumn


\begin{thebibliography}{99}
\bibitem{bib:c}{ In this Letter we use units with $c=1$ and denote both
mass and momentum with the units of energy.}
\bibitem{bib:kskl}{The efficiencies for the $B^{+(0)}\rightarrow
K^0\pi^{+(0)}$ and $K^0 K^+$ final states are corrected by $1/2$
to account for the approximately equal mixture of $K^0_S$ and $K^0_L$ states.
The $B\rightarrow K^0\bar{K}^0$ final state is corrected
by $1/2$ since the asymmetric $K^0_SK^0_L$ final state is forbidden.}
\bibitem{bib:errors}{All results are presented with the statistical
error first followed by the systematic error.}
\bibitem{bib:fpmfoo}{We assume an equal fraction of charged and
neutral $B$ meson pairs.}
\bibitem{bib:kspi0}{An upper limit is not calculated for the
$B^0\rightarrow K^0_S\pi^0$ mode since this mode was previously
reported as an observation in Ref. 11. }
\bibitem{bib:likelihood}{The $90\%$ confidence level in the signal
yield, $N$, is calculated by the relation $\int_{0}^{N}{{\cal L}(n)dn}
= 0.9\int_{0}^{\infty}{{\cal L}(n)dn}$.}
\end{thebibliography}
\end{document}